\documentclass[a4paper,11pt]{article}
\usepackage{a4wide,amsmath,amssymb,bbm}
\usepackage[english]{babel}

\makeatletter\AtBeginDocument{\let\@elt\relax}\makeatother  %line to resolve bug of Riccardo's compiler

\usepackage{color}
\usepackage{slashed}
\usepackage[normalem]{ulem}
\usepackage{cite}
\usepackage[bottom]{footmisc}
\usepackage{graphicx}

\usepackage[pdfencoding=auto,psdextra]{hyperref}
\usepackage{enumerate}
\hypersetup{
    colorlinks,
    linkcolor={black},
    citecolor={black},
    urlcolor={black}
}

\makeatletter
 \let\old@startsection=\@startsection
 \let\oldl@section=\l@section
 \renewcommand{\@startsection}[6]{\old@startsection{#1}{#2}{#3}{#4}{#5}{#6\mathversion{bold}}}
 \renewcommand{\l@section}[2]{\oldl@section{\mathversion{bold}#1}{#2}}
\makeatother

\makeatletter
\DeclareFontFamily{OMX}{MnSymbolE}{}
\DeclareSymbolFont{MnLargeSymbols}{OMX}{MnSymbolE}{m}{n}
\SetSymbolFont{MnLargeSymbols}{bold}{OMX}{MnSymbolE}{b}{n}
\DeclareFontShape{OMX}{MnSymbolE}{m}{n}{
    <-6>  MnSymbolE5
   <6-7>  MnSymbolE6
   <7-8>  MnSymbolE7
   <8-9>  MnSymbolE8
   <9-10> MnSymbolE9
  <10-12> MnSymbolE10
  <12->   MnSymbolE12
}{}
\DeclareFontShape{OMX}{MnSymbolE}{b}{n}{
    <-6>  MnSymbolE-Bold5
   <6-7>  MnSymbolE-Bold6
   <7-8>  MnSymbolE-Bold7
   <8-9>  MnSymbolE-Bold8
   <9-10> MnSymbolE-Bold9
  <10-12> MnSymbolE-Bold10
  <12->   MnSymbolE-Bold12
}{}

\let\llangle\@undefined
\let\rrangle\@undefined
\DeclareMathDelimiter{\llangle}{\mathopen}%
                     {MnLargeSymbols}{'164}{MnLargeSymbols}{'164}
\DeclareMathDelimiter{\rrangle}{\mathclose}%
                     {MnLargeSymbols}{'171}{MnLargeSymbols}{'171}
\makeatother

\usepackage{mathrsfs}

\usepackage{comment}
\usepackage{braket}

\newcommand{\Tr}{\mathrm{Tr}}
\newcommand{\STr}{\mathrm{STr}}
\newcommand{\be}{\begin{equation}}
\newcommand{\ee}{\end{equation}}
\newcommand{\alg}{\mathfrak}

\newcommand{\Ad}{\operatorname{ad}}
\newcommand{\AD}{\operatorname{Ad}}

\DeclareMathOperator*{\res}{Res}

\makeatletter
\renewcommand{\fnum@figure}[1]{\textbf{\footnotesize Figure~\thefigure :}}

\@addtoreset{equation}{section} \makeatother

\begin{document}
\begin{flushright}\footnotesize\ttfamily
DMUS-MP-22/09
\end{flushright}

\null

\vspace{30pt}

\begin{center}
{\huge{\bf  Semiclassical spectrum of a Jordanian deformation of $AdS_5 \times S^5$ }}

\vspace{80pt}

Riccardo Borsato,$^{a}$ \ \ Sibylle Driezen,$^{a}$ \ \ Juan Miguel Nieto Garc\'ia$\, ^{b}$ \ \ and \ \ Leander Wyss$\, ^{b}$

\vspace{15pt}

{
\small {{\it 
$^a$ Instituto Gallego de F\'isica de Altas Energ\'ias (IGFAE) and
Departamento de F\'\i sica de Part\'\i culas,\\[7pt]
Universidade de  Santiago de Compostela\\
\vspace{12pt}
$^b$ Department of Mathematics, University of Surrey, Guildford, GU2 7XH, UK\\
\vspace{12pt}
\texttt{riccardo.borsato@usc.es, sib.driezen@gmail.com,\\ j.nietogarcia@surrey.ac.uk, l.wyss@surrey.ac.uk}}}}\\

\vspace{100pt}

{\bf Abstract}
\end{center}
\noindent
We study a  Jordanian deformation of the $AdS_5 \times S^5$ superstring that preserves 12 superisometries. It is an example of homogeneous Yang-Baxter deformations, a class that generalises TsT deformations to the non-abelian case. Many of the attractive features of TsT carry over to this more general class, from the possibility of generating new supergravity solutions to the preservation of worldsheet integrability. In this paper, we exploit the fact that the deformed $\sigma$-model with periodic boundary conditions can be reformulated as an undeformed one with twisted boundary conditions, to discuss the construction of the classical spectral curve and its semi-classical quantisation. First, we find global coordinates for the deformed background, and identify the global time corresponding to the energy that should be computed in the spectral problem. Using the curve of the twisted model, we obtain the one-loop correction to the energy of a particular solution, and we find that the charge encoding the twisted boundary conditions does not receive an anomalous correction. Finally, we give evidence suggesting that the unimodular version of the deformation (giving rise to a supergravity background) and the non-unimodular one (whose background does not solve the supergravity equations) have the same spectrum at least to one-loop.

\pagebreak

\setcounter{page}{1}
\newcounter{nameOfYourChoice}

\tableofcontents

%%%%%%%%%%%%%%%%%%%%%%%%%%%%%%%%%%%%%%%%%%%%%%%%%%%%%%%%%%%%%%%%%%%%%%%%
\section{Introduction}

We currently know various examples of deformations of $\sigma$-models that preserve integrability, see~\cite{Hoare:2021dix} for a review. Important representatives are the $\lambda$-deformation~\cite{Sfetsos:2013wia,Hollowood:2014rla,Hollowood:2014qma}, the inhomogeneous deformation (also called $\eta$-deformation)~\cite{Klimcik:2002zj,Klimcik:2008eq,Delduc:2013fga,Delduc:2013qra}, and the class of homogeneous Yang-Baxter deformations~\cite{Kawaguchi:2014qwa,vanTongeren:2015soa}. These are being studied with various motivations, for example to clarify a possible link between integrability and renormalisability of the $\sigma$-models~\cite{Hoare:2019ark,Hoare:2019mcc,Hoare:2020fye}. 
 When these deformations are applied to string $\sigma$-models, they lead to deformations of the corresponding target-space backgrounds. By now, the conditions when they give rise to backgrounds that solve the supergravity equations are  well understood~\cite{Borsato:2016ose,Hoare:2018ngg,Sfetsos:2014cea,Appadu:2015nfa,Borsato:2021gma,Borsato:2021vfy}. Interestingly, they were also reformulated in the language of Double Field Theory (see e.g.~\cite{Sakamoto:2017cpu,Demulder:2018lmj,Catal-Ozer:2019tmm,Borsato:2021vfy}), which in turn can be used to identify their $\alpha'$-corrections, at least to the first order (i.e.~two loops in the $\sigma$-model) in the $\alpha'$-expansion~\cite{Borsato:2020bqo}, see also~\cite{Borsato:2020wwk,Hassler:2020tvz,Codina:2020yma} for the generalisation to other solution-generating techniques.
One reason why they have attracted so much attention is that they can be used to deform a very large class of string backgrounds, including the maximally symmetric ones with $AdS$ factors, which were paramount in the development of the AdS/CFT correspondence. The possibility of deforming such backgrounds (e.g.~$AdS_5\times S^5$ or $AdS_3\times S^3\times T^4$) while preserving the underlying integrability is an exciting direction to identify possible generalisations of the AdS/CFT correspondence, with new gauge theories that are potentially exactly solvable (in the same sense as $\mathcal N=4$ is exactly solvable in the large-$N$ limit). 

In this article, we will focus on homogeneous Yang-Baxter deformations (and for simplicity we will always omit the prefix  ``homogeneous'' from now on). This is a class of integrable deformations that can be understood as a generalisation of TsT deformations (which in the Double Field Theory language would be called $\beta$-shifts) \cite{Osten:2016dvf}. While TsT deformations \cite{Lunin:2005jy,Frolov:2005dj,Frolov:2005ty} can be applied whenever we have two commuting isometries of the ``seed'' undeformed background, Yang-Baxter deformations generalise this possibility to the non-abelian case. The central ingredient in the construction is a constant and antisymmetric $R$-matrix that must solve the classical Yang-Baxter equation on the Lie (super)algebra of isometries. When restricting to supercosets (for example the one of $AdS_5\times S^5$), this family of deformations has a rich list of possibilities, and everything is known about their supergravity embedding. In particular, the deformed background (which in general can have all NSNS and RR fields non-trivially turned on) is a solution of the type IIB supergravity equations if and only if the $R$-matrix solves a very simple and algebraic ``unimodularity'' condition~\cite{Borsato:2016ose,Borsato:2017qsx}.\footnote{See~\cite{Borsato:2018idb} for the generic treatment of a Green-Schwarz string with a semisimple Lie algebra of isometries. Recently, Yang-Baxter deformations have also been  constructed for symmetric space sigma models with non-semisimple symmetry algebra \cite{Idiab:2022dil}.}

On the other hand, not much has been done to extend  the methods of integrability, that were developed for the undeformed $AdS_5\times S^5$ case, to the Yang-Baxter deformed models.\footnote{We stress that in this paper we focus on \emph{homogeneous} Yang-Baxter deformations. For the case of the inhomogeneous deformation, the integrability methods have been pushed to the level of the thermodynamic Bethe ansatz~\cite{Arutyunov:2014wdg} and  Quantum Spectral Curve~\cite{Klabbers:2017vtw}.}
 For the very restricted subclass of ``diagonal'' TsT deformations (i.e.~combinations of TsT deformations that involve only the Cartan isometries of $AdS_5\times S^5$), the solution to the spectral problem is understood in terms of a simple deformation of the Bethe equations and Thermodynamic Bethe Ansatz~\cite{Beisert:2005if,deLeeuw:2012hp} or Quantum Spectral Curve~\cite{Kazakov:2018ugh}. But as soon as one goes beyond this special subclass, various obstructions do not yet allow for the application of the  integrability techniques. This happens already for the rather simple class of ``non-diagonal'' TsT deformations (i.e.~involving at least one non-Cartan isometry)~\cite{Guica:2017mtd}.

A natural strategy to try to extend the methods of integrability to the full class of Yang-Baxter deformations is to do what worked in the case of diagonal TsT deformations, where one replaces the study of the spectral problem of the \emph{deformed} $\sigma$-model with that of an \emph{undeformed} model which, however, has \emph{twisted} boundary conditions on the worldsheet. The two $\sigma$-models (the deformed one and the undeformed yet twisted one) are equivalent on-shell, i.e.~there is a map that relates solutions of the $\sigma$-model equations of motion on the two sides. In the case of TsT deformations, the twisted boundary conditions for the alternative picture are very easy to write down, essentially because they are linear in terms of the fields~\cite{Frolov:2005dj,Frolov:2005ty,Alday:2005ww}. For more generic Yang-Baxter deformations, writing down the equivalent twisted boundary conditions is more complicated. Previous expressions were written in terms of a path-ordered exponential, and the non-localities that this introduces were making it impossible to make any progress \cite{Matsumoto:2015jja,Vicedo:2015pna,vanTongeren:2018vpb}. This problem was solved in~\cite{Borsato:2021fuy}, where the twisted boundary conditions were rewritten in terms of the convenient degrees of freedom, namely those of the twisted model itself. While the twisted boundary conditions can lead to complicated non-linear relations among the fields---a fact which is related to the non-abelian nature of the deformations---the expressions are \emph{local}.

Among the methods of integrability that we can apply, the classical spectral curve  has the right balance of complexity and computational power. The  curve is defined as the $N$-sheeted Riemann surface obtained from the eigenvalue-problem of a monodromy matrix of size $N \times N$ \cite{Babelon:2003qtg}. Its power  was originally applied in the context of string theory in $AdS_5 \times S^5$ by reformulating the construction of classical string solutions as a Riemann-Hilbert problem \cite{Kazakov:2004qf,Beisert:2005bm,Dorey:2006zj,Vicedo:2007rp,Vicedo:2008ryn},  as all their information is encoded in terms of cuts and poles on the Riemann surface. It gained even more traction after it was realised that the classical curve can also be used to obtain 1-loop information related to these classical solutions \cite{Gromov:2007aq,Gromov:2007cd,Gromov:2008ie,Vicedo:2008jy,Gromov:2008ec}. This information is retrieved by adding microscopic cuts and poles to the Riemann surface, which will behave as quantum fluctuations around the classical solution. The classical spectral curve was soon generalised from $AdS_5 \times S^5$ to other $AdS_d$ backgrounds, see \cite{Zarembo:2010yz} and references therein, as well as  deformed backgrounds, such as the flux-deformed $AdS_3 \times S^3$ \cite{Lloyd:2013wza,Babichenko:2014yaa,Nieto:2018jzi} or the Schr\"odinger background obtained from TsT \cite{Ouyang:2017yko}.

In this article, we will use the equivalence between deformed models and undeformed yet twisted model to apply the method of the classical spectral curve and its semiclassical quantisation to  a particular example of a Yang-Baxter deformation of $AdS_5\times S^5$ of non-abelian type. It belongs to the so-called class of Jordanian deformations, which make use of an  $\alg{sl}(2,\mathbb R)$ subalgebra of the Lie algebra of superisometries \cite{Kawaguchi:2014qwa,vanTongeren:2015soa,Hoare:2016hwh,Orlando:2016qqu}. Importantly, the simplest version of Jordanian deformations (i.e. the one needing just the $\alg{sl}(2,\mathbb R)$) is non-unimodular. Therefore, the deformed background does not lead to a solution of the type IIB supergravity equations, but rather to one of the modified supergravity equations of~\cite{Arutyunov:2015mqj,Wulff:2016tju}. It is however possible to cure this problem by constructing ``extended'' Jordanian deformations \cite{vanTongeren:2019dlq} that exploit  a \emph{superalgebra} of isometries containing $\alg{sl}(2,\mathbb R)$. They lead to unimodular $R$-matrices, and to backgrounds that are solutions of the standard type IIB supergravity equations. Notice that the unimodular and non-unimodular versions share the same background metric and Kalb-Ramond field.\footnote{In the unimodular case these are supplemented by a dilaton and RR fluxes that solve the type IIB equations. In the non-unimodular case the RR fields are replaced by fields that do not satisfy the standard Bianchi identities, and the background is supplemented by a non-dynamical Killing vector that enters the modified supergravity equations.}
Interestingly, both cases are closely related to non-abelian T-duality \cite{Orlando:2016qqu,Hoare:2016wsk,Borsato:2016pas}: a non-vanishing deformation parameter can be rescaled away in the background by a simple coordinate transformation, and the solution is equivalent to the background resulting from doing non-abelian T-duality on the two-dimensional Borel subalgebra of $\alg{sl}(2,\mathbb R)$ (or of its supersymmetric extension if considering the unimodular case).

The article is organised as follows. In section~\ref{sec:back}, we review the construction of Yang-Baxter deformations of  semi-symmetric space $\sigma$-models, and we specify the particular Jordanian deformation which we will study. After displaying its deformed metric and Kalb-Ramond field, we identify in section \ref{sec:iso} the residual bosonic isometries that survive. In section \ref{sec:global} we then propose a  coordinate system for which we explicitly identify the time coordinate. We prove that this  system provides global coordinates of the deformed spacetime for every value of the deformation parameter by analysing its geodesic completeness. In section \ref{s:pointlikesolution} we derive a particular pointlike string solution of the deformed $\sigma$-model, which can be interpreted as the analog of the BMN solution of $AdS_5\times S^5$. 

In section \ref{sec:map}, we transfer from the deformed Yang-Baxter picture to the undeformed model with twisted boundary conditions. We review the results of \cite{Borsato:2021fuy} regarding the on-shell equivalence in section \ref{s:reviewtwist} and identify the symmetries of the twisted model in \ref{sec:symmtwisted}. In section \ref{sec:fact}, we show how the twist of the Jordanian model can be simplified. The transformation of our BMN-like solution to the twisted model is then performed in section \ref{sec:map-sol-full}, where we pay particular attention to the gauge ambiguities of the supercoset. Additionally, in this section we analyse also solutions of the twisted model in more generality and show, under certain assumptions, that they lead to  extended string solutions with profiles involving Airy functions.

We start the  program of the classical spectral curve and its quantisation in section \ref{sec:csc} with a review of the construction of the classical curve in terms of the quasimomenta of the (general) twisted  supercoset model on $PSU(2,2|4)$. We show in section \ref{sec:asymptotics} how the asymptotics of the quasimomenta give rise to local Cartan charges, involving the energy, of the symmetry algebra of our  Jordanian twisted model. We then construct the algebraic curve for our  BMN-like solution in section \ref{s:CSC-BMN}. In section \ref{sec:qCSC}, we study the semi-classical quantum corrections to the corresponding BMN-like quasimomenta by applying the recipe of \cite{Gromov:2007aq,Gromov:2008ec} to our twisted case. We identify the frequencies of all possible excitations whose resummation, which we perform in section \ref{sec:one-loop}, gives the one-loop correction to the energy of our string solution. In section \ref{a:anomalousQ} we furthermore show that the charge identifying the twist of the boundary conditions is protected---it does not receive any anomalous correction---at one-loop. Finally, in section \ref{sec:uni}, we discuss how our results regarding the classical spectral curve and its semi-classical quantisation are insensitive to whether we consider the unimodular or non-modular version of our Jordanian deformation.

We end in section \ref{sec:conclusions} with a conclusion and outlook. In appendix \ref{a:embedding}, we show how the global coordinates defined in section \ref{sec:global} are related to the embedding coordinates of (undeformed) $AdS$. For completeness, we discuss  in appendix \ref{a:poincare} why the geodesic incompleteness of the  Poincar\'e coordinate system persists along the deformation. In appendix \ref{a:CSA} we identify the possible Cartan subalgebras of the residual isometry algebra of the deformed model. In appendix \ref{StructureFluctuations} we present the details regarding the calculation of the frequencies of excitations presented in section \ref{sec:qCSC}. We cross-check these results in appendix \ref{QuadraticFluctuations} by computing explicitly, in the picture of the deformed model, the frequencies of the quadratic bosonic fluctuations around the classical solution.

%%%%%%%%%%%%%%%%%%%%%%%%%%%%%%%%%%%%%%%%%%%%%%%%%%%%%%%%%%%%%%%%%%%%%%%%
\section{A Jordanian-deformed background}\label{sec:back}

%%%%%%%%%%%%%%%%%%%%%%%%%%%%%%%%%%%%%%%%%%%%%%%%%%%%%%%%%%%%%%%%%%%%%%%%
\subsection{The deformed $\sigma$-model and the background fields}\label{sec:intro-def}
In order to set up the notation, let us review the construction of the homogeneous Yang-Baxter deformation of a string $\sigma$-model on a semi-symmetric space. The starting point is a Lie supergroup $G$ with a corresponding Lie superalgebra $\alg g$ that admits a $\mathbb Z_4$-graded decomposition, i.e.~$\alg g=\oplus_{i=0}^3 \alg g^{(i)}$ such that $[[\alg g^{(i)},\alg g^{(j)}]]\subset \alg g^{(i+j \text{ mod } 4)}$. Here we are using $[[\cdot,\cdot]]$ to denote the graded bracket on $\alg g$. The action of the string $\sigma$-model on the supercoset $G\setminus G^{(0)}$ can be written as 
\begin{equation}\label{eq:S-0}
S_0 = -\frac{\sqrt{\lambda}}{4\pi}\int d\tau d\sigma \ \Pi_{(-)}^{\alpha\beta}\  \STr \left( J_\alpha \ \hat d J_\beta\right),
\end{equation}
where $J=g^{-1}dg\in \alg g$ is the Maurer-Cartan form constructed from the group element $g(\tau,\sigma)\in G$ that depends on the worldsheet coordinates $\tau,\sigma$. The linear operator $\hat d:\alg g\to \alg g$ is a linear combination $\hat d=\tfrac12 P^{(1)}+ P^{(2)}-\tfrac12  P^{(3)}$ of projectors $P^{(i)}$, which by definition project onto the subspace $\alg g^{(i)}$.\footnote{For algebra elements $x\in\mathfrak{g}$ we will  use the notation $x^{(i)} = P^{(i)} x$. When there is no risk of ambiguity, we will always use a notation such that a linear operator on the Lie algebra acts on what sits to its right, e.g.~$\hat{d}J=\hat{d}(J)$.}  Moreover, $\STr$ denotes the supertrace on $\alg g$, which we use to obtain an ad-invariant graded-symmetric non-degenerate bilinear form on $\alg g$. The worldsheet indices $\alpha,\beta$ in the action are contracted with the projector  $\Pi_{(\pm)}^{\alpha\beta}=\tfrac12 (\sqrt{|h|}h^{\alpha\beta}\pm\epsilon^{\alpha\beta})$, where $h_{\alpha\beta}$ is the worldsheet metric and $\epsilon^{\tau\sigma}=-\epsilon^{\sigma\tau}=-1$. Finally, we use $\frac{\sqrt{\lambda}}{4\pi}$ to denote the string tension. 
 
 To construct the homogeneous Yang-Baxter deformation of the above action, one needs a linear operator $R:\alg g\to \alg g$ that is antisymmetric with respect to the supertrace
\begin{equation}\label{eq:antisymm-R}
\STr(Rx\ y)=-\STr(x\ Ry),\qquad\qquad \forall x,y\in \alg g,
\end{equation}
and that solves the classical Yang-Baxter equation (CYBE) on $\alg g$
\begin{equation}\label{eq:CYBE}
[[Rx,Ry]]-R\left([[Rx,y]]+[[x,Ry]]\right)=0,\qquad\qquad \forall x,y\in \alg g.
\end{equation}
If we choose a basis $\mathsf{T}_A$ for $\alg g$ such that  $[[\mathsf{T}_A,\mathsf{T}_B]]=f_{AB}{}^C \mathsf{T}_C$, then we can think of $R$ as a matrix identified by $R \mathsf{T}_A=R_A{}^B \mathsf{T}_B$.
Moreover, if $K_{AB}=\STr(\mathsf{T}_A \mathsf{T}_B)$ is the metric on $\alg g$ and $K^{AB}$ its inverse, we can define the dual generators $\mathsf{T}^A=\mathsf{T}_B K^{BA}$ that satisfy\footnote{Extra care must be taken when the elements are of odd grading, because in that case $\STr(xy)=-\STr(yx)$.} $\STr(\mathsf{T}_A \mathsf{T}^B)=\delta_A^B$.
We can then  equivalently think of $R$ as $r=-\tfrac12 R^{AB} \mathsf{T}_A\wedge \mathsf{T}_B\in \alg g\wedge\alg g$ where we use the graded wedge product $x\wedge y=x\otimes y-(-1)^{\text{deg}(x)*\text{deg}(y)}y\otimes x$, and $R_A{}^B=K_{AC}R^{CB}$, so that $Rx=\STr_2(r(1\otimes x))$ in which $\STr_2$ denotes the supertrace on the second factor of the tensor product.
Naturally, we only consider $R$-matrices that do not mix even and odd gradings, meaning that $r=-\tfrac12 R^{ab} \mathsf{T}_a\wedge \mathsf{T}_b-\tfrac12 R^{\hat\alpha\hat\beta} \mathsf{T}_{\hat\alpha}\wedge \mathsf{T}_{\hat\beta}$, where $a,b$ are indices of even (bosonic) generators while $\hat\alpha,\hat\beta$ are indices of odd (fermionic) generators.

Notice that~\eqref{eq:antisymm-R} implies that $R^{ab}$ is antisymmetric while $R^{\hat\alpha\hat\beta}$ is symmetric.
With these ingredients the deformed action is now given by~\cite{Delduc:2013qra}
\begin{equation}\label{eq:S-eta}
S_\eta = -\frac{\sqrt{\lambda}}{4\pi}\int d\tau d\sigma \ \Pi_{(-)}^{\alpha\beta}\  \STr \left( J_\alpha \ \hat d \frac{1}{1-\eta R_g\hat d}J_\beta\right),
\end{equation}
where $R_g=\AD_g^{-1} R \AD_g$ (with $\AD_gx=gxg^{-1}$) and $\eta$ is a deformation parameter. 
To have a real action, we take $\eta\in \mathbb R$ and $R^{AB}$ antihermitian, so that $R^{ab}$ is real while $R^{\hat\alpha\hat\beta}$ has imaginary components.

The space defined by the image of the $R$-matrix, $\alg f=\text{Im}(R)$, will play a central rôle in our computations. We can check that it is  a subalgebra of $\alg g$ as a consequence of the CYBE. We will denote its corresponding Lie group by $F$. Later we will use indices $I,J$ for generators $\mathsf{T}_I\in \alg f$, and indices $i,j$ when restricting to the bosonic subalgebra. Using the metric on $\alg g$ induced by the supertrace, we can also define the dual to $\alg f$, which we denote by $\alg f^*$. It is a subspace spanned by $\mathsf{T}^I=\mathsf{T}_AK^{AI}$ such that $\STr(\mathsf{T}_I \mathsf{T}^J)=\delta_I^J$.

So far we have reviewed the main ingredients --- in particular, eqs.~\eqref{eq:antisymm-R} and \eqref{eq:CYBE} --- that make the action~\eqref{eq:S-eta}  an integrable deformation of the seed supercoset action~\eqref{eq:S-0}~\cite{Klimcik:2008eq,Delduc:2013fga,Delduc:2013qra}. See later for more details.
In general, extra conditions are necessary in order to make sure that the deformed model can be interpreted also as a consistent string $\sigma$-model.
Since we are deforming a $\sigma$-model on a semi-symmetric space, the necessary and sufficient condition for the background fields of~\eqref{eq:S-eta} to satisfy the type IIB supergravity equations, is that $R$ also solves the unimodularity condition~\cite{Borsato:2016ose}
\begin{equation}
0=K^{AB}[[\mathsf{T}_A,R \mathsf{T}_B]]=R^{AB}f_{AB}{}^D \mathsf{T}_D.
\end{equation}
When this is not satisfied, the background fields solve the more general equations of ``modified supergravity''~\cite{Arutyunov:2015mqj,Wulff:2016tju}.

\vspace{12pt}

In this article, we are interested in deformations of the superstring on $AdS_5\times S^5$ and thus we will take $\alg g=\alg{psu}(2,2|4)$. The original supercoset is then $PSU(2,2|4)\setminus(SO(1,4)\times SO(5))$.\footnote{We will not need to give the full set of (anti)commutation relations of this superalgebra, nor provide an explicit matrix realisation of its elements. Conventions for these can be found in various places in the literature, see e.g.~\cite{Arutyunov:2009ga}.} Furthermore, we want to study deformations that are not interpretable as simple TsT transformations.
This implies that we must deform  the AdS factor of the background. Therefore, it will be sufficient  to state only the commutation relations for the $\alg{so}(2,4)\cong \alg{su}(2,2)$ conformal algebra, which is a subalgebra  of $\alg g$ and corresponds to the isometries of $AdS_5$. Given the Lorentz indices $\mu, \nu = 0, \ldots, 3$ in $3+1$ dimensions, we will use the Lorentz generators $M^{\mu\nu}$, the translations $p^\mu$, the special conformal transformations $k^\mu$, and the dilatation $d$ that close into the following commutation relations
\begin{equation}
\begin{aligned}
&[M^{\mu\nu}, p^\rho] = \eta^{\nu\rho} p^\mu - \eta^{\mu\rho} p^\nu , \qquad
&&[d, p^\mu] = +p^\mu , \\
&[M^{\mu\nu}, k^\rho] = \eta^{\nu\rho} k^\mu - \eta^{\mu\rho} k^\nu  , \qquad
&&[d, k^\mu] = -k^\mu , \\ 
&[M^{\mu\nu} , d] = 0 , \qquad
&&[p^\mu , k^\nu] = 2 M^{\mu\nu} + 2 \eta^{\mu\nu} d , \\
&[M^{\mu\nu}, M^{\rho \sigma}] = -\eta^{\mu\rho} M^{\nu \sigma} +\eta^{\nu \rho} M^{\mu\sigma}+\eta^{\mu  \sigma} M^{\nu\rho} - \eta^{\nu\sigma} M^{\mu\rho},
\end{aligned}
\end{equation}
with the Minkowski metric $\eta^{\mu\nu} = \text{diag}(-1, 1, 1, 1) $.

An interesting and rich class of deformations that deform AdS are given by Jordanian ones~\cite{Kawaguchi:2014qwa,vanTongeren:2015soa,Hoare:2016hwh,Orlando:2016qqu}. Here we choose to work with\footnote{Comparing to \cite{vanTongeren:2019dlq} one should rescale the deformation parameter there as $\eta \rightarrow - \eta/2 $.}~\cite{vanTongeren:2019dlq}
\begin{equation} \label{eq:rmatrix}
 r = \mathsf e \wedge \mathsf h +\tfrac{i}{2}\zeta (\mathsf Q_1\wedge \mathsf Q_1+\mathsf Q_2\wedge \mathsf Q_2)
\end{equation}
where\footnote{If we denote $\mathsf{T}_{\mathsf h}=\mathsf h, \mathsf{T}_+={\mathsf e}$ as generators of $\alg f$ then we can identify the duals $\mathsf{T}^{\mathsf h} = 2 \mathsf{T}_{\mathsf h}$ and $\mathsf{T}^+ = \frac{k^1-k^0}{2 \sqrt{2}} = \mathsf{T}_-$ as generators of $\alg f^*$, so that $\STr(\mathsf{T}^a \mathsf{T}_b)=\delta^a_b$. Moreover, we can check that they form an $\alg{sl}(2,\mathbb R)$ algebra, $[\mathsf{T}_{\mathsf h}, \mathsf{T}_\pm] =\pm \mathsf{T}_\pm$ and $[\mathsf{T}_+ , \mathsf{T}_-] = 2 \mathsf{T}_{\mathsf h}$. According to our conventions the $R$-matrix acts as $R(\mathsf{T}^{\mathsf h}) = \mathsf{T}_+$ and $R(\mathsf{T}^+) = -\mathsf{T}_{\mathsf h}$.
\label{foot:jord-sl2r}
}
\begin{equation}
\mathsf h = \frac{d+M^{01}}{2} , \qquad \mathsf e = \frac{p^0 + p^1}{\sqrt{2}}
\end{equation}
and they satisfy the commutation relation
\begin{equation}
[\mathsf h,\mathsf e]=\mathsf e.
\end{equation}
The supercharges $\mathsf Q_1,\mathsf Q_2$ complete the (anti)commutation relations to an $\mathcal N=1$ super Weyl algebra in one dimension. In this article, we will not need to specify them further, and more information can be found in~\cite{vanTongeren:2019dlq}.  For our convenience we have also introduced the parameter $\zeta$. It is worth to stress that $r$ solves the CYBE only for $\zeta=0$ or $\zeta=1$. When $\zeta=0$ the purely bosonic $r$-matrix  is \emph{not} unimodular, and it gives rise to background fields that solve only the modified supergravity equations; when $\zeta=1$, we have an additional ``fermionic tail'' such that $r$ is also unimodular, and therefore one can obtain a standard supergravity background~\cite{Borsato:2016ose,vanTongeren:2019dlq}. In this article we will find it interesting to consider both cases and compare them.
 
 Let us identify the bosonic part of~\eqref{eq:S-eta} (i.e.~when setting fermions to zero) with  $S_\eta=-\tfrac{\sqrt{\lambda}}{4\pi}\int d\tau d\sigma\ \Pi_{(-)}^{\alpha\beta}\ \partial_\alpha X^m\partial_\beta X^n(G_{mn}+B_{mn})$.  Obviously, the background metric $G_{mn}$ and the Kalb-Ramond field $B_{mn}$ do not depend on $\zeta$. In order to find an explicit expression, we can parameterise the group element as $g=g_{\alg a}g_{\alg s}$ where $g_{\alg a}\in SO(2,4)$ parameterises the AdS spacetime in the undeformed limit, and $g_{\alg s}\in SO(6)$   the sphere factor. We take
 \begin{equation} \label{eq:ge1}
g_{\alg a} = \exp(\theta M^{23})\exp(\rho p^3 + x^0 p^0 + x^1 p^1)\exp(\log(z) d) ,
\end{equation}
which gives\footnote{%\RB{It seems that when comparing to~\cite{vanTongeren:2019dlq} we have a different sign in front of $dz\wedge dx^-$ in the $B$-field.}
In terms of the usual Poincar\'e coordinates from $g_{\alg a} =  \exp(x^2 p^2 + x^3 p^3 + x^0 p^0 + x^1 p^1)\exp(\log(z) d)$ we would get the metric $z^{-2}(dz^2 + (dx^2)^2 + (dx^3)^2  - 2 dx^+ dx^- ) - 1/4 z^{-6}(\eta^2 (z^2 + (x^2)^2 + (x^3)^2) dx^-{}^2 $ for the deformed AdS factor. Instead, we  choose to use polar coordinates  $x^2 = \rho \sin \theta$ and $x^3 = \rho\cos\theta$ in the $2-3$ plane, such that  a $U(1)$ isometry corresponding to shifts of $\theta$, which survives the deformation, becomes manifest. In the next subsection, we will elaborate on the residual isometries of the deformed model.}
\begin{equation} \label{eq:defmetric1}
\begin{aligned}
ds^2 &= \frac{dz^2 + d\rho^2 + \rho^2 d\theta^2 - 2 dx^+ dx^- }{z^2} - \frac{\eta^2 (z^2 + \rho^2) dx^-{}^2}{4z^6} +ds_{S^5}^2,\\
B & = \tfrac{1}{2} B_{mn } dX^m\wedge dX^n = \frac{\eta}{2}\left(\frac{\rho\ d\rho\wedge dx^-}{z^4}+\frac{dz\wedge dx^-}{z^3}\right) , 
\end{aligned}
\end{equation}
with $x^\pm=(x^0\pm x^1)/\sqrt{2}$ and $ds_{S^5}^2$ the metric of the undeformed 5-sphere. The background is supported by an $F_3$  RR-flux, which vanishes for $\eta\rightarrow 0$,\footnote{The explicit expression for  $F_3$ can be found in~\cite{vanTongeren:2019dlq}.} and by the undeformed $F_5$ RR-flux of $AdS_5\times S^5$. Furthermore, the dilaton remains constant, $\Phi = \Phi_0$, in contrast to many other known  deformations.
 Let us recall that it is only for $\zeta=1$ that one can find this set of Ramond-Ramond fields and dilaton that complete the background to a standard supergravity one.

Notice that when $\eta\neq 0$, this deformation parameter can be reabsorbed by a simple coordinate redefinition, which is however singular in the $\eta\to 0$ limit. This property is in fact a consequence of the fact that the Jordanian deformation is related to a non-abelian T-duality transformation of the original background \cite{Hoare:2016wsk,Borsato:2016pas}.

%%%%%%%%%%%%%%%%%%%%%%%%%%%%%%%%%%%%%%%%%%%%%%%%%%%%%%%%%%%%%%%%%%%%%%%%
\subsection{Residual (super)isometries}\label{sec:iso}

As is well known, 
the (super)isometries are realised in the generic $\sigma$-model action \eqref{eq:S-0} as global transformations which act as $g \rightarrow g_L g$ with $g_L$  constant, in this case an element of $G=PSU(2,2|4)$.
The Lie algebra of bosonic isometries of the undeformed $AdS_5\times S^5$ background is $\mathfrak{so}(2,4)\oplus \mathfrak{so}(6) \cong \mathfrak{su}(2,2)\oplus \mathfrak{su}(4)$, corresponding respectively to the $AdS_5$ and $S^5$ factor of the spacetime.  The deformed action \eqref{eq:S-eta} is invariant only under a subset of these  global transformations, in particular those which are preserved by the operator $R$
\begin{equation}
\AD_{g_L}^{-1} R \AD_{g_L} = R \ .
\end{equation}
At the level of the algebra, this condition translates to 
\begin{equation} \label{eq:isomcond}
[[\mathsf{T}_{\bar A}, R(x) ]] = R[[\mathsf{T}_{\bar A},x]] , \qquad \forall x \in \mathfrak{g}
\end{equation}
where $\mathsf{T}_{\bar A}$ denotes a generator of a (super)isometry. The subset $\mathrm{span}(\mathsf{T}_{\bar A}) \subset \mathfrak{g}$ forms a subalgebra by means of the Jacobi identity. 
Using the above condition it is easy to check that the non-unimodular deformation ($\zeta=0$) preserves 16 superisometries, while the unimodular deformation ($\zeta=1$) preserves 12 superisometries.
In the non-unimodular case, since the $R$-matrix acts trivially on the sphere, all of the $\mathfrak{so}(6)$ isometries will be preserved by the deformation, and its Noether currents correspond as usual to the Maurer-Cartan forms $g_{\alg s}^{-1}d g_{\alg s}$, see e.g.~\cite{Arutyunov:2009ga}. In the unimodular case, instead, some of the $\mathfrak{so}(6)$ isometries are broken by the additional fermionic tail of the $R$-matrix. This is reflected, for example, at the level of the RR fields, which are not invariant under $\mathfrak{so}(6)$. Nevertheless, classical string configurations are insensitive to this fermionic tale, and at that level one can work as if there is $\mathfrak{so}(6)$ invariance.
More generally, in the deformed model the  Noether currents  corresponding to the residual  (super)isometries  are given by\footnote{Because of the factorisation of the AdS and sphere algebra, if we are interested only in the residual AdS isometries, we may simply substitute $g$ with $g_{\alg a}$ in the formulas above.}
\begin{equation} \label{eq:def-resnoethercurrents}
{\cal J}_{\bar A\pm} = \STr ( \mathcal J_\pm \mathsf{T}_{\bar A} ) ,\qquad\quad
\mathcal J_\pm=\AD_g\left(A^{(2)}_\pm \mp\tfrac{1}{2} (A^{(1)}_\pm-A^{(3)}_\pm)\right),
\end{equation}
where the worldsheet one-form $A=A_{+} d\sigma^+ + A_{-} d\sigma^-$, here written in terms of the worldsheet  lightcone coordinates  $\sigma^\pm = \frac{1}{2}(\tau \pm \sigma)$, is defined as
\begin{equation} \label{eq:def-current}
A_{\pm} = \frac{1}{1\pm \eta R_{g} \hat{d} } g^{-1}\partial_\pm g  \ .
\end{equation}
From these Noether currents one may also define the Noether charges
\begin{equation}\label{eq:Noet-ch}
Q_{\bar A} = \frac{1}{2\pi}\int_0^{2\pi} d\sigma\, \mathcal J_{\bar A\tau}.
\end{equation}
For a generic $\sigma$-model with background metric and Kalb-Ramond field, the  Killing vectors $k_{\bar a}$ associated to \emph{bosonic} isometries can then be found from solving (see e.g.~\cite{Rocek:1991ps})
\begin{equation}
{\cal J}_{ \bar{a} \pm} = k_{\bar a}^m (G_{mn} \mp B_{mn}) \partial_\pm X^n \pm  \omega_{\bar{a} n} \partial_\pm X^n , 
\end{equation}
where ${\cal L}_{k_{\bar a}} G = 0$ and ${\cal L}_{k_{\bar a}} B = d\omega_{\bar a}$, with ${\cal L}$ the Lie derivative and  $\omega_{\bar a}$ an arbitrary one-form. 
Since  the deformation   only acts non-trivially on  $AdS$, we will focus on this subsector for most of the remainder of this paper.

Solving \eqref{eq:isomcond} for the $R$-matrix \eqref{eq:rmatrix} in question then gives five residual isometries of $\mathfrak{so}(2,4)$, forming the subalgebra\footnote{From an algebraic perspective, we thus see that indeed $M^{23}$ is a residual $U(1)$ isometry that corresponds to shifts in an angular coordinate.}
\begin{equation} \label{eq:isoms}
\mathfrak{k}=\mathrm{span}( \mathsf{T}_{\bar a} , \bar{a}=1, \ldots, 5) = \mathrm{span}( d-M^{01} , p^0, p^1, k^0+k^1, M^{23} ) \ . 
\end{equation}
The corresponding Killing vectors in the (polar) Poincar\'e coordinates read
\begin{equation}
\begin{gathered}
k_{\bar 1}^m \partial_m = 2 x^- \partial_{x^-} + \rho \partial_\rho + z \partial_z \ , \qquad
k_{\bar 2}^m \partial_m= \frac{\partial_{x^+} + \partial_{x^-}}{\sqrt{2}} \ ,  \qquad 
k_{\bar 3}^m \partial_m = \frac{\partial_{x^+} - \partial_{x^-}}{\sqrt{2}} \ , \\
k_{\bar 4}^m \partial_m= \sqrt{2} (z^2+\rho^2) \partial_{x^+} + \sqrt{2}2 (x^-)^2 \partial_{x^-} + \sqrt{2}2 x^- \rho \partial_\rho + \sqrt{2}2 x^- z \partial_z \ , \quad
k_{\bar 5}^m \partial_m= \partial_\theta \ ,
\end{gathered}
\end{equation}
which shows that we have in particular  three translation isometries in $x^+$, $x^-$ and $\theta$.\footnote{One can also derive the one-forms $\omega_{\bar a}$ needed to relate the Noether currents ${\cal J}_{\bar a}$ to the Killing vectors $k_{\bar a}$. In this case, they are found to be non-vanishing but closed for generic values of $\eta$. The $B$-field itself is therefore isometric with respect to the residual isometries.}

%%%%%%%%%%%%%%%%%%%%%%%%%%%%%%%%%%%%%%%%%%%%%%%%%%%%%%%%%%%%%%%%%%%%%%%%
\subsection{Global coordinates} \label{sec:global}

In view of the applications of the Jordanian deformation of $AdS_5\times S^5$ in holography, it is important to identify global coordinates for the background for generic values of $\eta$. As it is well known, the Poincar\'e coordinates are globally incomplete already for the undeformed $AdS$ spacetime. Thus, our first task is to find an appropriate coordinate transformation. A first obvious attempt is to transform to the usual global  coordinates of $AdS$ and then check if they remain global coordinates also when $\eta\neq 0$. However, we find that the deformed metric  in these coordinates is complicated and does not exhibit any of the manifest residual isometries. We have therefore not pursued this coordinate system further, as it would not be useful for practical purposes. 

In fact, similar issues were found in \cite{Blau:2009gd}, which analysed  the geometry of Sch\"odinger spacetimes $Sch_z$. Such geometries possess  an anisotropic scale invariance $(t,x^i) \rightarrow (\lambda^z t, \lambda x^i)$ characterised by a critical exponent $z> 1$, where the case $z=1$ corresponds to usual $AdS$. A global coordinate system different from the usual global $AdS$ coordinates was then obtained for $z=2$.
Conveniently, the Jordanian deformed background \eqref{eq:defmetric1} is extremely reminiscent of the $Sch_2$ case and, in fact,  exhibits the same anisotropic  scale symmetry.\footnote{In particular, when defining  $(x^2, x^3) = (\rho\sin\theta , \rho\cos\theta)$,  the Jordanian background is invariant under $z\rightarrow \lambda z$, $x^2 \rightarrow \lambda x^2$, $x^3 \rightarrow \lambda x^3$, $x^- \rightarrow \lambda^2 x^-$ and $x^+\rightarrow x^+$, which translates to eq.~(2.1) of \cite{Blau:2009gd} by identifying $(z,x^2,x^3,x^-,x^+)$ with $(r,x^2,x^3,t,\xi)$ for the critical exponent $z=2$. Notice that the only difference between the Jordanian metric and the Schr\"odinger  $Sch_2$ metric, given in eq.~(1.1) of  \cite{Blau:2009gd}, is the term $\frac{\rho^2}{4 z^6} dx^-{}^2=\frac{(x^2)^2+(x^3)^2}{4 z^6} dx^-{}^2$.} It is therefore conceivable that a good candidate for a global coordinate system of our Jordanian background coincides with that found in \cite{Blau:2009gd}. The transformation from  polar Poincar\'e coordinates to the global ones $X^m=(T,V,P,\Theta,Z)$ is obtained from\footnote{Strictly speaking, the relation to the global Schr\"odinger coordinates of \cite{Blau:2009gd} is  obtained by transforming $(P,Z)$  to $(X^2, X^3) = (P\sin\Theta , P\cos\Theta)$.} 
\begin{equation} \label{eq:PtoSchr}
\begin{gathered}
x^+ = V + \frac{1}{2} (Z^2 + P^2) \tan T , \quad z = \frac{Z}{\cos T} , \quad \rho = \frac{P}{\cos T} , \\
\theta = \Theta, \quad x^- = \tan T .
\end{gathered}
\end{equation}

We can cover the full spacetime if we take the coordinates in the ranges $T,V\in ]-\infty,+\infty[$, $\Theta\in[0,2\pi[$ and $P,Z\in ]0,+\infty[$. The only coordinate that we allow to be periodically identified is $\Theta$, with period $2\pi$. These ranges have been derived from  the relation between these coordinates and the embedding coordinates of $AdS$, which we discuss in appendix~\ref{a:embedding}. In the new coordinates the background fields \eqref{eq:defmetric1} then read
\begin{equation}
\begin{aligned}
\label{lagglob}
ds^2 &= \frac{dZ^2 + dP^2 + P^2 d\Theta^2 -2 dT dV}{Z^2} - \frac{4Z^4 ( Z^2+P^2) + \eta^2 (Z^2 +  P^2) }{4 Z^6} dT^2 +ds_{S^5}^2 , \\
B &= \frac{\eta}{2} \left( \frac{P dP \wedge dT}{Z^4}+\frac{ dZ \wedge dT}{Z^3} \right) \ .
\end{aligned}
\end{equation}
Clearly, this coordinate system will be useful for practical purposes, since it manifestly displays three  translational isometries, namely in $T$, $V$, and $\Theta$. Because the Cartan subalgebra of the algebra of  residual isometries is three-dimensional (see appendix \ref{a:CSA}) this is in fact the maximal number of manifest translational isometries that a coordinate system could display.  Explicitly, the Killing vectors introduced earlier read
\begin{equation} \label{eq:globalKilling}
\begin{aligned}
k_{\bar 1}^m \partial_m&=\sin (2 T)\partial_T-\left(P^2+Z^2\right) \sin (2 T)\partial_V+\cos (2
   T)(P\partial_P+ Z \partial_Z ),\\
k_{\bar 2}^m \partial_m&=\frac{\cos ^2(T)}{\sqrt{2}}\partial_T-\frac{\left(P^2+Z^2\right) \cos (2 T)-2}{2
   \sqrt{2}}\partial_V-\frac{ \sin (2T)}{2\sqrt{2}}(P\partial_P+ Z \partial_Z ),\\
 k_{\bar 3}^m \partial_m&=-\frac{\cos ^2(T)}{\sqrt{2}}\partial_T+\frac{\left(P^2+Z^2\right) \cos (2 T)+2}{2
   \sqrt{2}}\partial_V+\frac{ \sin (2T)}{2\sqrt{2}}(P\partial_P+ Z \partial_Z ),\\
    k_{\bar 4}^m \partial_m&=2 \sqrt{2} \sin ^2(T)\partial_T+\sqrt{2} \left(P^2+Z^2\right) \cos (2 T)\partial_V+\sqrt{2} 
   \sin (2 T)(P\partial_P+ Z \partial_Z ),\\
       k_{\bar 5}^m \partial_m&=\partial_\Theta.
\end{aligned}
\end{equation}
Note that $(k^m_{\bar 2} + k^m_{\bar 3})\partial_m = \sqrt{2} \partial_V$ and $(2k^m_{\bar 2} -2 k^m_{\bar 3} + k^m_{\bar 4})\partial_m = 2\sqrt{2} \partial_T$.

Following the above proposal, in the rest of this section, we will prove that the coordinate system $X^m=(T,V,X^2,X^3,Z)$ with $(X^2,X^3) = (P\sin\Theta , P\cos\Theta)$ gives rise to a geodesically complete spacetime.

%%%%%%%%%%%%%%%%%%%%%%%%%%%%%%%%%%%%%%%%%%%%%%%%%%%%%%%%%%%%%%%%%%%%%%%%
\subsubsection{On the generators of time translations}

Before performing the full analysis of geodesic completeness of the deformed spacetime, let us first identify the  possible Hamiltonians (generators of  time translations) within the isometry algebra $\mathfrak{k}$ in~\eqref{eq:isoms}. We will look for  generators of this kind that are part of the Cartan subalgebra of $\mathfrak{k}$, because in that case one can find an adapted coordinate system where the  generator of time translations simply  shifts one of the coordinates, which can be identified as time.

The first natural question is thus to identify all the possible Cartan subalgebras of $\mathfrak{k}$, as explained in appendix \ref{a:CSA}. Up to automorphisms, there are two possible inequivalent choices of Cartan subalgebras, namely 
\begin{equation}
\begin{aligned}
\text{(I)}:&\ \text{span}\{d - M^{01} , \ p^0 +  p^1,\ M^{23}  \},\\
\text{(II)}:&\  \text{span}\{p^0 -  p^1+\alpha (k^0+k^1), \ p^0 +  p^1,\ M^{23} \},
\end{aligned}
\end{equation}
where we leave a possible $\alpha>0$ coefficient for later convenience.

Identifying the possible generators of time translations is now a very simple task. One can construct a generic linear combination of the Cartan generators and demand that the corresponding Killing vector $\bar{k}^m$ has a strictly negative norm in every region of spacetime. Notice, however, that  $p^0 + p^1$ and $M^{23}$  are not timelike and, being central elements, it is easy to remove their contribution by a spacetime-coordinate redefinition. We can therefore simply compute the norm of the Killing vectors associated to the two possible Cartans of the $\mathfrak{sl}(2,\mathbb{R})$ subalgebra of $\alg k$. We find
\begin{equation}
\begin{alignedat}{2}
\text{(I)}:&\ \bar{k}^m = k^m_{\bar 1}, \qquad &&\bar{k}^m \bar{k}_m = \frac{(P^2+Z^2)(4 Z^4 - \eta^2 \sin(2T)^2)}{4 Z^6} \\
\text{(II)}:&\   \bar{k}^m =k_{(\bar{2})}^m-k_{(\bar{3})}^m+\tfrac12 k_{\bar 4}^m, \qquad  &&\bar{k}^m \bar{k}_m = - \frac{(Z^2+P^2) (\eta^2 + 4 Z^4)}{2 Z^6} ,
\end{alignedat}
\end{equation}
where in the latter case we took $\alpha=1/2$. While other choices of $\alpha>0$ are possible, we make this choice here to match with the expression $(2k^m_{(2)} -2 k^m_{(3)} + k^m_{\bar 4})\partial_m = 2\sqrt{2} \partial_T$ found above for the generator of $T$ translations. Whilst in case (I) the norm of the Killing vector is not everywhere negative (set e.g. $T=0$), in case (II) we do have a strictly timelike Killing vector for generic values of the coordinates and of the deformation parameter $\eta$.  The relevant Cartan subalgebra is thus that of class (II) with $\alpha=1/2$. Concluding, the coordinate $T$ will be our preferred global time coordinate and the Hamiltonian is\footnote{In the literature, in similar setups, sometimes the Hamiltonian is identified with $d-M^{01} \sim h$. While this is of course possible when considering the complexified algebra (where any choice of Cartan can be brought to $d-M^{01}$, see e.g.~\eqref{eq:xtoh}), here we insist on respecting reality conditions. Notice that the square of $d-M^{01}$ has a positive trace, and thus cannot correspond to a timelike Killing vector. In fact, if we take for the sake of the argument the case of group manifolds, the Killing vectors can be identified by $k_a^m=\Tr(\mathsf{T}^b\AD_g^{-1} \mathsf{T}_a)\ell_b{}^m$ where $\ell_b{}^m$ is the inverse of $\ell_m{}^b$ and $dX^m\ell_\mu^a \mathsf{T}_a=g^{-1}dg$. When moving to the identity of the group (i.e.~at $g=1$), the norm $k_a^m k_{am}$ of the Killing vector reduces to $\Tr(\mathsf{T}_a \mathsf{T}_a)$. Therefore, a necessary condition for $k_a$ to be timelike is that $\Tr(\mathsf{T}_a \mathsf{T}_a)<0$.}
 \begin{equation} \label{eq:globaltimegenerator}
 \mathsf{H}=\tfrac{1}{\sqrt{2}}( p^0-p^1 +\tfrac12 (k^0+k^1)) \ .
 \end{equation} 
  Interestingly, when taking the undeformed limit $\eta\to 0$, the above Hamiltonian does not reduce to something equivalent to the BMN generator of time translations~\cite{Berenstein:2002jq}, 
 because it is possible to check that there is no inner automorphism of $SU(2,2)$ that relates $\mathsf{H}$ to $p_0+k_0$.

%%%%%%%%%%%%%%%%%%%%%%%%%%%%%%%%%%%%%%%%%%%%%%%%%%%%%%%%%%%%%%%%%%%%%%%%
\subsubsection{Analysis of geodesic completeness}\label{sec:geocompl}

In this section we will analyse whether  the coordinate system $X^m=(T,V,P,\Theta,Z)$, which describes $Sch_2$ spacetimes globally, are also well-defined global  coordinates for the Jordanian deformation. This analysis is necessary because when comparing the $Sch_2$ metric, given in eq.~(3.18) of \cite{Blau:2009gd}, and the Jordanian metric, given in \eqref{lagglob}, we see that the Jordanian case has an additional term $\frac{P^2}{4Z^6} dT^2$, which may lead to pathologies.

There  are typically two common ways to analyse whether coordinates are global or not. First, one can try to find the embedding  in a higher dimensional spacetime\footnote{In the case of undeformed $AdS_5$, this is of course $\mathbb{R}^{2,4}$, which is invariant under the conformal algebra $SO(2,4)$. We show the relation between the coordinate system $X^m=(T,V,P,\Theta,Z)$ and the $AdS$ embedding coordinates in appendix \ref{a:embedding}.} and argue whether or not the embedding coordinates cover the full hypersurface. Another possibility is to consider geodesics $\gamma (t)$,  parameterised by a geodesic parameter $t \in \mathbb{R}$, and investigate geodesic completeness of the coordinate system. We will follow the latter method. Before doing so explicitly, let us recall the criterions concerning geodesic (in)completeness and singular spacetimes.

A coordinate system is called geodesically complete if all geodesics $\gamma (t)$ are defined for all values of the geodesic parameter $t \in \mathbb{R}$, within the ranges of coordinates that parameterise the manifold.
A spacetime is said to be geodesically incomplete if there exists a \emph{finite} value $t_0$ of the geodesic parameter such that the geodesic $\gamma(t_0)$ hits an extremal value allowed by the range of spacetime coordinates, and the geodesic $\gamma(t)$ cannot be extended to $t>t_0$.
Hitting an (apparent) metric singularity would not be considered pathological --- and therefore would not classify as geodesic incompleteness --- if the geodesic ``bounces back'' (i.e.~if it can be continued to allowed ranges of the spacetime coordinates at later $t$) or if the singularity is reached only in the limit $t\to \infty$.\footnote{When at least one geodesic ends in a point $p$ of the manifold, then $p$ is a singularity. In this case, one should further study whether $p$ is a physical singularity or just a coordinate singularity. It is only in the latter case that one speaks of geodesic incompleteness.
To distinguish the nature of the singularity, one typically computes curvature scalars or tidal forces. The point $p$ is an honest curvature singularity when at least one of these become infinite at $p$. If all of them remain finite, and the geodesics appear to end at $p$, then the coordinate system is considered to be incomplete.
 A fully general criterion about whether or not an apparent singularity is honest, however, does not exist. Fortunately, we will not find such singularities and therefore we will not need to perform this analysis.}

Rather than solving the five second-order geodesic equations for $X^m(t)$, we can make use of the five residual Killing isometries to perform a \emph{single} integration and simplify the calculations considerably. In particular, recall that for each Killing vector field $k^m_{\bar a}$, there is an independent charge 
\begin{equation} \label{eq:geoconstant}
Q_{\bar a} = k^m_{\bar a} G_{mn} \dot{X}^n \ , 
\end{equation}
 which is conserved along the geodesics.\footnote{Here the dot refers to derivation with respect to the geodesic parameter $t$.} In other words, since the full system of geodesic equations can be rewritten as $\dot{Q}_{\bar a} = 0$ and we have as many Killing vectors as independent coordinates,  the geodesics can be analysed by studying the system \eqref{eq:geoconstant} rather than the usual geodesic equations. To do so, it will be useful to introduce  the conserved quantities  related to the manifest translational isometries in $T$, $V$ and $\Theta$,\footnote{The relation to the previously defined Noether charges is $Q_{\bar 2} = \frac{1}{4} \left(2 \sqrt{2} Q_T+ 2 \sqrt{2} Q_V - Q_{\bar 4} \right)$, $Q_{\bar 3} = \frac{1}{4} \left(- 2 \sqrt{2}Q_T+ 2 \sqrt{2} Q_V + Q_{\bar 4} \right)$ and $Q_{\bar 5} = Q_\Theta$. \label{f:chargesrot}}
 \begin{equation} \label{eq:manifestisocharges}
\begin{alignedat}{2} 
k_T &=\partial_T: \qquad Q_T &&= - \frac{4Z^4 ( Z^2+P^2) + \eta^2 (Z^2 +  P^2) }{4Z^6} \dot{T} - \frac{1}{Z^2} \dot{V }  \ , \\
k_V &=\partial_V: \qquad Q_V &&= - \frac{1}{Z^2} \dot{T}  \ , \\
k_\Theta &=\partial_\Theta: \qquad Q_\Theta &&= \frac{P^2}{Z^2} \dot{\Theta}  .
\end{alignedat}
 \end{equation}
 An interesting feature now emerges when analysing the system  \eqref{eq:geoconstant}: instead of five first order differential equations for $X^m(t)$, one finds that \eqref{eq:geoconstant} decouples into four first order differential equations and one algebraic relation among the coordinates, which restricts the geodesics to a hypersurface. We choose to decouple \eqref{eq:geoconstant} as
 \begin{equation} \label{eq:firstordergeos}
\begin{aligned}
\dot{T}(t) &= - Q_V Z(t)^2 \ , \\
\dot{V}(t) &= \frac{Q_V}{4} (\eta^2 + 4Z(t)^4)  \left( 1+ \frac{P(t)^2}{Z(t)^2} \right) - Q_T Z(t)^2 \ ,\\
\dot{\Theta}(t) &= \frac{Z(t)^2}{P(t)^2} Q_\Theta \ , \\
\dot{P}(t) &= \frac{Z(t)}{P(t)} \left(\frac{Z(t) \left(Q_{\bar 1} + \sin 2 T(t) \left( Q_V ( Z(t)^2 + P(t)^2 ) - Q_T  \right)   \right) }{\cos 2 T(t)}  - \dot{Z}(t)\right) \ ,
\end{aligned}
\end{equation}
subjected to the algebraic relation, which we will  call the ``hypersurface equation'',
\begin{equation} \label{eq:hypersurface}
2 Q_V P^2(t) = 2 Q_T - (2Q_T-\sqrt{2} Q_{\bar 4}) \cos 2 T(t) - 2  Q_{\bar 1} \sin 2 T(t) - 2 Q_V Z(t)^2  \ .
\end{equation}
In fact, this equation  defines a circle in the $(P,Z)$ plane with a varying (but bounded) radius depending on $T(t)$, i.e.~$P(t)^2 + Z(t)^2 = R^2 (T(t))$.\footnote{There are a number of conditions that we can derive for the values of the charges such that $R^2\geq 0$, which should be interpreted as reality conditions for $Z$ and $P$. They should be further subjected to a condition ensuring also that $T$ is real when $T$ remains constant along the geodesic.} 
In addition, there is another useful constant of motion along the geodesics, namely
\begin{equation} \label{eq:causalgeo}
\epsilon= G_{mn} \dot{X}^m \dot{X}^n  \ ,
\end{equation}
with $\epsilon=0,-1,1$ for null, timelike, or spacelike paths. Substituting the system \eqref{eq:firstordergeos}  into \eqref{eq:causalgeo} leads, in general, to a complicated differential equation for $\dot{Z}$ subjected to the hypersurface equation. \\

Let us now analyse potential pathological regions of the deformed background. Notice that we can completely exclude  the coordinates $T,V,\Theta$ from our analysis, as translations in these directions are manifest isometries and, thus, will not give rise to pathologies. From the background fields \eqref{lagglob} we see the only regions that can  potentially be pathological are $Z, P = \{0, +\infty \}$, as the (inverse) spacetime metric appears to blow up there.

First, let us consider the regions $Z\to\infty$ or $P\to \infty$, in which case  the hypersurface equation plays an important role in the analysis. In particular,  the hypersurface bounds  the geodesic motion in $Z$ and $P$, and forces both to a maximal absolute value, $\lvert Z \rvert , \ \lvert P \rvert  \leq \lvert R \rvert$, which is always finite for a finite value of the conserved charges $Q_{\bar a}$ and for any finite value of $t\in \mathbb{R}$. The only exception to this is when the charge $Q_V = 0$, for which the dependence of $P$ and $Z$ drops out of \eqref{eq:hypersurface} (equivalently, the radius $R$ becomes infinite). However, we can argue that also in this case there is no pathology. First,  for $Q_V=0$ the hypersurface equation forces $T(t)$ to be a constant in $t$, which is consistent with its geodesic equation $\dot{T} = - Q_V Z^2$. Hence, the coordinate time does not evolve in the geodesic parameter. This should therefore be an extreme spacelike path, as we also see by analysing the equation for $\dot{Z}$ obtained from \eqref{eq:causalgeo}. Remarkably, for $Q_V =0$, this equation in fact greatly simplifies to $\dot{Z}^2(t) = \epsilon Z(t)^2$ which is immediately solved by
\begin{equation}
Z(t) = A e^{\pm \sqrt{\epsilon}t} \ ,
\end{equation}
where $A$ is an integration constant. Reality conditions thus tell us that paths with $Q_V=0$ are only possible when they are spacelike or null (the latter being trivial in that case). The spacelike paths can in fact reach $Z = \infty$ (as well as $Z=0$, see also later), but they  do so only at an infinite value of the geodesic parameter $t$. We can thus  conclude that $Z\to \infty$ is not a pathological region. Concerning the behaviour at $P\to \infty$, one finds that taking $Q_V =0$ would lead to paths that are  not defined due to reality conditions. 
Given that for generic $Q_V$ the motion in $P$ is bounded, we can again conclude that also the region $P\rightarrow \infty$ is not pathological. In fact, even though $P$ can become very large for small $Q_V$, the region $P\rightarrow \infty$ will never be reached by any geodesic in finite $t$.

Next, let us consider the $Z\rightarrow 0$ region and analyse the small $Z$ behaviour. Note that we can just focus on the system $T,P,Z$ (the solutions for $\Theta$ and $V$ will be given once solutions for $T,P,Z$ are found). We will write $Z(t) = \lambda \hat{Z}(t)$ and expand the equations around $\lambda = 0$ up to ${\cal O}(\lambda)$, unless stated otherwise. From the first equation in~\eqref{eq:firstordergeos} and \eqref{eq:hypersurface}, we will find that $T$ and $P$ are constant in $t$ up to linear order in $\lambda$. In particular\footnote{Recall that $P\in ]0,\infty[$.}
\begin{equation} \label{eq:PTleading}
\begin{aligned}
T(t) &= T_0 + {\cal O}(\lambda^2), \\ 
P(t) &=  \sqrt{R^2(T_0)} + {\cal O}(\lambda^2) \\
&=  \sqrt{\frac{2Q_T - (2Q_T - \sqrt{2}Q_{\bar 4}) \cos 2 T_0 - 2 Q_{\bar 1} \sin 2T_0}{2 Q_V}} + {\cal O}(\lambda^2) \ ,
\end{aligned}
\end{equation}
with $T_0$ the constant value of time.\footnote{Note that if $T_0$ is such that $R^2(T_0) = 0$, then \eqref{eq:hypersurface} implies that $P(t)$ will have a linear in $\lambda$ contribution, namely $P(t) = \sqrt{-\hat{Z}(t)^2} \lambda + {\cal O}(\lambda^2) $. However, due to reality conditions, the latter is only possible when $\hat{Z}=0$ and thus $P={\cal O}(\lambda^2) $. Hence, this is  consistent with \eqref{eq:PTleading} at $R^2(T_0) = 0$. \label{f:zeroradius}}  Now let us analyse the equation for $\dot{Z}$ obtained from \eqref{eq:causalgeo} and \eqref{eq:firstordergeos} for small $Z$. Up to subleading order in $\lambda$, where $T(t) = T_0 + \lambda^2 {\cal T}(t)+ {\cal O}(\lambda^3)$ and $P^2(t) = R^2(T(t)) - \lambda^2 \hat{Z}^2(t)+ {\cal O}(\lambda^3) = R^2(T_0) + \lambda^2 ( \partial_{T_0} R^2(T_0) {\cal T}(t) - \hat{Z}^2(t)) + {\cal O}(\lambda^3)$, where ${\cal T}(t)$ is such that $\dot{\cal T}(t) = -Q_V Z(t)^2$ and ${\cal T}(t=0)=0$, we find
\begin{equation}
\epsilon = \frac{1}{\lambda^2} \left( \frac{\eta^2 Q_V^2 R^2(T_0) + \dot{Z}(t)^2}{\hat{Z}(t)^2} \right) + \left(\frac{\dot{Z}(t)^2}{R^2(T_0)} + \frac{\eta^2 Q_V^2 \partial_{T_0} R^2(T_0) {\cal T}(t)}{\hat{Z}(t)^2} \right) + {\cal O}(\lambda) \ ,
\end{equation}
which one can analyse order by order in $\lambda$. The leading term thus gives
\begin{equation} \label{eq:ZdotfromEps}
\dot{Z}(t)^2 = - \eta^2 Q_V^2 R(T_0)^2 + {\cal O}(\lambda^2) \ ,
\end{equation}
which in general would give an imaginary solution for $Z$, i.e. $Z(t) = Z_0 \pm t \sqrt{-\eta^2 Q_V^2 R(T_0)^2} + {\cal O}(\lambda^2) $. This means that we cannot get arbitrarily close to $Z=0$ except if \textit{(i)} $\eta=0$ (the undeformed limit, in which case the spacetime is geodesically complete, see also appendix \ref{a:embedding}), \textit{(ii)} $Q_V = 0$ (a case which we already analysed in generality above and where it was found that  geodesics can reach $Z=0$   only  in an infinite amount of the geodesic time), and \textit{(iii)} $R^2(T_0) = 0$ (in which case  $P={\cal O}(\lambda^2)$ and $\hat{Z}=0$ (see also footnote \ref{f:zeroradius}) and thus all the coordinates are constant in $t$). 
Finally, we must analyse consistency with the equation for $\dot{P}$ given in \eqref{eq:firstordergeos}, which can be interpreted as $R \dot{R} - Z \dot{Z}=P \dot{P} $. 
 From the leading solution for $P$ in \eqref{eq:PTleading} we know that $\dot{P} =  {\cal O}(\lambda^2)$. Therefore also $P \dot{P} = {\cal O}(\lambda^2)$ and we can rewrite $R \dot{R} - Z \dot{Z}=P \dot{P} $ as
\begin{equation}
\lambda^2 \frac{\hat{Z}(t)^2 (Q_{\bar 1} + \sin 2T_0 (Q_V R^2(T_0) - Q_T))}{\cos 2T_0} - \lambda \hat{Z}(t) \dot{Z}(t) = {\cal O}(\lambda^2)
\end{equation}
i.e. 
$\dot{Z}(\tau) =  {\cal O}(\lambda)$. This is  consistent with what we have found previously, i.e.~eq.~\eqref{eq:ZdotfromEps}, requiring either \textit{(i)} $\eta=0$, \textit{(ii)} $Q_V=0$, or \textit{(iii)} $R(T_0)^2=0$. We can thus conclude that the geodesics will never reach $Z=0$ by an evolution in $t$ and that this  region is  not pathological. Finally, analysing the region $P\rightarrow 0$ using a similar strategy does not give us a  conclusive answer. However, this is of course only the apparent pathology of polar coordinates: neither the metric nor its inverse is singular for $X^2=X^3=0$.

The arguments above prove that the coordinate system $X^m=(T,V,X^2,X^3, Z)$ is globally well-defined for any value of the deformation parameter. We will however prefer to work in terms of $X^m=(T,V,P, \Theta, Z)$, in which  the shift isometry in $\Theta$ is manifest. 
 This concludes our discussion and, combined with the previous section, shows that the coordinate $T$ is our \textit{global} time coordinate.

%%%%%%%%%%%%%%%%%%%%%%%%%%%%%%%%%%%%%%%%%%%%%%%%%%%%%%%%%%%%%%%%%%%%%%%%
\section{BMN-like solution of the $\sigma$-model}  \label{s:pointlikesolution}

In this section, we give the derivation of a particular pointlike string solution to the equations of motion of our Jordanian deformation. As we will show in section~\ref{sec:csc},  it will lead to relatively simple quasimomenta. For us it will be the analog of the BMN solution valid in $AdS_5\times S^5$, and we will call it a BMN-like solution.

For a general bosonic string configuration $X^m(\tau, \sigma)$,  recall that the equations of motion of the action $S_\eta=-\tfrac{\sqrt{\lambda}}{4\pi}\int d\tau d\sigma\ \Pi_{(-)}^{\alpha\beta}\ \partial_\alpha X^m\partial_\beta X^n(G_{mn}+B_{mn})$ can be written in conformal gauge as 
\begin{equation} \label{eq:smeoms}
\eta^{\alpha\beta}\partial_\alpha \partial_\beta X^m + \left(\eta^{\alpha\beta} \Gamma^m_{nl} + \frac{1}{2}\epsilon^{\alpha\beta} H^m{}_{nl} \right) \partial_\alpha X^n  \partial_\beta X^l = 0 \ ,
\end{equation}
which is a generalisation of the geodesic equation of a point particle. Here,  $\Gamma^m_{nl}$ are the usual Christoffel symbols,  and $H_{mnl}$ are the components of the torsion 3-form $H=dB$. Interestingly, it can be checked that these equations evaluated for the Jordanian deformation  admit a consistent truncation on $x_2(\tau,\sigma) = x_3(\tau,\sigma) = 0$, or equivalently $P(\tau,\sigma) = \Theta(\tau,\sigma) = 0$,\footnote{In the latter case, the limit $P(\tau,\sigma) \rightarrow 0$  on the equations of motion (in particular, the equation of motion for $\Theta$) should be taken with care. First one must set $P(\tau,\sigma)$ to a constant $P$ to then find that in the resulting equations of motion the limit $P\rightarrow 0 $ can indeed be taken smoothly. } for any value of the deformation parameter.

Let us now  consider the following $\sigma$-independent  ansatz on the  coordinates: 
\begin{equation} \label{eq:pointansatz}
X^m = a_m \tau + b_m \ ,
\end{equation}
with $a_m$ and $b_m$  constant variables which will be determined upon imposing \eqref{eq:smeoms}. Furthermore, we will work on the consistent truncation $P(\tau,\sigma)=\Theta(\tau,\sigma)=0$, equivalently we set $a_P=a_\Theta=b_P=b_\Theta =0$. Since there is no $\sigma$-dependence, this pointlike string will not couple to the Kalb-Ramond two-form, and its equations of motion are therefore simply the geodesic equations.
 Solving them for the variables $a_m$ and $b_m$ while imposing that fields are real singles out the solution\footnote{When one does not impose the consistent truncation $P(\tau,\sigma)=\Theta(\tau,\sigma)=0$, but still assumes the pointlike ansatz \eqref{eq:pointansatz}, there is one other real solution namely $a_Z=a_P=0$, $a_V = - \frac{\eta^2 a_T (b_Z^2 + b_P^2)}{4 b_Z^4}$ and $a_\Theta = a_T \sqrt{\eta^2 + 4 b_Z^4}{2b_Z^2}$ while the other variables remain free. Carrying the analysis of this solution through (again using the isometries to set $b_T=b_V=b_\Theta=0$) one will find the associated quasimomenta to be quite complicated and we will therefore not consider this possibility further.}
\begin{equation}
 a_Z=0 , \quad \text{and} \quad a_V = -\frac{\eta^2 a_T}{4 b_Z^2} ,
 \end{equation} 
 while the other variables remain free. Because of the manifest isometries in $T$, $V$ and $\Theta$ we can however set $b_T=b_V=b_\Theta=0$ without loss of generality. In summary, our pointlike string solution thus evolves as
\begin{equation} \label{eq:BMNlikesol}
T = a_T \tau, \;\;\; V = -\frac{\eta^2 a_T}{4b_Z^2} \tau , \;\;\; Z = b_Z ,\;\;\; P=0,\;\;\; \Theta = 0  \ .
\end{equation}
This solution   is in fact exactly the same as the one studied for the Schr\"odinger $Sch_5\times S^5$ background (at least, in the deformed AdS subsector) in \cite{Guica:2017mtd}. The reason is that on the consistent truncation $P=0$, the Jordanian metric given in \eqref{eq:defmetric1} coincides precisely with the Schr\"odinger metric. Given that this solution is pointlike,  the isometry charges defined in  \eqref{eq:geoconstant} will coincide  with the Noether charges defined in~\eqref{eq:Noet-ch}.   In terms of \eqref{eq:geoconstant} and \eqref{eq:manifestisocharges},  we get
\begin{equation} \label{eq:bmn-isom-charges}
Q_T = - a_T, \qquad Q_V = - a_T b_Z^{-2} , \qquad Q_\Theta =0 , \qquad Q_{\bar 1} =0 , \qquad Q_{\bar 4} = - \sqrt{2} a_T \ .
\end{equation}
Note that these charges are independent of the deformation parameter. As a consistency check one can furthermore verify that  the hypersurface equation \eqref{eq:hypersurface} is satisfied, as it should.

The solution we described here is further subjected to the Virasoro constraints which will couple the deformed $AdS$  with the undeformed sphere subsector. For Yang-Baxter deformations the Virasoro constraints, obtained by varying the action \eqref{eq:S-eta} with respect to the worldsheet metric, can be written as~\cite{Delduc:2013qra}
 \begin{equation}\label{eq:Vira-YB}
\STr \left( A_{(\pm)}^{\alpha(2)} A_{(\pm)}^{\beta(2)} \right) = 0 \ , 
\end{equation}
where  we defined the worldsheet projections $A^\alpha_{(\pm)} = \Pi^{\alpha\beta}_{(\pm)} A_{\beta}$   of the one-form $A$ defined in~\eqref{eq:def-current}. In conformal gauge, the Virasoro constraints would read $\STr ( A_{\pm}^{(2)} A_{\pm}^{(2)} ) = 0 $, and  parametrising the group element as before as $g=g_{\alg a} g_{\alg s}$, we can  identify an $AdS$ and sphere contribution as 
\begin{equation} 
\STr \left( A_{{\alg a}\pm}^{(2)} A_{{\alg a}\pm}^{(2)} \right) + \STr \left( A_{{\alg s}\pm}^{(2)} A_{{\alg s}\pm}^{(2)} \right) = 0 
\iff
\Tr \left( A_{{\alg a}\pm}^{(2)} A_{{\alg a}\pm}^{(2)} \right) = \Tr \left( A_{{\alg s}\pm}^{(2)} A_{{\alg s}\pm}^{(2)} \right) \ ,
\end{equation}
where we used that the supertrace is defined as the trace in the AdS algebra while it is \emph{minus} the trace in the sphere algebra, and
where $A_{{\alg a}\pm}$  and $A_{{\alg s}\pm}$ are the projections of $A$ on the AdS and sphere subalgebra respectively.  On our  solution, \eqref{eq:BMNlikesol} the Virasoro constraints then require the following relation between sphere and $AdS$ variables:
\begin{equation} \label{eq:vir-gen}
\left(1 - \frac{\eta^2}{4 b_Z^4} \right) a_T^2=-\Tr \left( A_{{\alg s}\pm}^{(2)} A_{{\alg s}\pm}^{(2)} \right)  \ .
\end{equation}

%%%%%%%%%%%%%%%%%%%%%%%%%%%%%%%%%%%%%%%%%%%%%%%%%%%%%%%%%%%%%%%%%%%%%%%%
\section{Map to an undeformed yet twisted model} \label{sec:map}

Starting from this section, for simplicity we specify our discussion to the non-unimodular case. We therefore consider a deformation generated by an $R$-matrix as in~\eqref{eq:rmatrix}  with $\zeta=0$. We remind that this case gives rise to background fields that do not satisfy the type IIB supergravity equations, but only the ``modified'' supergravity equations. This study will be however preparatory  to the unimodular case, to which we will go back in section~\ref{sec:uni}.

%%%%%%%%%%%%%%%%%%%%%%%%%%%%%%%%%%%%%%%%%%%%%%%%%%%%%%%%%%%%%%%%%%%%%%%%
\subsection{Review of the twist} \label{s:reviewtwist}

Homogeneous Yang-Baxter deformed $\sigma$-models are known to be on-shell equivalent to undeformed yet twisted models, see e.g~\cite{Vicedo:2015pna, vanTongeren:2018vpb}. This is due to the fact that it is possible to identify the Lax connections of the two models, while achieving a map of the equations of motion of the two sides. In other words, there is an on-shell relation between the $\sigma$-model on the $\eta$-dependent deformed background and the $\sigma$-model on the undeformed one, and there is a one-to-one map of the solutions to  the equations of motion of the two sides.
This map, however, implies that if we choose to have \emph{periodic} boundary conditions on the worldsheet for the \emph{deformed} model (which may be motivated by the choice of studying closed strings on the deformed background), then we must generically have \emph{twisted}  boundary conditions for the \emph{undeformed} one. This situation is a generalisation of what happens for TsT transformations, where the twisted boundary conditions can be derived and written explicitly in a local and linear way~\cite{Frolov:2005ty,Frolov:2005dj}. From now on, we will use tildes on all objects of the undeformed yet twisted model. For example, if $g\in G$ denotes the group element used to construct the periodic Yang-Baxter model, we will use $\tilde g\in G$ to denote the group element of the $\eta$-independent twisted model.
The map between the two sets of degrees of freedom can be written as
\begin{equation} \label{eq:gtogtilde}
g=\mathcal F\, \tilde g \, h,
\end{equation}
where $\mathcal F\in F$ is called the ``twist field'' and  is responsible for translating between the two formulations. Recall that $F$ is the Lie group of $\alg f=\mathrm{Im}(R)$. 
In these expressions we are also allowing for  a possible gauge transformation, implemented as a right-multiplication by $h\in G^{(0)}$. This possibility is always present in the (super)coset case, given the local $G^{(0)}$-invariance of the $\sigma$-model action. 
The twist field $\mathcal F$ is fixed by demanding that the Lax connections of the deformed and undeformed models can be identified (see also~\eqref{eq:Lax}). This amounts to require that the ``modified current'' $A_\pm$ of the deformed model defined in~\eqref{eq:def-current} can be identified with the Maurer-Cartan current $\tilde J_\pm=\tilde g^{-1}\partial_\pm\tilde g$ of the undeformed model, \emph{up to} a possible gauge transformation $h$, so that   
\begin{equation}\label{eq:osm}
A_\pm  =  h^{-1}\tilde{J}_\pm  h + h^{-1} \partial_\pm h.
\end{equation}
It is important to remark that the Virasoro constraints are compatible with the on-shell identification. In fact, it is easy to see that using~\eqref{eq:osm}  the Virasoro conditions for the deformed model written in~\eqref{eq:Vira-YB} give $\STr \left( \tilde J_{(\pm)}^{\alpha(2)} \tilde J_{(\pm)}^{\beta(2)} \right) = 0$, which are precisely the Virasoro conditions for  the undeformed model. This is actually a crucial point to claim on-shell equivalence in the case of a string $\sigma$-model.
 
The fact that $g$ has periodic worldsheet boundary conditions now indeed implies that $\tilde{g}$  is not periodic in $\sigma\in [0,2\pi]$ but rather satisfies the following twisted boundary conditions
\begin{equation} \label{eq:tbc}
\tilde g(2\pi)=W\, \tilde g(0)\,  h(0)\,  h^{-1}(2\pi),
\end{equation}
where $W = {\cal F}^{-1}(2\pi) {\cal F}(0)$ is called the twist.\footnote{Notice that although $g,\tilde g, h, \mathcal F$ depend on both worldsheet coordinates $\tau$ and $\sigma$,  when writing the boundary conditions we will omit the explicit $\tau$-dependence and write down only the values that $\sigma$ takes, to have a lighter notation.} Both the twist $W$ and the twist field $\mathcal{F}$ are gauge-invariant under the right-multiplication by $h$. Moreover, it is clear that it is always possible to use a compensating gauge transformation to fix $h=1$ in~\eqref{eq:gtogtilde}. In this sense, only $W$ controls the physically relevant twisted boundary conditions, and we can talk about a \emph{left} twist only.
At the same time, we want to include the possibility of  having a non-trivial $h\neq 1$ because it may be important in certain calculations. It becomes  crucial, for example,  if we want to insist on using the same parametrisation for $g$ and $\tilde g$ in terms of coordinates.\footnote{For example, as we will do later, this happens if we parametrise $g$ as in~\eqref{eq:ge1} in terms of coordinates $\{x^0,x^1,\rho,\theta,z\}$ and $\tilde g$ in the same way but with coordinates $\{\tilde x^0,\tilde x^1,\tilde \rho,\tilde \theta,\tilde z\}$ (or equivalently in terms of the corresponding global coordinates under the transformation \eqref{eq:PtoSchr}). In this case the on-shell identification condition requires $h\neq 1$. A gauge transformation can be used to reabsorb $h$ in $g$ or $\tilde g$, but that would imply that the parameterisation of the coset representative would not be given anymore by~\eqref{eq:ge1}.} The reader may worry that the need of determining $h$ will introduce unnecessary complications in the calculations,  but in section~\ref{sec:map-sol} we will explain how to efficiently by-pass it and, if needed, easily derive the relevant $h$.

By identifying the Lax connections, it is in principle straightforward to write down an expression for ${\cal F}$ and $W$ in terms of the degrees of freedom $g$ of the deformed model, see~\cite{Vicedo:2015pna, vanTongeren:2018vpb}. In addition to being an unnatural expression for the twisting of the undeformed model, that solution is however written  in terms of a path-order exponential. It is therefore plagued by non-localities that make it unusable for practical purposes. In~\cite{Borsato:2021fuy} this problem was solved by rewriting the solution for $W$ in terms of the degrees of freedom $\tilde g$ instead. Depending on the choice of the $R$-matrix, the twisted boundary conditions may correspond to complicated non-linear relations among the coordinates evaluated at $\sigma=2\pi$ and $\sigma=0$ (a fact which is related to the non-abelian nature of the deformation) but $W$ itself takes a simple expression in terms of $\tilde g$ that only involves single (as opposed to nested) integrals.
Importantly, $W$ is constant and, therefore, it is written in terms of conserved charges of the twisted $\sigma$-models. These do not need to be Noether charges, as they may correspond to some hidden symmetries.
We refer to~\cite{Borsato:2021fuy} for a more detailed explanation on how to construct $W$ from a given $R$-matrix, where the derivation was done for the case of the PCM as well as that of (super)cosets on (semi)symmetric spaces. In fact, the generalisation from the former to the latter only involves some decorations of the formulas with certain linear combinations of projectors on the $\mathbb Z_n$-graded subspaces of $\alg g$ (with $n=2,4$). In particular, the derivation of the Jordanian twist of section 4.3 of~\cite{Borsato:2021fuy} is still valid also in the supercoset case, and it is therefore given by\footnote{Compared to~\cite{Borsato:2021fuy}, here we use the notation $\mathbf Q=\log Q_A$ and $\mathbf q=Q_B$.} 
\begin{equation}\label{eq:jord-twist}
W=\exp\left(\mathbf Q(\mathsf h-\mathbf q \ \mathsf e)\right),
\end{equation}
which takes values in the Lie subgroup $F$, and where we have
 the following expressions for the conserved charges
\begin{equation} \label{eq:Jordaniancharges}
\mathbf Q \equiv \log\frac{1-\eta\,   Y_+(\tau, 0)}{1-\eta\,   Y_+(\tau, 2\pi)},\qquad\qquad
\mathbf q \equiv \frac{Y_{\mathsf h}(\tau, 2\pi)-Y_{\mathsf h}(\tau, 0)}{Y_+(\tau, 2\pi)-Y_+(\tau, 0)},
\end{equation}
with $Y_{\mathsf h},Y_+$  projections of the Lie-algebra valued field
\begin{equation}\label{eq:Y-supcos}
\begin{aligned}
Y(\tau, \sigma) &= Y_{\mathsf h} (\tau, \sigma) \mathsf{T}^{\mathsf h} + Y_+ (\tau, \sigma) \mathsf{T}^+ \\
&= P^T \left(\int^\sigma_0 d\sigma' \hat{d}_{\tilde{g}^{-1}} \left( \partial_\tau \tilde{g} \tilde{g}^{-1} \right) (\tau, \sigma') + \int^\tau_0 d\tau' \hat{d}_{\tilde{g}^{-1}} \left( \partial_\sigma \tilde{g} \tilde{g}^{-1} \right)(\tau', 0) \right) + Y(0,0) .
\end{aligned}
\end{equation}
As anticipated, the above expression differs from the one valid in the PCM case (which can be found in  \cite{Borsato:2021fuy}) just by the insertion of the linear operator $\hat{d}_{\tilde{g}^{-1}}=\AD_{\tilde g}\hat d \AD_{\tilde g}^{-1}$, where $\hat d$ was defined in section~\ref{sec:intro-def}. Note that also $Y$ is gauge-invariant under the local right-multiplication of $\tilde{g}$ by $h$.
The above formula for the Jordanian twist $W$ will be our starting point for the construction of the classical spectral curve of our model, as well as for the computation of the one-loop quantum corrections to its spectrum.

%%%%%%%%%%%%%%%%%%%%%%%%%%%%%%%%%%%%%%%%%%%%%%%%%%%%%%%%%%%%%%%%%%%%%%%%
\subsection{Manifest symmetries of the twisted model} \label{sec:symmtwisted}
Before continuing with the study of the Jordanian deformation and the corresponding twisted model, it is useful to analyse the manifest symmetries of the latter and understand how they are related to those of the former, which were discussed in section~\ref{sec:iso}. The following discussion is valid for the whole family of homogeneous Yang-Baxter deformations, so we do not need to specify to the Jordanian case. We give the presentation for the supercoset, but the same considerations can be made already in the simpler PCM setup.

Let us define 
\begin{equation}\label{eq:Qhat}
\hat Q=\frac{1}{2\pi}\int_0^{2\pi}d\sigma \ \hat{\mathcal J}_\tau,
\end{equation}
where\footnote{In conformal gauge, it  reads  $\hat{\mathcal  J}_\tau=\AD_{\tilde g}\left(\tilde J^{(2)}_\tau-\tfrac12(\tilde J^{(1)}_\sigma-\tilde J^{(3)}_\sigma)\right)$. }
\begin{equation}\label{eq:JNoether}
\hat{\mathcal J}_\alpha=\AD_{\tilde g}\left(\tilde J^{(2)}_\alpha-\tfrac12\gamma_{\alpha\beta}\epsilon^{\beta\gamma}(\tilde J^{(1)}_\gamma-\tilde J^{(3)}_\gamma)\right).
\end{equation} 
In the undeformed and periodic (i.e.,~$W=1$) supercoset $\sigma$-model, \eqref{eq:Qhat} are the conserved  Noether charges and~\eqref{eq:JNoether}  the Noether currents corresponding to the (left) $G$ invariance of the action~\cite{Arutyunov:2009ga}.
In fact, the Noether charges $ Q$ of the deformed model defined in~\eqref{eq:Noet-ch} agree with $\hat Q$ in the limit $\eta\to 0$.
In the twisted case ($W\neq 1$),  $\hat Q$ in~\eqref{eq:Qhat} is not necessarily a conserved charge. In fact, although the Noether current is still conserved ($\partial_\alpha\hat{\mathcal J}^\alpha=0$) simply as a consequence of the equations of motion, the twisted boundary conditions may in general break the constancy in time of $\hat Q$, since
\begin{equation}
\partial_\tau \hat Q=\frac{1}{2\pi} \int_0^{2\pi}d\sigma\ \partial_\tau \hat{\mathcal J}_\tau=\frac{1}{2\pi}\int_0^{2\pi}d\sigma\ \partial_\sigma \hat{\mathcal J}_\sigma=(\AD_{W}-1)(\hat{\mathcal J}_\sigma)|_{\sigma=0},
\end{equation}
where \eqref{eq:tbc} was used. In particular, if we consider the projection of $\hat Q$ along a given generator $\mathsf{T}_A\in \alg g$ as $\hat Q_A=\STr(\hat Q \mathsf{T}_A)$, we have  $\partial_\tau\hat Q_A=\STr[(\hat{\mathcal J}_\sigma)|_{\sigma=0}(W^{-1} \mathsf{T}_AW-\mathsf{T}_A)]$. It is now clear that if some $\mathsf{T}_{\tilde A}\subset \mathsf{T}_{A}$ commutes with the whole subalgebra $\alg f$, then $\hat Q_{\tilde A}$ is a conserved charge for the twisted model. Note that $\mathrm{span}(\mathsf{T}_{\tilde A})$ forms a subalgebra of $\mathfrak{g}$. For the specific Jordanian deformation we consider in this paper, we have that the manifest bosonic symmetries are 
\begin{equation} \label{eq:symmtwist}
\mathsf{T}_{\tilde a} = \{d-M^{01} , p^0-p^1, k^0+k^1 , M^{23} \} \ .
\end{equation}
 Importantly, if $\mathsf{T}_{\tilde A}$ commutes with $\alg f$, then it is also a symmetry of the $R$-matrix and of the deformed model because it automatically solves~\eqref{eq:isomcond}.\footnote{In fact, $[[\mathsf{T}_A,RX]]=R[[\mathsf{T}_A,X]]$ because the left-hand-side is obviously 0, and because  using $Rx=-\tfrac12 R^{IJ}\STr(x \mathsf{T}_J) \mathsf{T}_I$ with $\mathsf{T}_I \in \mathfrak{f}$ the right-hand-side is $-\tfrac12 R^{IJ}\STr([\mathsf{T}_A,X]\mathsf{T}_J) \mathsf{T}_I=-\tfrac12 R^{IJ}\STr([\mathsf{T}_J,\mathsf{T}_A]X) \mathsf{T}_I=0$.} Therefore, manifest symmetries of the twisted model correspond to isometries of the YB model.

It is worth  stressing that the opposite is not always true: one may have an isometry of the YB model satisfying~\eqref{eq:isomcond}, which is not a manifest symmetry of the twisted model because $\mathsf{T}_A$ does not commute with $\alg f$. In our case of a Jordanian R-matrix, an example of this kind is given by $\mathsf e=(p^0+p^1)/\sqrt{2}$. In fact,  $\mathsf{e}$ is a symmetry of the $R$-matrix and therefore an isometry of the YB model, but it does not commute with $\alg f$, so $\hat Q_{\mathsf{e}}=\STr(\hat Q \mathsf{e})$ is not a conserved charge (in particular $W^{-1}\mathsf{e}W\neq \mathsf{e}$).

Notice that when $\mathsf{T}_A$ commutes with $\alg f$, the expression for the charges of the twisted model agree with those of the Noether charges of the deformed one, and they are related by the on-shell map~\eqref{eq:osm}. In fact, let us start from
\begin{equation} \label{eq:isomcharge}
Q=\frac{1}{2\pi}\int_0^{2\pi}d\sigma\ \mathcal J_\tau,
\end{equation}
where $\mathcal J$ was defined in~\eqref{eq:def-resnoethercurrents}, so that $ Q_A=\STr( Q \mathsf{T}_A)$ is a  conserved Noether charge of the deformed model already defined in~\eqref{eq:Noet-ch}  if $\mathsf{T}_A$ satisfies~\eqref{eq:isomcond}. Then it is straightforward to see that
\begin{align}
[[\mathsf{T}_A,\alg{f}]]=0\quad\implies\quad
\hat Q_A&=\frac{1}{2\pi}\int_0^{2\pi}d\sigma \ \STr[\tilde g[\tilde J^{(2)}_\tau-\tfrac12(\tilde J^{(1)}_\sigma-\tilde J^{(3)}_\sigma)]\tilde g^{-1} \mathsf{T}_A] \nonumber\\
&=\frac{1}{2\pi}\int_0^{2\pi}d\sigma \ \STr[\mathcal F^{-1} g[A^{(2)}_\tau-\tfrac12(A^{(1)}_\sigma-A^{(3)}_\sigma)]g^{-1}\mathcal F \mathsf{T}_A]\nonumber\\
&=\frac{1}{2\pi}\int_0^{2\pi}d\sigma \ \STr[g[A^{(2)}_\tau-\tfrac12(A^{(1)}_\sigma-A^{(3)}_\sigma)]g^{-1}\mathcal F \mathsf{T}_A\mathcal F^{-1} ]\nonumber\\
&=\frac{1}{2\pi}\int_0^{2\pi}d\sigma \ \STr[g[A^{(2)}_\tau-\tfrac12(A^{(1)}_\sigma-A^{(3)}_\sigma)]g^{-1} \mathsf{T}_A ]\nonumber\\
&= Q_A,
\end{align}
where we used $P^{(i)} ( \tilde{J} ) = h P^{(i )}( A ) h^{-1} $ for $i\neq 0$.
According to the discussion above, all manifest symmetries $\hat Q_{\tilde A}$ can be written as isometries $ Q_{\bar A}$ of the deformed model, but not all isometries $ Q_{\bar A}$ can be written as a manifest symmetry $\hat Q_{\tilde A}$, so that schematically $\{\hat Q_{\tilde A}\}\subset \{ Q_{\bar A}\}$. Because of the classical equivalence between the two models, we expect that the isometries of the deformed model that are not manifest symmetries of the twisted model should correspond to a more general family of hidden symmetries, which is always there for integrable models.

%%%%%%%%%%%%%%%%%%%%%%%%%%%%%%%%%%%%%%%%%%%%%%%%%%%%%%%%%%%%%%%%%%%%%%%%
\subsection{A simpler twist}\label{sec:fact}

In general, there is not a unique way to write the twist $W$ that controls the boundary conditions. We should rather think in terms of \emph{equivalent classes} of twists. For example, if $\tilde g$ satisfies the twisted boundary conditions \eqref{eq:tbc} and we define a new field $\tilde g'$ related to $\tilde g$ by the simple field redefinition $\tilde g'=u\tilde g$ with $u\in G$ constant, then it follows that the new field satisfies the boundary conditions
\begin{equation}
\tilde g'(2\pi)=W'\tilde g'(0) h(0)h^{-1}(2\pi),\qquad\qquad
W'=uWu^{-1}.
\end{equation}
The fields $\tilde g$ and $\tilde g'$ are equivalent  representations of the same model in terms of different variables, so that we are free to choose whether we want to work with $\tilde g$ or $\tilde g'$. Therefore, we should not think of the twisted boundary conditions as being identified by a unique expression $W$, since $W'=uWu^{-1}$ belonging to the same equivalence class actually describes the same physics.

Notice that if $W\in F$ then in general $W'$ belongs to $F'=uFu^{-1}$, in other words it belongs to an adjoint orbit of $G$ on $F$. The relation to the degrees of freedom of the YB model now is $g=\mathcal Fu^{-1} \tilde g'h$. In~\cite{Borsato:2021fuy} the solution for the twist field was constructed by fixing the  initial condition $\mathcal F(0,0)$, and while the above field redefinition breaks this choice, it can be restored by the compensating field redefinition $g'=ug$, so that we can write
\begin{equation}
g'=\mathcal F' \tilde g'h,\qquad\qquad
\mathcal F'=u\mathcal Fu^{-1}.
\end{equation}
Notice that in general the YB model written in terms of these new degrees of freedom $g'$ will be constructed not in terms of $R$ but rather by the $R$-matrix $R'=\AD_uR\AD_u^{-1}$. This is obviously a physically equivalent antisymmetric solution of the CYBE.\footnote{While the above considerations are valid for a generic $u\in G$, if $u$ is also a symmetry of $R$ then $R'=R$, and both $\mathcal F'$ and $W'$ are still elements of the unprimed $F$ subgroup, see also~\cite{Borsato:2021fuy}. It turns out that the $u$ that we are about to consider  is of this type.}
It is worth remarking that the Lax connections constructed out of $\tilde g$ and $\tilde g'$ are equal because $u$ is assumed to be constant, and therefore also the two corresponding monodromy matrices (and not just their eigenvalues) are equal to each other.

We now want to exploit the above possibility to define a new  twist $W'$ for the Jordanian deformation, that has the advantage of having a simpler expression compared to $W$. We start from~\eqref{eq:jord-twist} and we rewrite it in a ``factorised'' form by using the following identities coming from the Baker-Campbell-Hausdorff (BCH) formula:
\begin{equation}
\exp(A)\exp(B) = \exp\left(A+\frac{s}{1-e^{-s}}B\right),\quad\text{and}\quad
\exp(A)\exp(B) \exp(-A)=\exp\left(e^sB\right),
\end{equation}
which hold when $[A,B]=sB$ for   $s\neq 2\pi i n$. First, we parameterise $W$ in~\eqref{eq:jord-twist} as
\begin{equation}
W=\exp\left(A+\frac{s'}{1-e^{-s'}}B'\right) \ ,
\end{equation}
 where we have to identify
\begin{equation}
A=\mathbf Q \mathsf h,\qquad
B'=\mathbf q(e^{-\mathbf Q}-1)\mathsf e,\qquad
s'=\mathbf Q,
\end{equation}
which is consistent with $[\mathsf h,\mathsf e]=\mathsf e$. Now, using the above identities we can write
\begin{equation}\label{eq:fact-id}
\begin{aligned}
\exp\left(A+\frac{s'}{1-e^{-s'}}B'\right)
&=\exp(s'B')\exp(A)=\exp((e^s-1)B)\exp(A)\\
&=\exp(-B)\exp(e^sB)\exp(A)=\exp(-B)\exp(A)\exp(B),
\end{aligned}
\end{equation}
where we defined $B\equiv e^{s'}B'/(e^s-1)$. For consistency with $[A,B]=sB$, we must have $s=s'$.\footnote{In fact, as soon as we have $[A,B]=sB$ and we define $B'=\alpha B$ (for any non-vanishing $\alpha$) while requiring that $[A,B']=s'B'$,  it is then obvious that we must have $s'=s$.}
Notice that in our case we simply have $B=-\mathbf q\mathsf e$. To conclude, the twist can be written as
\begin{equation}
W=u^{-1}W'u,\qquad
\text{where } W'=\exp(\mathbf Q \mathsf h),
\text{ and } u=\exp(-\mathbf q\mathsf e).
\end{equation}
This is remarkable, because the new twist is identified by the Lie algebra element $\mathsf h$ only, and all information regarding $\mathsf e$ is lost in the $\tilde g'$-model. The above arguments also imply that the spectrum of both the $\tilde g$- and the $\tilde g'$-models do not depend on the charge $\mathbf q$, at least at the classical and one-loop level. Naively, the charge $\mathbf q$ controls the boundary conditions of the twisted model, but as we just showed, this dependence can be removed by a simple field redefinition. In the next section, we will see this independence of the spectrum on $\mathbf q$  more explicitly.
Notice that in the derivation of factorising the twist there is a subtlety when $\mathbf Q\to 0$, because then some formulas diverge. We will come back to this point in section~\ref{sec:uni}, where we will discuss the factorisation of the twist in a more general set-up that encompasses also the case of unimodular deformations.

%%%%%%%%%%%%%%%%%%%%%%%%%%%%%%%%%%%%%%%%%%%%%%%%%%%%%%%%%%%%%%%%%%%%%%%%
\subsection{Classical solution in the twisted model}\label{sec:map-sol-full}

\subsubsection{Map from the classical solution of the deformed model}\label{sec:map-sol}

In this section, we return to our discussion of the particular point-like solution \eqref{eq:BMNlikesol} of the Jordanian deformed model  and show how it can be mapped to a twisted solution of undeformed $AdS_5$. This means that we need to apply the reverse  logic of \cite{Borsato:2021fuy}  and derive the twist field ${\cal F}$ starting from the deformed variables. As we mentioned in \ref{s:reviewtwist}, its expression is then in general a complicated path-ordered exponential, but on our specific simple solution \eqref{eq:BMNlikesol} it can however be evaluated explicitly. In this case the easiest way to obtain $\mathcal F$ is actually to work with the associated differential system, namely
\begin{equation} \label{eq:diff-eqs-F}
\partial_\pm {\cal F} = V_\pm {\cal F} , \qquad V_\pm = \pm \eta R \AD_g \hat{d} A_\pm ,
\end{equation}
which is obtained from identifying the Lax connections of the two models. When using the initial condition ${\cal F}(0,0)=1$, we have ${\cal F}\in F$ and thus
in full generality we can write
\begin{equation}
{\cal F} = \exp \left(   f_{\mathsf h}  (\tau, \sigma) \mathsf h + f_{\mathsf e}  (\tau, \sigma) \mathsf e \right) .
\end{equation}
 Solving the differential equations \eqref{eq:diff-eqs-F} on the solution \eqref{eq:BMNlikesol} is then rather straightforward, and after imposing  ${\cal F}(0,0) = 1$, we obtain
\begin{equation}
f_{\mathsf h}(\tau , \sigma) =  \frac{\eta a_T \sigma}{b_Z^2}  , \qquad f_{\mathsf e}(\tau, \sigma) =  \frac{ \eta^3 a_T^2     \tau \sigma }{ 4\left( 1- e^{\frac{\eta a_T \sigma}{b_Z^2}} \right)b_Z^4} \ .
\end{equation}
Note that the twist field ${\cal F}$  reduces to the identity element in the undeformed $\eta\rightarrow 0$ limit, as it should.

Given the solution for ${\cal F}$, the transformation to the undeformed variables is done in principle by employing the map $g= {\cal F} \tilde{g} h$, although this may be complicated by possible gauge transformations. In fact, in our case, if we insist on parameterising both  $g$ and $\tilde{g}$ in the same way, then we cannot relate the two models if $h=1$. Guessing the appropriate gauge transformation  is not straightforward, if not unsatisfactory, and thus the translation between deformed and twisted variables begs a gauge-invariant procedure in the case of (super)coset models. To do this, we introduce the gauge-invariant objects 
\begin{equation}
\mathrm G = g K g^t \ , \qquad \mathrm{ \tilde{G} }= \tilde{g} K\tilde{g}^t \ ,
\end{equation}
where $t$ denotes usual matrix transposition and $K$ satisfies $h K h^t = K $ for $h \in SO(1,4)$ (see e.g.~§1.5.2 of \cite{Arutyunov:2009ga}).\footnote{In the matrix realisation of $\alg{su}(2,2|4)$ that is typically used (see e.g.~\cite{Arutyunov:2009ga}) one takes
\begin{equation}
K= \begin{pmatrix}
J_2 & 0\\0 & J_2
\end{pmatrix} , \qquad J_2 = \begin{pmatrix}
0& -1 \\ 1 &0
\end{pmatrix} .
\end{equation}
} Then indeed $\mathrm G$ and $\mathrm{\tilde{G}}$ are invariant under the local right-multiplications $g\rightarrow g h $ and $\tilde{g} \rightarrow \tilde{g}h$ respectively. Using these objects, the map $g= {\cal F} \tilde{g} h$ can then be rewritten in the following gauge-invariant way:
\begin{equation} \label{eq:gauge-invariant-mapping}
\mathrm G = {\cal F}\, \mathrm{\tilde{G}}\,  {\cal F}^t \ ,
\end{equation}
and also the twisted boundary conditions \eqref{eq:tbc} can  be rewritten as  
\begin{equation} \label{eq:gauge-invariant-tbc}
\mathrm{\tilde{G}} (2\pi) = W\,  \mathrm{\tilde{G}}(0)\,  W^t \ .
\end{equation}
None of these formulae now suffer from any possible complications arising from gauge ambiguities.

Using \eqref{eq:gauge-invariant-mapping}, it is now very easy to translate to the undeformed variables. We find, for $\tilde{g}$ parametrised in terms of the global coordinates,   that \eqref{eq:BMNlikesol} is mapped to
\begin{equation} \label{eq:BMNlikesoltwisted}
\tilde{T} = a_T \tau , \qquad \tilde{V} = 0 , \qquad \tilde{Z} = \exp \left( -\frac{\eta a_T \sigma}{2 b_Z^2} \right) b_Z, \qquad \tilde{P} = 0 ,
\end{equation}
while $\tilde{\Theta}$ remains free. The latter is however only a redundancy of the chosen parametrisation and we may set $\tilde{\Theta}=0$ without loss of generality. Having already determined $g,\tilde g,\mathcal F$, the unknown gauge transformation $h$ can now be simply derived as $h = \tilde{g}^{-1} {\cal F}^{-1} g$. Furthermore, knowing ${\cal F}$ we can also obtain the twist $W$ of the solution, i.e.
\begin{equation}
W = {\cal F}^{-1}(2\pi) {\cal F}(0) = \exp\left(- \frac{2\pi \eta a_T}{b_Z^2} \mathsf{h} \right) \ .
\end{equation}
 Consequently, we can identify the Jordanian charges $\mathbf Q$ and $\mathbf q$ from \eqref{eq:jord-twist} giving precisely
\begin{equation} \label{eq:BMNlikeQAQB}
\mathbf{Q}  = - \frac{2\pi \eta a_T}{b_Z^2}   , \qquad \mathbf{q} = 0 \ .
\end{equation}
Thus, for  this particular classical solution, there is no need to simplify the twist following section~\ref{sec:fact}, because $\mathbf q$ is already zero. Translating the twisted boundary conditions for $\tilde g$ in terms of explicit expressions for the associated coordinates, we see that all of them are periodic except $\tilde{Z}$, which satisfies $\tilde{Z}(2\pi) = \exp (\mathbf{Q}/2) \tilde{Z} (0)$.

There are now several consistency checks of the calculations just performed. First, one can verify that \eqref{eq:gauge-invariant-tbc} holds, while the twisted boundary conditions for $\tilde g$ have a non-trivial gauge transformation $h$ in their expression.
Secondly, knowing the solution $\tilde{g}$, one can calculate the gauge-invariant object $Y$ defined in \eqref{eq:Y-supcos}. We find 
\begin{equation}
Y_{\mathsf{h}}(\tau, \sigma) = - \frac{\eta a_T \tau}{4 b_Z^2}  , \qquad Y_+(\tau, \sigma) = \eta^{-1} \left( 1-  \exp \left( \frac{\eta a_T \sigma}{b_Z^2} \right) \right) \ ,
\end{equation}
which satisfies $Y(0,0)=0$, consistent with ${\cal F}(0,0)=1$.
Comparing now the Jordanian charges $\mathbf{Q}$ and $\mathbf{q}$ obtained from \eqref{eq:Jordaniancharges} with the expressions \eqref{eq:BMNlikeQAQB} obtained directly from the twist field, we can check that the two results agree. At last, one can verify that \eqref{eq:BMNlikesoltwisted} is  a solution to the equations of motion of the undeformed $\sigma$-model.

\subsubsection{More general classical solutions of the twisted model}

We now want to explore the question of finding  more general solutions of the undeformed twisted model. Although the boundary conditions will depend on the deformation parameter $\eta$, the equations of motion will not, and thus one may hope to find more solutions than  \eqref{eq:BMNlikesoltwisted}. To maintain some level of simplicity, however, we will continue to work on the consistent truncation $\tilde{P}(\tau, \sigma) = \tilde{\Theta}(\tau, \sigma) = 0$ and assume that the global coordinate time $\tilde{T}$ evolves linearly with the worldsheet time $\tau$, i.e.~as before $\tilde{T}(\tau, \sigma) = a_T \tau$. 

On these assumptions, let us first determine the most general solution to the twisted boundary conditions \eqref{eq:gauge-invariant-tbc} for the  Jordanian twist $W$ \eqref{eq:jord-twist} in terms of the coordinates $\tilde{Z}(\tau, \sigma)$ and $\tilde{V}(\tau, \sigma)$. It is not difficult to show that in general they must satisfy  
\begin{equation} \label{eq:gen-tbc}
\tilde{Z}(\tau, 2\pi) = \exp(\mathbf Q/2) \tilde{Z}(\tau, 0) , \qquad \tilde{V}(\tau, 2\pi) = \exp(\mathbf Q) \tilde{V}(\tau, 0) + (1-\exp(\mathbf Q))\mathbf q \ .
\end{equation}
As before one will have, however, $\tilde{g} (2\pi)\neq W \tilde{g}(0)$ implying that the group element used to construct the Lax connection of the twisted model must be dressed with a gauge transformation.

Next, we compute the equations of motion of the undeformed 
 model, and find that they reduce to the following  three  differential  equations:
\begin{align}
0 &= \partial_\tau \tilde{Z}   , \label{eq:tEOM1} \\
0 &= 2 a_T \partial_\tau \tilde{V} - (\partial_\sigma \tilde{Z})^2 + \tilde{Z} \partial_\sigma^2 \tilde{Z}  , \label{eq:tEOM2}\\
 0 &= 2 \partial_\sigma \tilde{V} \partial_\sigma \tilde{Z}  - \tilde{Z} \partial_\sigma^2 \tilde{V}  \ . \label{eq:tEOM3}
\end{align}
From \eqref{eq:tEOM1} and \eqref{eq:tEOM2} we find that $\tilde{V}$ must be linear in $\tau$ as  $\tilde{V} = \frac{1}{2 a_T} \tilde{V}_1 \tau + \tilde{V}_0$ with  
\begin{equation}
\tilde{V}_1 \equiv   (\partial_\sigma \tilde{Z})^2 - \tilde{Z} \partial_\sigma^2 \tilde{Z}  \ ,
\end{equation}
and $\tilde{V}_0 \equiv \tilde{V}_0 (\sigma)$ an unknown function of $\sigma$.
Now \eqref{eq:tEOM3} holds two equations, one at ${\cal O}(\tau^0)$ and one at ${\cal O}(\tau)$, namely
\begin{align}
-2 \partial_\sigma \tilde{V}_0 \partial_\sigma \tilde{Z} + \tilde{Z} \partial_\sigma^2 \tilde{V}_0 =0 \  , \quad \text{and} \quad -2 \partial_\sigma \tilde{V}_1 \partial_\sigma \tilde{Z} + \tilde{Z} \partial_\sigma^2 \tilde{V}_1 =0
\ ,
\end{align}
respectively.
They can be rewritten as
\begin{equation}
2 \partial_\sigma \log\tilde{Z} = \partial_\sigma \log \tilde{V}'_0 = \partial_\sigma \log \tilde{V}'_1 \ ,
\end{equation}
where the the prime denotes derivation to $\sigma$,  $\tilde{V}'_0 \equiv \partial_\sigma \tilde{V}_0$ and $\tilde{V}'_1 \equiv \partial_\sigma \tilde{V}_1$. 
Since all the objects involved only depend on $\sigma$, we can  perform one integration to find
\begin{equation} \label{eq:tEOMint1}
\tilde{V}'_0 = c_0 \tilde{Z}^2 , \qquad \tilde{V}'_1 = c_1 \tilde{Z}^2 , 
\end{equation}
where $c_0$ and $c_1$ are integration constants. Only the latter equation of \eqref{eq:tEOMint1} is now  a differential equation for $\tilde{Z}$, and once we obtain its solution, the solution for $\tilde{V}_0$ (and consequently of $\tilde{V}$) is obtained immediately by performing a straightforward integration of the first equation of \eqref{eq:tEOMint1}. First note that we can rewrite the expression for $\tilde{V}'_1$ as
\begin{equation}
\tilde{V}'_1 = \tilde{Z}' \tilde{Z}'' - \tilde{Z} \tilde{Z}''' = -\tilde{Z}^2 \left( \frac{\tilde{Z}''}{\tilde{Z}} \right)' \ .
\end{equation}
 Then the second equation of \eqref{eq:tEOMint1} becomes $\left(\tilde{Z}'' \tilde{Z}^{-1} \right)' = -c_1$ which means we can do another simple integration to find
\begin{equation} \label{eq:stokeseq}
\tilde{Z}'' + (c_1 \sigma + c_2) \tilde{Z} = 0 \ ,
\end{equation}
with $c_2$ another integration constant. Here the experienced reader may recognise the Airy (or Stokes) differential equation that has as linearly independent solutions the Airy functions $\mathrm{Ai}(x)$ and $\mathrm{Bi}(x)$. Concluding, within our assumptions given in the beginning of this section, the most general solution for the variable $\tilde{Z}$ of the undeformed model is
\begin{equation}
\tilde{Z}(\tau , \sigma) = \alpha \mathrm{Ai} \left( - \frac{c_1 \sigma + c_2}{(-c_1)^{2/3}} \right)  + \beta \mathrm{Bi} \left( - \frac{c_1 \sigma + c_2}{(-c_1)^{2/3}} \right) \ ,
\end{equation}
\emph{before} imposing any boundary condition. These are oscillatory functions that at a certain turning point become exponential. In principle, we can now also write down an explicit solution for $\tilde{V}$, as mentioned before, however we will refrain from doing so because it is not particularly enlightening. Nevertheless, note that, because of the additional integration performed to obtain $\tilde{V}_0$, the system has one additional integration constant (say $c_3$). Before imposing boundary conditions, the full solution thus has in total 7 free parameters, namely $a_T$, $c_0$, $c_1$, $c_2$, $c_3$, $\alpha$ and $\beta$.

 Imposing the twisted boundary conditions \eqref{eq:gen-tbc} gives three equations for $\tilde{Z}, \tilde{V}_1$ and $\tilde{V}_0$, obtained order by order in $\tau$. The equation for $\tilde{Z}$ fixes an expression for $\mathbf{Q}$ while the one for $\tilde{V}_0$ fixes $\mathbf{q}$. The remaining condition on $\tilde{V}_1$ becomes a constraint between the parameters,\footnote{On the solution for $\tilde{Z}$ and $\tilde{V}_1$ the constraint simply reads $\tilde{V}_1(2\pi) \tilde{Z}(0)^2 = \tilde{V}_1(0) \tilde{Z}(2\pi)^2$. } reducing the number of free parameters to 6. Two further constraints between the parameters may arise when comparing the expressions of $\mathbf{Q}$ and $\mathbf{q}$ with those obtained from \eqref{eq:Jordaniancharges} and \eqref{eq:Y-supcos}, as these must of course coincide for consistency. We have not looked into this further.

We close this section by discussing the special $c_1=0$ case. From \eqref{eq:stokeseq} it is obvious that the most generic solution for $\tilde{Z}$ then is simply
\begin{equation}
\tilde{Z}(\sigma) = \alpha \exp (\sqrt{-c_2} \sigma) + \beta \exp(-\sqrt{-c_2} \sigma) .
\end{equation}
Depending on the sign of $c_2$ these are exponential  or oscillatory functions.  The simple solution given in \eqref{eq:BMNlikesoltwisted}, which is the one we study in the remainder of this paper, falls into the former class. It has $\alpha = 0$, $\beta = b_Z$ and $\sqrt{-c_2} =  \frac{\eta a_T}{2b_Z^2}$. In general for $\tilde{V}$ we then get $\tilde{V} = - c_0\frac{b_Z^2 \exp \left(- \frac{\eta a_T \sigma}{b_Z^2}\right)}{\eta a_T} + c_3$ which seems more general than \eqref{eq:BMNlikesoltwisted}. Consistency with the boundary conditions \eqref{eq:gen-tbc} however sets $c_0=0$ and $c_3=\mathbf{q}$. Here we see explicitly that a possible contribution from the $\mathbf{q}$ charge can be eliminated by a field redefinition, which  is in fact a translational isometry in the $V$ coordinate.

%%%%%%%%%%%%%%%%%%%%%%%%%%%%%%%%%%%%%%%%%%%%%%%%%%%%%%%%%%%%%%%%%%%%%%%%
\section{Classical spectral curve}\label{sec:csc}

Knowing a Lax connection that is flat on the $\sigma$-model equations of motion allows one to apply the methods of classical integrability. While in principle it is possible to work from the point of view of the Yang-Baxter deformed $\sigma$-model, we find it more convenient to work from the point of view of the undeformed yet twisted $\sigma$-model, which is on-shell equivalent to the former. In fact, this is in analogy to how the classical and quantum spectrum of TsT deformations is obtained, see e.g.~\cite{Frolov:2005ty,Beisert:2005if}. Here, we want to give a brief summary of facts that are valid in the undeformed and periodic case and that carry over also to the twisted one. For more details, we refer to the original literature developed in the periodic case, see e.g.~\cite{Beisert:2005bm}.

The Lax connection of the twisted $\sigma$-model on the $AdS_5\times S^5$ supercoset is\footnote{In conformal gauge, this is just ${\mathcal L}_\pm (z) = \tilde J_\pm^{(0)} + z \tilde J_{\pm}^{(1)}+ z^{\mp 2} \tilde J_{\pm}^{(2)}+ z^{- 1} \tilde J_{\pm}^{(3)} $. The Lax connection of the Yang-Baxter deformed model is obtained simply by replacing $\tilde J$ by $A$.}
\begin{equation}\label{eq:Lax}
\mathcal L_\alpha=\tilde J^{(0)}_\alpha+\frac12\left(z^2+\frac{1}{z^2}\right)\tilde J^{(2)}_\alpha
-\frac12\left(z^2-\frac{1}{z^2}\right)\gamma_{\alpha\beta}\epsilon^{\beta\gamma}\tilde J^{(2)}_\gamma
+z\ \tilde J^{(1)}_\alpha
+\frac{1}{z}\ \tilde J^{(3)}_\alpha,
\end{equation}
which agrees with the Lax connection of the undeformed and periodic case~\cite{Bena:2003wd,Alday:2005gi} because boundary conditions play no role in its construction. Here, $\tilde J^{(i)}=P^{(i)}\tilde J$ are projections of the currents on the $\mathbb Z_4$-graded subspaces of $\alg{psu}(2,2|4)$, and $\gamma^{\alpha\beta}=\sqrt{|h|}h^{\alpha\beta}$.
From the Lax connection one can construct the monodromy matrix
\begin{equation}
\Omega (z,\tau) = {\cal P} \exp \left( - \int^{2\pi}_0 d\sigma' {\cal L}_\sigma(z,\tau,\sigma') \right) ,
\end{equation}
with $z\in \mathbb C$ the spectral parameter.
In the undeformed and periodic case, the eigenvalues $\lambda(z)$ of $\Omega(z,\tau)$ are conserved because $\mathcal L_\tau(2\pi)=\mathcal L_\tau(0)$ as a consequence of  the periodic boundary conditions. When considering twisted boundary conditions as in~\eqref{eq:tbc}, $\mathcal L_\tau$ remains periodic as long as we take into account the compensating gauge transformation. Therefore, the eigenvalues $\lambda(z)$ of $\Omega(z,\tau)$ are again time-independent, and encode the infinite number of conserved quantities (when expanding in powers of the spectral parameter $z$), rendering the model classically integrable. The eigenvalues depend analytically on $z$ except at $z=0,\infty$ (where the poles of the Lax connection imply essential singularities for $\lambda(z)$) and at the points where two (or more) eigenvalues degenerate. Instead of working with the eigenvalues, it is simpler to work with the quasimomenta $p(z)$, defined as $\lambda=e^{ip}$, as the essential singularities of $\lambda$ at $z=0,\infty$ become poles. For the quasimomenta, we use the same notation as for the undeformed and periodic case
\begin{equation}
\{\hat p_1(z),\hat p_2(z),\hat p_3(z),\hat p_4(z)||\tilde p_1(z),\tilde p_2(z),\tilde p_3(z),\tilde p_4(z)\} ,
\end{equation}
where $\hat p_i(z)$ with $i=1,\ldots,4$ are the quasimomenta of the (deformed/twisted) $AdS$ factor corresponding to $SU(2,2)\subset PSU(2,2|4)$ and $\tilde p_i(z)$ with $i=1,\ldots,4$ are those of the (in this case undeformed/untwisted) sphere factor corresponding to $SU(4)\subset PSU(2,2|4)$.
When two eigenvalues of the same type (i.e.~both $\hat p$ or both $\tilde p$) degenerate, they give rise to branch points of square-root cuts, that correspond to collective \emph{bosonic} excitations. When two eigenvalues of different type (i.e.~one $\hat p$ and the other $\tilde p$) degenerate, they give rise to poles, that correspond to \emph{fermionic} excitations.
The quasimomenta are everywhere analytic except at these bosonic branch points and fermionic poles, and at $z=0,\infty$.
Because of its definition, the quasimomentum has a $2\pi \mathbb Z$ ambiguity, but this ambiguity is lost when considering $dp$, which can be thought of as an abelian differential. The spectral problem can therefore be reformulated in terms of the classification of the admissible algebraic curves with abelian differential $dp$. These curves can be understood as a collection of $4+4$ sheets (one for each of the quasimomenta) connected by the above cuts and poles.
We follow the usual conventions and employ an alternative useful parameterisation of the spectral parameter, taking $z=\sqrt{\frac{1+x}{1-x}}$ so that the points $z=0,\infty$ correspond to $x=-1,+1$.

Let us continue with the list of the facts that are valid in the undeformed and periodic $AdS_5\times S^5$ case and that continue to be valid also in the twisted one. The Lax connection is supertraceless, which implies that the superdeterminant of the monodromy matrix is 1. In turn, this implies the ``supertraceless condition''\footnote{In principle one can modify the above equation by adding a shift by $2\pi n$ with $n\in \mathbb Z$. This can be removed by the freedom of shifting the quasimomenta by any integer multiple of $2\pi$. In the periodic case one can use this freedom so that at large $x$ the quasimomenta go like $p(x)\sim \mathcal O(1/x)$. In the twisted case  the twist can introduce additional finite terms in the strict $x\to\infty$ limit. However, equation~\eqref{eq:suptrl} remains valid because these finite shifts cancel each other in the above equation, as a consequence of $\det W=1$.}
\begin{equation}\label{eq:suptrl}
\sum_i\hat p_i=\sum_i\tilde p_i.
\end{equation}
Moreover, the $\mathbb Z_4$ automorphism of $\alg{psu}(2,2|4)$ is still implemented by sending $z\to i z$ or equivalently $x\to 1/x$ at the level of the Lax connection. We will call this the ``inversion symmetry''. At the level of the quasimomenta, it implies~\cite{Beisert:2005bm}
\begin{equation} \label{eq:inversion-symm}
\hat p_k(1/x) = -\hat p_{k'}(x),\qquad\qquad
\tilde p_k(1/x) = -\tilde p_{k'}(x)+2\pi m\epsilon_k,
\end{equation}
where $k=(1,2,3,4)$ and $k'=(2,1,4,3)$. Notice that this permutation is related to the choice of charge conjugation matrix in $\alg{psu}(2,2|4)$. For the sphere quasimomenta, the inversion symmetry allows for a possible shift with $\epsilon_k=(1,1,-1,-1)$ and $m\in \mathbb Z$, which is related to winding around the $S^5$ space. For the $AdS$ quasimomenta, this possibility is not allowed by the requirement of absence of winding for the time coordinate.

Combining the supertraceless condition, the inversion symmetry, and the Virasoro constraints,  one has a synchronisation\footnote{When we later study one-loop corrections, we will slightly relax this synchronisation of the poles to $p(x) \sim \frac{ \mathrm{diag} \left(  \alpha^{(\pm)} , \alpha^{(\pm)}, \beta^{(\pm)}, \beta^{(\pm)} |  \alpha^{(\pm)} , \alpha^{(\pm)}, \beta^{(\pm)}, \beta^{(\pm)} \right)}{x \pm 1} + {\cal O}(x\pm 1 )^0$, because in that case one should consider also fermionic contributions and the tracelessness condition of $\alg{su}(2,2)$ and $\alg{su}(4)$ of the purely bosonic classical case is no longer enforced. \label{f:synchronization}} of the poles of the quasimomenta at $x=\pm 1$ also in the twisted case
\begin{equation}
p(x) = \frac{ \mathrm{diag} \left(  \alpha^{(\pm)} , \alpha^{(\pm)}, -\alpha^{(\pm)}, -\alpha^{(\pm)} |  \alpha^{(\pm)} , \alpha^{(\pm)}, -\alpha^{(\pm)}, -\alpha^{(\pm)} \right)}{x \pm 1} + {\cal O}(x\pm 1 )^0 \ .
\end{equation}
The bosonic cuts and the fermionic poles must respect the inversion symmetry, which means that in both cases they must come in even number. For example,  we can write the gluing condition of two quasimomenta on a cut (or more generally on a collection of cuts) $\hat{\mathcal C}_{(k,l)}$ or $\tilde{\mathcal C}_{(k,l)}$ as
\begin{equation}\label{eq:ppn}
\begin{aligned}
\hat p_k(x+i\epsilon)- \hat p_l(x-i\epsilon)=2\pi \hat n_{(k,l)},\qquad \text{for } \ x\in \hat{\mathcal C}_{(k,l)} ,\\
\tilde p_k(x+i\epsilon)- \tilde p_l(x-i\epsilon)=2\pi \tilde n_{(k,l)},\qquad \text{for } \ x\in \tilde{\mathcal C}_{(k,l)} ,
\end{aligned}
\end{equation}
and similar equations must hold for the cuts which are the images of $\hat{\mathcal C}_{(k,l)}$ and $\tilde{\mathcal C}_{(k,l)}$ under $x\to 1/x$. Similarly, if there is a fermionic pole at $x^*_{(k,l)}$, by inversion symmetry there will be another pole for quasimomenta at $1/x^*_{(k,l)}$. Because of this property, we can declare a ``physical" region as the one for which $|x|>1$. The cuts in the physical region (i.e.~half of the cuts in the pair of cuts connected by inversion symmetry) will be considered the ``fundamental'' ones.

To conclude, let us also say that we can still define the filling fractions for sphere cuts $\tilde{\mathcal C}_{(k,l)}$ connecting the sheets $\tilde k$ and $\tilde l$ as
\begin{equation}
\tilde K_{(k,l)} = -\frac{\sqrt{\lambda}}{8\pi^2i}\oint dx\left(1-\frac{1}{x^2}\right)\tilde p_k
=\frac{\sqrt{\lambda}}{8\pi^2i}\oint \left(x+\frac{1}{x}\right)d\tilde p_k
=-\frac{\sqrt{\lambda}}{8\pi^2i}\oint \left(x+\frac{1}{x}\right)d\tilde p_l \ ,
\end{equation}
where the integration path is defined by a closed loop that encircles the \emph{fundamental} cut $\tilde{\mathcal C}_{(k,l)}$. Similarly, we define the filling fractions  for $AdS$ cuts $\hat{\mathcal C}_{(k,l)}$ connecting the sheets $\hat k$ and $\hat l$ as
\begin{equation}
\hat K_{(k,l)} = -\frac{\sqrt{\lambda}}{8\pi^2i}\oint dx\left(1-\frac{1}{x^2}\right)\hat p_l
=\frac{\sqrt{\lambda}}{8\pi^2i}\oint \left(x+\frac{1}{x}\right)d\hat p_l
=-\frac{\sqrt{\lambda}}{8\pi^2i}\oint \left(x+\frac{1}{x}\right)d\hat p_k. \label{ffAdS} \ ,
\end{equation}
where the integration path is defined by a closed loop that encircles the \emph{fundamental}  cut $\hat{\mathcal C}_{(k,l)}$.
In the periodic case these filling fractions are identified as the action variables of the integrable model \cite{Dorey:2006zj}, and they are taken to be quantised as integers in a semiclassical treatment \cite{Beisert:2005bm}. In the twisted case we can do the same, because the identification of the definition of the filling fractions comes from the position of the poles of the Lax connection, and these are still at $x=\pm 1$. The filling fractions can also be written in terms of the so-called twist function~\cite{Delduc:2012qb}, and this is invariant under an homogeneous Yang-Baxter deformation~\cite{Vicedo:2015pna}.

In the periodic case, the expansion of the monodromy matrix around $z=1$ (or equivalently $x=\infty$) yields (at the lowest non-trivial order) the conserved Noether charges coming from the $\alg{psu}(2,2|4)$ superisometry. In the twisted case, this picture is modified by the presence of $W$. In fact, after considering a gauge transformed Lax connection $\mathcal L'_\sigma= \tilde g\mathcal L_\sigma \tilde g^{-1} - \partial_\sigma \tilde g \tilde g^{-1}$, one finds\footnote{For simplicity, we are writing these formulas in the case that the twisted boundary conditions are just $\tilde g(2\pi)=W\tilde g(0)$, i.e.~$h=1$ in~\eqref{eq:tbc}. If $h\neq 1$, one simply has to use a gauge transformation by $\tilde gh$ rather than just $\tilde g$, and the right-hand-side of~\eqref{eq:epsilon-Omega} is still given by the same expression.}
\begin{equation}
\tilde g(0)\Omega(z)\tilde g(0)^{-1} =W^{-1}\mathcal P\exp\left(-\int_0^{2\pi}{d\sigma} \ \mathcal L_\sigma'\right),
\end{equation}
which, when writing $z=1+\epsilon$, leads to 
\begin{equation}\label{eq:epsilon-Omega}
\tilde g(0)\Omega(z=1+\epsilon)\tilde g(0)^{-1} =W^{-1}+4\pi\epsilon \ W^{-1}\hat Q +\mathcal O(\epsilon^2),
\end{equation}
where $\hat Q$ was defined in~\eqref{eq:Qhat}.  In the periodic case ($W=1$), $\hat Q$ is the Lie-algebra valued conserved Noether charge, and one can use a $PSU(2,2|4)$ automorphism to put it in the Cartan subalgebra. After doing that, the monodromy matrix is diagonal at least to order $\epsilon$ and the eigenvalues are therefore written in terms of the Cartan charges. In the twisted case, the monodromy matrix is non-trivial already at order zero in the $\epsilon$-expansion, and the quasimomenta will depend also on the eigenvalues of $W$. Generically, $\hat Q$ on its own will not be conserved in the twisted case. As we will see in the next section, when computing the eigenvalues of the above expansion  they are not  necessarily written in terms of the standard Cartan charges of $\alg{psu}(2,2|4)$: there is a ``polarisation'' induced by the presence of $W$, so that instead the quasimomenta are written in terms of the conserved Cartan charges of the algebra of symmetries of the twisted model (which we discussed in section \ref{sec:symmtwisted}). The symmetries of the twisted model form a subalgebra of $\alg{psu}(2,2|4)$, and the corresponding choice of Cartan subalgebra may not be equivalent to the  one of the periodic case. We will now be more explicit about the computation of the eigenvalues of the monodromy matrix around $z=1$.

Let us emphasise the great advantage of making the above discussion from the point of view of the \emph{twisted} rather than \emph{deformed} model. In the latter case the expansion of the monodromy matrix around $z=1$ would give rise to complicated non-local expressions that would be very hard to work with.

%%%%%%%%%%%%%%%%%%%%%%%%%%%%%%%%%%%%%%%%%%%%%%%%%%%%%%%%%%%%%%%%%%%%%%%%
\subsection{Asymptotics of the quasimomenta}\label{sec:asymptotics}

To obtain the expression for the quasimomenta around $z=1$, we must compute the eigenvalues of the expression \eqref{eq:epsilon-Omega}. Since the Jordanian twist $W$ given in~\eqref{eq:jord-twist} is diagonalisable, we can restrict our discussion to diagonalisable matrices. 
The explanation that we are about to give is actually valid for a generic diagonalisable twist.

The diagonalisation of~\eqref{eq:epsilon-Omega} to order $\epsilon$ follows the same route as the diagonalisation of a Hamiltonian in perturbation theory of quantum mechanics. Let us rephrase that procedure in a different language.
First, consider the diagonalisation of $W^{-1}$ as $D = S^{-1} W^{-1} S$. To be as general as possible, we allow eigenvalues to be degenerate
\begin{equation}
D = \mathrm{diag} ( d_1 , \ldots , d_1 , d_2 , \ldots , d_2 , \ldots ) ,
\end{equation}
where the eigenvalue $d_i$ comes with multiplicity $m_i$. We will introduce projectors $P^{[i]} = \mathrm{diag} (0,0,\ldots , 0 , 1, 1, \ldots , 1 , 0 , 0 , \ldots , 0)$ that have only $m_i$ entries with value $1$ in correspondence with the eigenvalues $d_i$, and $0$ otherwise. Note that $P^{[i]} D = D P^{[i]} = d_i P^{[i]}$.
We can then write for $\Omega'(z) = \tilde{g}(0) \Omega(z) \tilde{g}(0)^{-1}$
\begin{equation}
\Omega' (z=1+\epsilon)= S (D+\epsilon X ) S^{-1} + {\cal O}(\epsilon^2)  , \qquad X \equiv 4\pi D S^{-1} \hat{Q} S \ .
\end{equation}
From simple reasoning in linear algebra, we know that to compute the $\epsilon$ correction to the eigenvalues of $D$, we must first project
\begin{equation}
X^{[i]}\equiv P^{[i]}XP^{[i]},
\end{equation}
and then compute the eigenvalues of $X^{[i]}$. If there is a block corresponding to eigenvalues all of which have multiplicity 1, then the above is equivalent to projecting on this non-degenerate block, and the diagonal part of this block will give the correction to the eigenvalues.

With this procedure, it is easy to compute the expansion of  the eigenvalues. In our case, we find in the  $AdS$ sector
\begin{equation}
\begin{aligned}
d_1&=1+2\pi\epsilon(-C_1-C_3)+\mathcal O(\epsilon^2),\\
d_2&=e^{+\mathbf{Q}/2}(1+2\pi\epsilon(C_1-C_2))+\mathcal O(\epsilon^2),\\
d_3&=e^{-\mathbf{Q}/2}(1+2\pi\epsilon(C_1+C_2))+\mathcal O(\epsilon^2),\\
d_4&=1+2\pi\epsilon(-C_1+C_3)+\mathcal O(\epsilon^2), \\
\end{aligned}
\end{equation}
where 
\begin{equation}
\begin{aligned}
   C_1&=i\hat Q_{{[M^{23}]}} ,\\
   C_2&=\hat Q_{[d+M^{01}]}- \sqrt{2} \mathbf{q} \hat Q_{[p^0+p^1]},\\
   C_3&=\sqrt{(\hat Q_{[d-M^{01}]})^2+ \hat Q_{[-p^0+p^1]}\hat Q_{[k^0+k^1]}
   },
   \end{aligned}
\end{equation}
and $\hat{Q}_i = \STr ( \hat{Q} \mathsf{T}_i)$.
The explicit expression for $\mathbf{q}$ given in \eqref{eq:Jordaniancharges} can be written precisely as $\mathbf{q} = \hat{Q}_{\mathsf h}/\hat{Q}_{\mathsf e} = \hat{Q}_{[d+M^{01}]}/(\sqrt{2}\hat{Q}_{[p^0+p^1]}) $  which implies  $C_2=0$. This fact is very important. It means that the only components of $\hat{Q}$ that contribute to the eigenvalues of the monodromy matrix are those which correspond to the \emph{symmetries of the twisted model}, see \eqref{eq:symmtwist}. In fact, we managed to get rid of both $d+M^{01}$ and $p^0+p^1$, which do not have this property. We remind that all of the symmetries of the twisted model are also isometries of the YB model, and in terms of the isometric charges ${Q}_{\bar{a}} = \STr (\mathsf{T}_{\bar{a}} {Q})$ with ${Q}$ defined in \eqref{eq:isomcharge} and the isometric generators $\mathsf{T}_{\bar{a}}$ in \eqref{eq:isoms}, we can write
\begin{equation}
   C_1=i{Q}_{\bar{5}},\qquad\qquad
   C_3=\sqrt{({Q}_{\bar{1}})^2+ (-{Q}_{\bar{2}}+{Q}_{\bar{3}}){Q}_{\bar{4}}}.
\end{equation}
Again, the combination ${Q}_{\bar{2}}+{Q}_{\bar{3}}$ does not appear: it would be an isometry of the deformed model, but it is not a symmetry  of the twisted model.

The appearance of a square root is archetypal in the computations of eigenvalues, and in fact it appears also in the undeformed periodic case. It is, however, an awkward feature for practical computations, especially when computing semi-classical corrections to the conserved charges of the model. Nevertheless, as in the periodic case see e.g.~\cite{Kazakov:2004qf}, one can use an automorphism to get rid of the square root and in fact  transform the expressions to ones which only involve Cartans of the symmetry algebra. First, as done in appendix~\ref{a:CSA},  we identify $\mathsf h= \tfrac12 (d-M^{01}),\mathsf e_+= \tfrac12 (p^0-p^1),\mathsf e_-=-\tfrac12 (k^0+k^1)$, which close into a $\mathfrak{sl}(2, \mathbb{R})$ subalgebra of the algebra of isometries $\mathfrak{k}$. We can construct dual generators $\mathsf h^{\star}
=\tfrac12 \mathsf h, \mathsf e^\pm=\mathsf e_\mp$ spanning a dual $\mathfrak{sl}(2,\mathbb R)$ algebra. We can then construct $Q_{\mathfrak{sl}(2,\mathbb R)}=Q_{\mathsf h} \mathsf h^{\star}+Q_+\mathsf e^++Q_-\mathsf e^- = ({Q}_{\bar{1}} \mathsf h^{\star} + ({Q}_{\bar{2}} - {Q}_{\bar{3}}  ) \mathsf e^+ - {Q}_{\bar{4}} \mathsf e^- )/2$. It is easy to check that $2\STr[(Q_{\mathfrak{sl}(2,\mathbb R)})^2]=({Q}_{\bar{1}})^2+ (-{Q}_{\bar{2}}+{Q}_{\bar{3}}){Q}_{\bar{4}}$ and thus
\begin{equation}
C_3 = \sqrt{2\STr[(Q_{\mathfrak{sl}(2,\mathbb R)})^2]} \ .
\end{equation}
Now recall that the generator of global time translations is \eqref{eq:globaltimegenerator}, which we can write as $\mathsf H = \frac{1}{\sqrt{2}} (2 \mathsf e_+ - \mathsf e_-)$. The global energy is thus $E = \STr({Q_{\mathfrak{sl}(2,\mathbb R)}} \mathsf H)=\frac{1}{\sqrt{2}} ( {Q}_{\bar{2}} - {Q}_{\bar{3}} + \frac{1}{2} {Q}_{\bar{4}})$. The other generators in $\mathfrak{sl}(2,\mathbb{R})$ can then be rotated such that they close into an $\mathfrak{so}(2,1)\sim \mathfrak{sl}(2,\mathbb{R})$ algebra. Indeed, define $\mathsf{T}_E \equiv \mathsf H/2$, $\mathsf{T}_P \equiv (\mathsf e_+ + \mathsf e_-/2 )/\sqrt{2}$, then $\mathsf{T}_J \equiv \mathsf h$  $[\mathsf{T}_J,\mathsf{T}_E]=\mathsf{T}_P$, $[\mathsf{T}_J,\mathsf{T}_P]=\mathsf{T}_E$ and $[\mathsf{T}_E,\mathsf{T}_P]=\mathsf{T}_J$ hold.\footnote{This is the algebra of $AdS_2$, or the Lorentz algebra in three dimensions. Hence, $\mathsf{T}_P$ can be interpreted as a momentum generator, and $\mathsf{T}_J$ as a boost.} We can now define the dual generator $\mathsf H^\star$ of global time translations as $\STr(\mathsf H^{\star} \mathsf{T}_P) = \STr(\mathsf H^{\star} \mathsf{T}_J) = 0$ and $\STr(\mathsf H^{\star} \mathsf H)=1$, which fixes 
\begin{equation}
\mathsf H^\star = - \frac{\mathsf e_+}{\sqrt{2}} + \frac{\mathsf e_-}{2\sqrt{2}} = - \frac{\mathsf{T}_{\bar{2}} - \mathsf{T}_{\bar{3}}}{2\sqrt{2}}- \frac{\mathsf{T}_{\bar{4}}}{4\sqrt{2}} \ .
\end{equation}
At this point we use the classification of adjoint orbits in $\mathfrak{sl}(2,\mathbb{R})$ (see e.g.~appendix~\ref{a:CSA}). Depending on the algebra element, the adjoint action may bring it to something proportional to $\mathsf h, \mathsf e_+$ or $\mathsf e_+-  \mathsf e_-$. When discussing the classical spectral curve, we always assume that we are considering classical solutions that correspond to highest weight states belonging to the same orbit as the generator of time translations $\mathsf H$. In other words, we assume that by an inner automorphism, we can transform $Q_{\mathfrak{sl}(2,\mathbb R)}$ to the form $Q_{\mathfrak{sl}(2,\mathbb R)}= E \mathsf H^\star$. In that case, one has   $\STr[(Q_{\mathfrak{sl}(2,\mathbb R)})^2] = -E^2/2$ such that we simply get
\begin{equation}
C_3 = i E \ .
\end{equation}

The $AdS$ quasimomenta in the $x$ variable now simply expand as 
\begin{equation}
\begin{aligned}
\hat{p}_1&\sim  -\frac{2\pi( {Q}_\Theta+ E)}{x}+\mathcal O(x^{-2}),\\
\hat{p}_2&\sim -\frac{i}{2} \mathbf{Q}+  \frac{2\pi{Q}_\Theta}{x}+\mathcal O(x^{-2}),\\
\hat{p}_3&\sim +\frac{i}{2}\mathbf{Q}+ \frac{2\pi{Q}_\Theta}{x}+\mathcal O(x^{-2}),\\
\hat{p}_4&\sim  \frac{ 2\pi( -{Q}_\Theta+ E)}{x}+\mathcal O(x^{-2}).
\end{aligned} \label{pasymptotic}
\end{equation}
where we defined ${Q}_\Theta \equiv {Q}_{\bar{5}}$,  since it is the angular momentum charge related to rotations by the angle $\Theta$. Notice that the asymptotics of the $AdS$ quasimomenta depend only on three conserved charges, namely the energy $E$ and the angular momentum ${Q}_\Theta$ at $\mathcal O(x^{-1})$, and the charge $\mathbf Q$ at order $\mathcal O(x^0)$, which  controls the twisted boundary conditions. This situation should be compared to the three charges (energy and two spins) that appear in the asymptotics of the quasimomenta in the periodic case at $\mathcal O(x^{-1})$.

%%%%%%%%%%%%%%%%%%%%%%%%%%%%%%%%%%%%%%%%%%%%%%%%%%%%%%%%%%%%%%%%%%%%%%%%

\subsection{Classical algebraic curve for the BMN-like solution} \label{s:CSC-BMN}

Let us now consider a specific classical solution and derive the corresponding quasimomenta. We will do this for the BMN-like solution that we have constructed.
We can compute the quasimomenta  using either the deformed or undeformed description.  On the one hand, in the deformed picture (when the solution is given by~\eqref{eq:BMNlikesol}), the Lax connection is clearly $\sigma$-independent, so that the path-ordered exponential in the definition of the monodromy matrix  reduces to a simple matrix exponential of ${\cal L}_\sigma$. On the other hand, in the twisted picture (when the solution is given by~\eqref{eq:BMNlikesoltwisted}), the solution depends on $\sigma$, making the computation potentially more involved. However, this dependence can be eliminated via a gauge transformation and leads to the same result. This is in fact not surprising, since the two models share the same Lax connection, so that the computation is bound to be the same. Given that we can ignore the path-ordering in the exponential,  to obtain the quasimomenta we can then simply compute the eigenvalues of ${\cal L}_\sigma$ itself. In terms of the spectral parameter $x$, we find
\begin{equation}\label{eq:res-p}
\begin{aligned}
\hat{p}_1(x) =-\hat{p}_4(x) =   \frac{2\pi  a_T \sqrt{\frac{  x^2-\beta^2}{(x-1)^2}}}{
   (x+1)} , \qquad
\hat{p}_2(x) = - \hat{p}_3(x) =\frac{2\pi  a_T x \sqrt{\frac{ 1-\beta^2 x^2}{(x-1)^2}}}{
   (x+1)}  ,
\end{aligned}
\end{equation}
where we defined $\beta\equiv\frac{\eta}{2 b_Z^2}$, and we remind that $a_T$ and $b_Z$ are parameters that enter the classical solution, see e.g.~\eqref{eq:BMNlikesol}.  In the vanishing deformation limit, which now corresponds to the $\beta\rightarrow 0$ limit, they reduce to the meromorphic quasimomenta associated to the BMN solution of undeformed $AdS_5\times S^5$ \cite{Gromov:2007aq} and in that sense, the solution \eqref{eq:BMNlikesol} (equivalently \eqref{eq:BMNlikesoltwisted}) is a ``BMN-like" solution.  The deformation, however, introduces  a cut between sheet $\hat{1}$ and $\hat{4}$ and a cut between sheet $\hat{2}$ and $\hat{3}$. Note that the inversion symmetry \eqref{eq:inversion-symm} is satisfied. For later convenience, let us write the quasimomenta in the  physical region $|x|>1$
\begin{equation}\label{eq:res-p2}
\begin{aligned}
\hat{p}_1(x) =-\hat{p}_4(x) = \frac{2\pi a_T \sqrt{x^2 - \beta^2}}{x^2-1} , \qquad \hat{p}_2(x) = - \hat{p}_3(x) = \frac{2\pi a_T x\sqrt{1 - x^2 \beta^2}}{x^2-1} \ .
\end{aligned}
\end{equation}
Matching these explicit expressions with the general asymptotics of the quasimomenta \eqref{pasymptotic} gives us the following expressions of the charges involved: 
\begin{equation}
{Q}_\Theta=0, \qquad E = Q_T= - a_T , \qquad \mathbf{Q} = - 4\pi a_T  \beta \ ,
\end{equation}
with matches with \eqref{eq:bmn-isom-charges}  and \eqref{eq:BMNlikeQAQB}.

In addition to the kinematics in the deformed $AdS_5$ space, we will take our solution to have non-trivial kinematics in the $S^5$ space as well. Specifically, we consider a pointlike string travelling with constant velocity around a big circle of $S^5$, $\phi=\omega \tau$.  Here, $\omega$ is proportional to the conserved charge associated to an angular momentum on $S^5$. As the $S^5$ part of our space is not deformed, we can borrow the quasimomenta from the usual BMN solution~\cite{Gromov:2007aq}
\begin{equation}\label{eq:sphere-qm}
    \tilde{p}_1=\tilde{p}_2=-\tilde{p}_3=-\tilde{p}_4= \frac{2\pi \omega x}{x^2-1} \ .
\end{equation}

In the solution that we are considering, the Virasoro condition \eqref{eq:vir-gen} reads
\begin{equation}\label{eq:vir-spec}
    \omega=a_T \sqrt{1-\beta^2} \ .
\end{equation}
This implies that if want both $\omega$ and $a_T$ to be real, we must have $|\beta|\leq 1$.
Notice that, as we commented above, the Virasoro constraint can be reinterpreted as a synchronisation condition on the poles of the quasimomenta at $x=\pm 1$.

%%%%%%%%%%%%%%%%%%%%%%%%%%%%%%%%%%%%%%%%%%%%%%%%%%%%%%%%%%%%%%%%%%%%%%%%
\section{Quantum corrections to the BMN-like solution} \label{sec:qCSC}

In this section, we study the quantum corrections to the quasimomenta related to  our specific BMN-like   solution. Specifically, we apply the recipe presented in \cite{Gromov:2007aq,Gromov:2008ec} to compute the one-loop correction to its energy. Despite the fact that those articles  originally present this technique for undeformed $AdS_5 \times S^5$, it has been successfully applied also to deformed backgrounds, such as the flux-deformed $AdS_3 \times S^3$ \cite{Lloyd:2013wza,Babichenko:2014yaa,Nieto:2018jzi} or the Schrödinger background \cite{Ouyang:2017yko}.

%%%%%%%%%%%%%%%%%%%%%%%%%%%%%%%%%%%%%%%%%%%%%%%%%%%%%%%%%%%%%%%%%%%%%%%%
\subsection{Corrections to the classical quasimomenta} \label{CorrectionClassical}

Although we are not working on undeformed $AdS_5 \times S^5$, most of the road map to construct quantum fluctuations described in \cite{Gromov:2007aq,Gromov:2008ec} applies to our deformed background.
The main idea behind this computation is to introduce (quantum) excitations in the form of cuts so small that we can treat them as poles. As in the macroscopic case, if the microscopic excitation we are adding connects two sheets associated to hatted quasimomenta or two sheets associated to a tilded quasimomenta,  the excitation is bosonic. If the excitation we are adding connects a tilded quasimomenta with a hatted quasimomenta, the excitation is fermionic.

These new cuts will modify the quasimomenta we computed in the previous section, but the properties that they must fulfil are restrictive enough to fully constrain how they are altered. The starting point is the fact that the corrections to the quasimomenta cannot alter the gluing condition on a cut (\ref{eq:ppn}), that is
\begin{equation}
    (p_k + \delta p_k) (x+i\epsilon) - (p_l + \delta p_l) (x-i \epsilon ) =2\pi n_{(k,l)},\qquad \text{for } \ x\in \hat{\mathcal C}_{(k,l)} 
    \quad\text{or } \ x\in \tilde{\mathcal C}_{(k,l)} \ .
\end{equation}
This condition is true  for the macroscopic cuts associated to the classical solution, which imposes the following condition on the perturbations:
\begin{equation}
    \delta p_k (x+i\epsilon) - \delta p_l (x-i \epsilon ) =0 ,\qquad \text{for } \ x\in \hat{\mathcal C}_{(k,l)} \quad\text{or } \ x\in \tilde{\mathcal C}_{(k,l)} \ .
\end{equation}
A similar condition is imposed for the \emph{microscopic cuts} that we are introducing to compute the quantum corrections, and this fixes the positions, $x_n^{kl}$, where these cuts can be placed by
\begin{equation}
    p_k (x_n^{kl}) - p_l (x_n^{kl}) =2\pi n \ , \label{PolePosition}
\end{equation}
where the mode number $n$ takes integer values.  Notice that, due to inversion symmetry (\ref{eq:inversion-symm}), we only focus on solutions with $|x_n^{kl}|>1$. We  say that an excitation is in the \emph{physical region} if the position of the pole fulfils that property.

Now that we know where to place the new cuts/poles, we also need information regarding their residues at $x_n^{kl}$. We will not discuss the explicit expression of the residues at this point, but instead   will only talk about the relative sign between the residues in the two sheets. We will provide more details on the residues in appendix~\ref{StructureFluctuations}. 
For bosonic excitations, the poles are actually infinitesimally small square-root cuts connecting two sheets. This means that the residues on the two sheets have to have opposite signs. For fermionic excitations, the poles are actual poles, and they have to have the same residue. This can be summarised as
\begin{align}
    \res_{x=x^{ij}_n} \hat{p}_k &= (\delta_{i\hat{k}} - \delta_{j\hat{k}} )\alpha (x_n^{ij} ) N^{ij}_n \ , & \res_{x=x^{ij}_n} \tilde{p}_k &= (\delta_{j\tilde{k}} - \delta_{i\tilde{k}} ) \alpha (x_n^{ij} ) N^{ij}_n \ ,
\end{align}
with $i<j$ taking values $\hat{1}$, $\hat{2}$, $\hat{3}$, $\hat{4}$, $\tilde{1}$, $\tilde{2}$, $\tilde{3}$, $\tilde{4}$. Here, $N^{ij}_n$ indicates the number of quantum excitations in the physical region that connect $p_i$ with $p_j$ and have mode number $n$, and $\alpha(x_n^{ij})$ is the residue of the pole associated to that excitation. A more detailed explanation on the origin of these signs can be found in \cite{Beisert:2005bm}.

The number of excitations $N^{ij}_n$ are not generically free: they have to satisfy the  level-matching condition
\begin{equation}
\sum_n n \sum_{\text{all } ij} N^{ij}_n = 0 \ .
\end{equation}
This condition can be understood  from a mathematical perspective as a consequence of the Riemann bilinear identity \cite{Beisert:2005bm}. Physically,  because the new poles modify the value of the filling fractions (see also footnote \ref{f:normalisation-filling-N}), the $N^{ij}_n$ can be interpreted as the amplitudes of the physical modes, which are related to mode numbers  through conventional string level-matching.

In addition to these pieces of information, these corrections have some important features that they inherit from properties of the classical quasimomenta. Among those, the most relevant for us are  the inversion symmetry  of \eqref{eq:inversion-symm}
\begin{equation}
    \delta p_k (x)= -\delta p_{k'} (1/x) \ , \label{inversioncorrections}
\end{equation}
and the synchronisation of the poles at $\pm 1$
\begin{equation}\label{eq:poles-synch}
\delta p(x) \sim \frac{ \mathrm{diag} \left(  \delta\alpha^{(\pm)} , \delta\alpha^{(\pm)}, \delta\beta^{(\pm)}, \delta\beta^{(\pm)} |  \delta\alpha^{(\pm)} , \delta\alpha^{(\pm)}, \delta\beta^{(\pm)}, \delta\beta^{(\pm)} \right)}{x \pm 1} + \dots \ .
\end{equation}
Notice that we are using the relaxed version of the synchronisation of poles as commented on in footnote~\ref{f:synchronization}, because  we will have to consider also fermionic excitations when computing quantum corrections.

The final piece of information that we need to compute the perturbations is their asymptotic behaviour for large values of $x$. This can be obtained by analysing the asymptotic behaviour of the quasimomenta (\ref{pasymptotic}). 
We require\footnote{In all the following expressions we   rescale our conserved charges with a factor of $\sqrt{\lambda}$, so that the reader can compare them with the previous literature more easily. We want to stress that the normalisation factor $\frac{4\pi}{\sqrt{\lambda}}$ can be modified in the sense that it does not affect the computation of the anomalous contribution to the energy $\delta \Delta$, at the condition that we also modify the normalisation of the residues $\alpha(x_n^{ij})$ consistently. However, the natural way to fix this normalisation factor is by demanding that the correction to the filling fractions $K_{(i,j)}$ is given by $N_{ij}$, i.e., $-\frac{\sqrt{\lambda}}{8\pi^2i}\oint dx\left(1-\frac{1}{x^2}\right) \left( p_i + \delta p_i \right)= K_{(i,j)}+N_{ij}$. In this sense, the normalisation chosen for the filling fractions fixes also the normalisation of the asymptotics of $\delta p$. \label{f:normalisation-filling-N}}
\begin{equation}
	\begin{pmatrix}
	\delta \hat{p}_1 \\
	\delta \hat{p}_4 \\
	\delta \tilde{p}_1 \\
	\delta \tilde{p}_2 \\
	\delta \tilde{p}_3 \\
	\delta \tilde{p}_4
	\end{pmatrix} = \frac{4\pi}{x \sqrt{\lambda}} \begin{pmatrix}
	+\frac{\delta \Delta}{2} + N_{\hat{1} \hat{4}} + N_{\hat{1} \hat{3}} + N_{\hat{1} \tilde{3}} + N_{\hat{1} \tilde{4}}   \\
	-\frac{\delta \Delta}{2} - N_{\hat{1} \hat{4}} - N_{\hat{2} \hat{4}} - N_{\tilde{2} \hat{4}} - N_{\tilde{1} \hat{4}} \\
	- N_{\tilde{1} \tilde{3}} - N_{\tilde{1} \tilde{4}} - N_{\tilde{1} \hat{3}} - N_{\tilde{1} \hat{4}} \\
	- N_{\tilde{2} \tilde{3}} - N_{\tilde{2} \tilde{4}} - N_{\tilde{2} \hat{3}} - N_{\tilde{2} \hat{4}}  \\
	+ N_{\tilde{1} \tilde{3}} + N_{\tilde{2} \tilde{3}} + N_{\hat{1} \tilde{3}} + N_{\hat{2} \tilde{3}} \\
	+ N_{\tilde{1} \tilde{4}} + N_{\tilde{2} \tilde{4}} + N_{\hat{1} \tilde{4}} + N_{\hat{2} \tilde{4}}
	\end{pmatrix} + \mathcal{O} (x^{-2}) \ , \label{asymptotics}
\end{equation}
where $\delta \Delta$ is the anomalous correction of the energy and where $N_{ij} = \sum_{n} N^{ij}_n$ is the total number of poles in the physical region connecting sheet $i$ and $j$. 
The above formula can be justified in the same way as in the undeformed periodic case. In particular, one can assume that the addition of quantum excitations produces integer shifts in the charges that appear at order $1/x$ in the large-$x$ asymptotics of the quasimomenta~\eqref{pasymptotic}, which justifies the presence of $N_{ij}$. Here we are also assuming that only the energy is allowed to receive anomalous corrections, and we will see that the results will be consistent with this. 
Notice that above we have not included the asymptotics of $\hat{p}_2$ and $\hat{p}_3$. The reason is that the classical asymptotic behaviour~\eqref{pasymptotic} suggests that their form is non-standard (i.e.~they might be finite in the $x\to\infty$ limit) as a consequence of the twist. We will actually not need them to compute $\delta \Delta$, and we refer to~\cite{Ouyang:2017yko} for a similar strategy applied in the case of the Schrödinger background. Nevertheless, we discuss the asymptotic behaviour of $\hat{p}_2$ in detail in section~\ref{a:anomalousQ} and, importantly, we will show that the charge $\mathbf{Q}$ does not get any anomalous correction.

Combining all the properties we have enumerated, we possess enough information about the analytic structure of these corrections to be able to reconstruct them. As this is a tedious and sometimes repetitive process, we have relegated the details regarding this reconstruction to appendix~\ref{StructureFluctuations}, while here we collect the final results. If we denote each contribution to $\delta \Delta$ as
\begin{equation}
    \delta \Delta = \sum_n \Omega^{ij} (x_n) N^{ij}_n   \ , 
\end{equation}
we find the following expressions:
\begin{align}
    &\Omega^{\tilde{1} \tilde{3}} (x_n) = \Omega^{\tilde{1} \tilde{4}} (x_n) = \Omega^{\tilde{2} \tilde{3}} (x_n) =\Omega^{\tilde{2} \tilde{4}} (x_n) =  2 \frac{K(1)}{x_n^2-1} \ , \\
    &\Omega^{\hat{1} \hat{4}} (x_n)= 2\frac{x_n^2 K(1/x_n) }{x_n^2 -1 } -2 \ ,  \\
    &\Omega^{\hat{2} \hat{3}} (x_n)= \frac{2 K(x_n)}{x_n^2 -1} \ ,  \\
    &\Omega^{\hat{1} \hat{3}} (x_n)=\Omega^{\hat{2} \hat{4}} (x_n)= \frac{x_n^2 K(1/x_n) + K(x_n)}{x_n^2-1}-2  \ , \\
    &\Omega^{\hat{1} \tilde{3}} (x_n)=\Omega^{\hat{1} \tilde{4}} (x_n)=\Omega^{\tilde{1} \hat{4}} (x_n)=\Omega^{\tilde{2} \hat{4}} (x_n)= \frac{x_n^2 K(1/x_n) +K(1)}{x_n^2-1} -1 \ , \\
    &\Omega^{\hat{2} \tilde{3}} (x_n)=\Omega^{\hat{2} \tilde{4}} (x_n)=\Omega^{\tilde{1} \hat{3}} (x_n)=\Omega^{\tilde{2} \hat{3}} (x_n)= \frac{K(x_n) +K(1)}{x_n^2-1} \ ,
\end{align}
where $K(x)=\sqrt{ 1 -  x^2\beta^2}$.

The final step is to use equation~(\ref{PolePosition}) to find the value of the spectral parameter at which we have to place the microscopic cuts, $x_n$, and substitute it in the above equations. As the process is the same for all the cases \textit{mutatis mutandis}, we will illustrate it in the case $\hat{1} \tilde{3}$. The algebraic equation $\hat{p}_1 (x_n) - \tilde{p}_3 (x_n) = 2\pi n$ has two solutions
\begin{equation}
    x_n = \frac{a_T \sqrt{1-\beta^2} \pm \sqrt{a_T^2 +n^2}}{n} \ .
\end{equation}
Because $\sqrt{a_T^2 +n^2}\geq a_T \geq a_T \sqrt{1-\beta^2}$ for real values of $n$ and $|\beta|\leq 1$ (as required from reality conditions, see below \eqref{eq:vir-spec})  the solution with minus sign has modulus smaller than one, and we will disregard it. Substituting this result into $\Omega^{\hat{1} \tilde{3}} (x_n)$, we find
\begin{equation}
    \Omega^{\hat{1} \tilde{3}} (x_n)= \frac{x_n^2 K(1/x_n) +K(1)}{x_n^2-1} -1= \sqrt{\frac{a_T^2 + n^2}{a_T^2}}-1 \ .
\end{equation}
The remaining contributions take the following form:
\begin{align}
    &\Omega^{\tilde{1} \tilde{3}} (x_n) = \Omega^{\tilde{1} \tilde{4}} (x_n) = \Omega^{\tilde{2} \tilde{3}} (x_n) =\Omega^{\tilde{2} \tilde{4}} (x_n) = -\sqrt{1-\beta^2} + \sqrt{1 - \beta^2 +\frac{n^2}{a_T^2}} \ , \label{SFrec} \\
    &\Omega^{\hat{1} \hat{4}} (x_n)= -2 + \sqrt{2 + \frac{n^2}{a_T^2} + 2 \sqrt{1 + \frac{n^2}{a_T^2} (1-\beta^2)}} \ , \label{AdSFrec1} \\
    &\Omega^{\hat{2} \hat{3}} (x_n)= \sqrt{2 + \frac{n^2}{a_T^2} - 2 \sqrt{1 + \frac{n^2}{a_T^2} (1-\beta^2)}} \ , \label{AdSFrec2} \\
    &\Omega^{\hat{1} \hat{3}} (x_n)=\Omega^{\hat{2} \hat{4}} (x_n)= -1 + \sqrt{1+ \beta^2 + \frac{n^2}{a_T^2}} \label{AdSFrec3} \ , \\
    &\Omega^{\hat{1} \tilde{3}} (x_n)=\Omega^{\hat{1} \tilde{4}} (x_n)=\Omega^{\tilde{1} \hat{4}} (x_n)=\Omega^{\tilde{2} \hat{4}} (x_n)=-1 + \sqrt{1+\frac{n^2}{a_T^2}}\ , \\
    &\Omega^{\hat{2} \tilde{3}} (x_n)=\Omega^{\hat{2} \tilde{4}} (x_n)=\Omega^{\tilde{1} \hat{3}} (x_n)=\Omega^{\tilde{2} \hat{3}} (x_n)= -\sqrt{1-\beta^2} + \sqrt{1+\frac{n^2}{a_T^2}}\ .
\end{align}
In order to have an independent check of our results, we have computed the bosonic $AdS$ contributions using the method of quadratic fluctuations in the picture of the deformed model. The details of that computation are collected in appendix~\ref{QuadraticFluctuations}, and the final results are consistent with the ones obtained using the classical spectral curve of the twisted model.

Another non-trivial check that our results pass is the fact that we recover the usual BMN contributions in the undeformed limit (see, for example, \cite{Frolov:2003tu}), i.e.~$\Omega=-1+\sqrt{1+\frac{n^2}{\omega^2}}$, upon using the Virasoro constraint \eqref{eq:vir-spec}.

%%%%%%%%%%%%%%%%%%%%%%%%%%%%%%%%%%%%%%%%%%%%%%%%%%%%%%%%%%%%%%%%%%%%%%%%

\subsection{Anomalous correction to the twist} \label{a:anomalousQ}

Before computing the one-loop correction to the dispersion relation of the BMN-like solution, we want to make some comments on the asymptotic behaviour of $\delta \hat{p}_2$  and its effect on the charge $\mathbf{Q}$ of the twist. In particular, we will show that there is no contribution to the $\mathcal{O} (x^0)$ term of the expansion.
To that end, we will use the inversion symmetry~(\ref{inversioncorrections}) to compute $\delta \hat{p}_2$ from the different $\delta \hat{p}_1$ excitations that we constructed in appendix~\ref{StructureFluctuations}.

Let us begin with the excitation $\tilde{2} \tilde{3}$. Using inversion symmetry on
\begin{equation}
    \delta \hat{p}_1 (x) = \frac{4\pi}{\sqrt{\lambda}} \frac{K(1)}{K(1/x)} \frac{x_n}{x_n^2-1} \frac{2x}{x^2-1} \ ,  
\end{equation}
we can check that
\begin{equation}
\delta \hat{p}_2 (x) = -\delta \hat{p}_1 \left( \frac{1}{x}\right) \approx \frac{8\pi}{x\sqrt{\lambda}} \frac{K(1)}{\sqrt{-\beta^2}} \frac{x_n}{x_n^2-1} + {\cal O}(x^{-2}) \ .
\end{equation}
Despite getting a non-trivial ${\cal O}(x^{-1}) $ term, we can check that it actually vanishes when we impose the level-matching condition. The other three excitations associated to the $S^5$ have the same behaviour, as $\delta \hat{p}_1$ has a similar form for all four.

The situation is very similar for both the excitations associated to $\hat{1} \hat{4}$ and $\hat{1} \tilde{3}$. In particular, we have
\begin{align}
    \delta \hat{p}_2 &\approx \sum_n \frac{K(1/x_n)}{\sqrt{-\beta^2}} \frac{\alpha (x_n) N_n}{x_n x} + {\cal O}(x^{-2}) \ , \\
    \delta \hat{p}_2 &\approx \sum_n \frac{\alpha (x_n)}{ x_n \sqrt{-\beta^2}} \frac{K(1/x_n) + K(1)}{2x } N_n + {\cal O}(x^{-2}) \ ,
\end{align}
respectively. In both cases, the $1/x$ term is proportional to the pole-fixing condition (\ref{PolePosition}) and, thus, proportional to $n$. This means that both the $x^0$ and $x^{-1}$ term vanish once the level-matching condition is imposed.

The situation for the excitations associated to $\hat{1} \hat{3}$ and $\hat{2} \tilde{3}$ is slightly different. For these kinds of excitations, we find that
\begin{align}
    \delta \hat{p}_2 &\approx - \sum_n 
    \left( \frac{4\pi}{\sqrt{\lambda}} + \frac{\alpha (x_n)}{x_n} \frac{K(x_n) + K(1/x_n)}{\sqrt{-\beta^2}} \right) \frac{N_n}{2x} + {\cal O}(x^{-2}) \ , \\
    \delta \hat{p}_2 &\approx \sum_n \left( \frac{4\pi}{\sqrt{\lambda}} + \frac{\alpha (x_n)}{x_n} \frac{K(x_n) + K(1) }{\sqrt{-\beta^2}} \right) \frac{N_n}{2x} + {\cal O}(x^{-2}) \ ,
\end{align}
respectively. The analysis in both cases is the same: the $x^{0}$ contribution vanishes automatically, while the second term in the $x^{-1}$ contribution vanishes due to the level-matching condition. However, in both cases, we have a non-vanishing $\mp \sum_n \frac{2\pi N_n}{\sqrt{\lambda}}$ at order $x^{-1}$.

The behaviour of the remaining excitations can be inferred using the different tricks described in appendix~\ref{StructureFluctuations}, e.g.~the composition of excitations described in \ref{a:composition} or the fact that the classical solution has pairwise symmetric quasimomenta \eqref{symmetricquasimomenta}.

Let us summarise and compare our results with what we would expect from the asymptotic behaviour of the quasimomenta \eqref{pasymptotic}. First, we find that the $x^{0}$ terms of the expansion of $\delta \hat{p}_2$ for large values of $x$ vanish in all the cases. In \eqref{pasymptotic}, these terms corresponds to the conserved charge $\mathbf{Q}$, associated to the twist. Hence, we find that $\mathbf{Q}$ does not receive any anomalous correction, $\delta \mathbf{Q}=0$. 
On the other hand, depending on the excitation we are considering, the $x^{-1}$ terms either vanish (after imposing the level-matching condition) or give a constant contribution. In particular, $\hat{1} \hat{3}$, $\hat{2} \hat{4}$, $\hat{2} \tilde{3}$, $\hat{2} \tilde{4}$, $\tilde{1} \hat{3}$ and $\tilde{2} \hat{3}$  fall in this second category, where the two bosonic excitations contribute with $-N_n/2$ and the four fermionic excitations with $+N_n/2$.\footnote{Despite not being mentioned in the analysis of \cite{Ouyang:2017yko}, a similar behaviour can be found in the case of the Schrödinger background.} If we try to interpret this from the lens of undeformed $AdS_5 \times S^5$, the appearance of, e.g., a contribution proportional to $N^{\hat{1} \hat{3}}_n/2$ in $\delta\hat{p}_2$ is surprising. Due to the presence of the twist, however, the contribution of certain excitations to the asymptotics of $\delta\hat{p}_2$ and $\delta\hat{p}_3$ do in fact mix.   Note in particular that the only non-vanishing excitations are those which connect one sheet affected by the twist with a sheet that is not affected by the twist.  Comparing with \eqref{pasymptotic} the new asymptotics suggest a shift of the spin charge $Q_\Theta$ as 
\begin{equation}
Q_\Theta \rightarrow Q_\Theta - N_{\hat{1}\hat{3}} - N_{\hat{2}\hat{4}}  + N_{\hat{2} \tilde{3}} + N_{\hat{2} \tilde{4}} + N_{\tilde{1} \hat{3}} + N_{\tilde{2} \hat{3}}  \ .
\end{equation}
Let us emphasise that this contribution is non-dynamical, as it only depends on the number of excitations $N_n^{ij}$ but not on the position of the poles $x_n^{ij}$. Similar shifts of spin charges are found also in undeformed  $AdS_5 \times S^5$, see e.g.~\cite{Gromov:2007aq}, where the spins  get corrections  in which the contributions of fermionic excitations are weighted with a factor of $1/2$ relative to those of bosonic excitations.
It would be interesting to understand this behaviour for twisted models in more generality.

%%%%%%%%%%%%%%%%%%%%%%%%%%%%%%%%%%%%%%%%%%%%%%%%%%%%%%%%%%%%%%%%%%%%%%%%
\subsection{One-loop correction to the energy} \label{sec:one-loop}

Now we have enough information to compute the one-loop correction to the energy of our classical string. This is done by considering that the excitations we are introducing behave as harmonic oscillators, and computing their ground state energy. Mathematically, this means that
\begin{equation}
    E\approx E_{\text{class.}} +E_{1-\text{loop}}=E_{\text{class.}} + \frac{1}{2} \sum_{n\in \mathbb{Z}} \sum_{ij} (-1)^{F_{ij}} \Omega^{ij} (x_n) \ ,
\end{equation}
where we have $F_{ij}=0$ for bosonic and $F_{ij}=1$ for fermionic contributions, respectively, and where $E_{\text{class.}} = -E = \sqrt{\lambda} a_T$ (recall that we rescale charges with $\sqrt{\lambda}$).

Before attempting to perform the sum over $n$, we should check if the series of the addends converges. We can check that it is the case, as
\begin{equation}
    \sum_{ij} (-1)^{F_{ij}} \Omega^{ij} (x_n) = -\frac{2\beta^2}{n^3} +\mathcal O(n^{-4})\ ,
\end{equation}
meaning that the partial sums over the mode number have to converge, and we are allowed to compute the sum over integer numbers.

First of all, we should notice that the mode number $n$ always appears divided by $a_T$. We can safely assume that $a_T \gg 1$ because it is related to the energy of our classical solution. This allows us to approximate our sum over $n\in \mathbb Z$ by an integral over $n\in \mathbb R$.\footnote{Here, we can use the integration method proposed in \cite{Schafer-Nameki:2006dtt} to argue that the error in our approximation has to be exponentially small in $a_T$, as none of our contributions have a branch cut for real values of $n$.} Then
\begin{equation}
   \frac{1}{2}  \sum_{n\in \mathbb{Z},ij} (-1)^{F_{ij}} \Omega^{ij} (x_n) 
   \approx \frac{a_T }{2} \int_{-\infty}^\infty dn\sum_{ij} (-1)^{F_{ij}} \Omega^{ij} (x_{a_T n}) =a_T \int_{0}^\infty dn\sum_{ij} (-1)^{F_{ij}} \Omega^{ij} (x_{a_T n})  \ ,
\end{equation}
where by $x_{a_T n}$ we mean that we are rescaling $n$ by a factor $a_T$. Notice that, to get the second equality, we have used that the frequencies only depend on $n$ quadratically and we are therefore dealing with an even integrand.

At this point, it is useful to divide our contributions into two different types
\begin{align}
\Omega_{\text{nested}} &= \Omega^{\hat{1} \hat{4}} (x_n)+ \Omega^{\hat{2} \hat{3}} (x_n)= \sqrt{2} \sqrt{2+n^2+\sqrt{n^2 (n^2 + 4 \beta^2)}} \ , \\
\Omega_{\text{other}} &= -8 \sqrt{1+n^2} + 4\sqrt{1+n^2-\beta^2}+2\sqrt{1+n^2+\beta^2}\ .
\end{align}
We do so because the nested square root structure makes the integration of $\Omega_{\text{nested}}$ a bit involved.
Let us first address the integral of $\Omega_{\text{other}}$. Instead of integrating from 0 to $\infty$, we shall integrate only up to a positive cut-off, $\Lambda$. We do so because, while the total series is convergent, the two separate contributions $\Omega_{\text{nested}}$ and $\Omega_{\text{other}}$ are not. Obviously, they have divergences that later will cancel each other. After some algebra, we find
\begin{equation}
\begin{aligned}
\frac{I_{\text{other}}}{a_T} ={}& 2 \lim_{\Lambda \rightarrow \infty} \int_{0}^{\Lambda} dn \, \Omega_{\text{other}} =-2 \Lambda ^2 - 2(1+\beta^2)\log (2 \Lambda )  \\
& -\left[ (1+ \beta ^2)    + (1+ \beta ^2) \log \left( 1 + \beta ^2 \right)  +2\left( 1 -\beta ^2\right) \log \left(1-\beta ^2\right) \right] 
\end{aligned}
\end{equation}

The integral over $\Omega_{\text{nested}}$ is more intricate, but it can be simplified using a change of variables inspired by the one proposed in Appendix C of \cite{Minahan:2005qj}. In particular, after we change into a variable $y(n)$ that eliminates the nested integral, i.e. $ \Omega_{\text{nested}} = \sqrt{4+y(n)^2} $, we can perform the integral as
\begin{equation}
   \lim_{\Lambda \rightarrow \infty} \int_{0}^{\Lambda} dn \, \Omega_{\text{nested}}  \\
= \lim_{\Lambda \rightarrow \infty} \int_{0}^{\Lambda'} dy \, \sqrt{4+y^2} \frac{y^3+8 y \beta^2}{2 \sqrt{y^2 + 4 \beta^2}^3} \ ,
\end{equation}
where $\Lambda' = 2 \Lambda + \frac{\beta^2}{\Lambda} + \mathcal{O} (\Lambda^{-3})$. Notice that, although we are interested in the limit of very large $\Lambda$, we have kept the next-to-leading order of the map between the two cut-offs. We have done that because the integral will diverge as $\Lambda^{\prime 2}$, which means that this subleading order of the transformation will give rise to a non-trivial contribution to the integral. After evaluating the integral, and with a healthy dose of algebra, we arrive at
\begin{equation}
\begin{aligned}
\frac{I_{nested}}{a_T} ={}& 2 \lim_{\Lambda \rightarrow \infty} \int_{0}^{\Lambda} dn \, \Omega_{\text{nested}} =2 \Lambda ^2 + 2(1+\beta^2)\log (2 \Lambda )  \\
& +1+2\beta - \beta^2 -(1+\beta ^2) \log \left(1-\beta ^2\right) + \left( 1 + \beta ^2\right) \log \left(\frac{1-\beta }{1+\beta }\right) \ .
\end{aligned}
\end{equation}
Putting both integrals together, it is immediate to check that the divergent parts perfectly cancel, giving us
\begin{equation}
\begin{aligned}
    \frac{2E_{1-\text{loop}}}{a_T} ={}&\frac{I_{nested}+I_{\text{other}}}{a_T}=2 \beta -2 \beta^2 -(3-\beta ^2) \log \left(1-\beta ^2\right)  \\
    &-(1+ \beta ^2) \log \left( 1 + \beta ^2 \right) +\left( 1 + \beta ^2\right) \log \left(\frac{1-\beta }{1+\beta }\right) \ .
\end{aligned}
\end{equation}

As a check of our expression, we can examine two particularly interesting values of $\beta$. On the one hand, we can check that $E_{1-\text{loop}}=0$ for $\beta=0$. This is consistent with the fact that we recover the BMN solution in the undeformed limit, which does not receive corrections. On the other hand, in the $\beta=1$ limit, the form of $\Omega_{\text{nested}}$ simplifies, as we can get rid of the nested square root
\begin{equation}
    \lim_{\beta \rightarrow 1} \Omega_{\text{nested}} = \sqrt{4+2n^2+2\sqrt{n^2 (n^2 + 4)}}= \sqrt{n^2}+\sqrt{n^2 + 4} \ .
\end{equation}
Thus, in this limit all the contributions have the form of a square root and there is no need to separate the integral into two divergent contributions, avoiding the possible issues that a cut-off may introduce. When taking $\beta\to 1$, the limit of the integral and the integral of the limit coincide and give $E_{1-\text{loop}}=-a_T \log (8)$.

To end this section, we want to elaborate on our choice of cut-off. As the deformation changes the masses associated to the different quantum fluctuations differently, we may worry that an issue with the choice of regularisation similar to the one present in $AdS_4 \times \mathbb{CP}^3$ may arise, see e.g.~\cite{Gromov:2008fy,Bandres:2009kw}, where  a unique summation prescription does not exist and one needs further information to choose the correct one. While for $AdS_5$ all the quantum excitations have the same mass, for the less-symmetrical $AdS_4$ they can take two possible values, which allows one to separate them into ``heavy'' or ``light'' modes. As  the  number of bosonic and fermionic heavy modes is the same, and idem for light modes, one is allowed to  choose a different cut-off for each type leading to a different result. Although our situation may look similar at first glance, the mass of the excitations in our background are associated to their origin, i.e.~if they are bosonic modes associated to the sphere, or to the deformed $AdS$ space, or to fermions. Consequently, although a different regularisation scheme may exist, it would be highly non-trivial.

%%%%%%%%%%%%%%%%%%%%%%%%%%%%%%%%%%%%%%%%%%%%%%%%%%%%%%%%%%%%%%%%%%%%%%%%
\section{The unimodular case and the spectral equivalence}\label{sec:uni}
So far we have considered the non-unimodular deformation, i.e.~we have set $\zeta=0$ in~\eqref{eq:rmatrix}. Here, we wish to show that these results can be  extended  to  unimodular deformations ($\zeta=1$) thanks to the observation that even in this case the twist can be factorised as explained in section~\ref{sec:fact} for the non-unimodular case. 
First, an extended Jordanian $R$-matrix of the type~\eqref{eq:rmatrix} with $\zeta=1$  solves the classical Yang-Baxter equation if the odd elements satisfy the (anti)commutation relations~\cite{Borsato:2016ose,vanTongeren:2019dlq,progressJord} 
\begin{equation}
\label{eq:QQ}
[\mathsf Q_i,\mathsf e]=0,\qquad
[\mathsf h,\mathsf Q_1]=\tfrac12 \mathsf Q_1-\xi \mathsf Q_2,\qquad
[\mathsf h,\mathsf Q_2]=\tfrac12 \mathsf Q_2+\xi \mathsf Q_1,\qquad
\{\mathsf Q_j,\mathsf Q_k\}=-i\delta_{jk}\mathsf e,
\end{equation}
where $\xi$ is a free real parameter.
From~\cite{Borsato:2021fuy}, we know that for a generic homogeneous Yang-Baxter deformation the twist can be taken to be in the subgroup $F$, and therefore, it can be written as $W^{-1} =\exp(\eta R\mathcal Q)$, where $\mathcal Q$ is a conserved charge that takes values in the dual (with respect to the bilinear form of $\alg g$) of $\alg f$.  That means that we can write it as
\begin{equation}\label{eq:twist-uni-gen}
\begin{aligned}
W^{-1} &=\exp(\eta R\mathcal Q)
=\exp\left[\eta\left(-\mathcal Q_{\mathsf e} \mathsf h+\mathcal Q_{\mathsf h} \mathsf e-i (\mathcal Q_1 \mathsf Q_1+\mathcal Q_2 \mathsf Q_2)\right)\right],
\end{aligned}
\end{equation}
where  $\mathcal Q_A$ are projections of $\mathcal Q$. We will not calculate them explicitly because this will not be needed for the following argument, but notice that, when setting $\zeta=1$, in principle all of them can receive contributions from the fermionic degrees of freedom. In other words, when going from the non-unimodular to the unimodular case, the difference is not only the presence of the new (fermionic) charges $\mathcal Q_1,\mathcal Q_2$, because we can also have different expressions for $\mathcal Q_{\mathsf h},\mathcal Q_{\mathsf e}$.

First notice that $\mathcal Q_1,\mathcal Q_2$ are Grassmann variables. We are therefore working with the Grassmann enveloping algebra, in which case anticommutators  become standard commutators. In particular, notice that if we define $\mathsf Z=- i\eta (\mathcal Q_1\mathsf Q_1+\mathcal Q_2\mathsf Q_2)$, then $[\mathsf Z,\mathsf Z]=0$, because the only non-vanishing anticommutator in~\eqref{eq:QQ} is when we take the same odd generator, but then $\mathcal Q_i^2=0$ because it is Grassmann. The non-unimodular twist is recovered by formally setting $\mathsf Z=0$.

When $\xi=0$ in the commutation relations, the generators $\mathsf Q_1,\mathsf Q_2$ do not mix and we can repeat the argument of section~\ref{sec:fact} to show that $W=v^{-1}W'v$, where $W'=\exp(\eta\mathcal Q_{\mathsf e} \mathsf h)$. In fact, first we can write the twist in the form
$
W^{-1}=\exp\left(A+\frac{s}{1-e^{-s}}B'\right),
$
and repeat the steps in~\eqref{eq:fact-id}
if we now identify
\begin{equation}
A=\eta(-\mathcal Q_{\mathsf e}\mathsf h+\mathcal Q_{\mathsf h}\mathsf e), \qquad \frac{s}{1-e^{-s}}B'=\mathsf Z.
\end{equation}
Using that $[\mathsf h,\mathsf Z]=\tfrac12 \mathsf Z$ and imposing $[A,B']=s B'$ one finds $s=-\eta \mathcal Q_{\mathsf e}/2$. This shows that the unimodular twist is related by a $\mathsf Z$-dependent similarity transformation to $\exp[\eta(-\mathcal Q_{\mathsf e}\mathsf h+\mathcal Q_{\mathsf h}\mathsf e)]$. This has the same form as the non-unimodular twist and, as explained in section~\ref{sec:fact}, it then follows that it is also related by a similarity transformation to $W'$.

To prove the factorisation of the twist for generic values of $\xi$, we will now consider a different and even simpler method, that works both for $\zeta=0$ and $\zeta=1$ and for any $\xi$. For definiteness, we will provide the discussion in the $\zeta=1$ case.
First, let us construct
\begin{equation}
v=\exp(\alpha_{\mathsf e}\mathsf e)\exp(\alpha_1\mathsf Q_1)\exp(\alpha_2\mathsf Q_2),
\end{equation}
where $\alpha_1,\alpha_2$ are Grassmann, and compute
\begin{equation}
v^{-1}\mathsf h v=\mathsf h+\alpha_{\mathsf e}\mathsf e + (\tfrac12\alpha_1+\xi \alpha_2)\mathsf Q_1+ (\tfrac12\alpha_2-\xi \alpha_1)\mathsf Q_2 -i \xi \alpha_1 \alpha_2 \mathsf{e} 
\end{equation}
The above formula implies that
\begin{equation}
v^{-1}\exp(c\mathsf h)v=\exp\left[c\left(\mathsf h+(\alpha_{\mathsf e} -i \xi \alpha_1 \alpha_2 )\mathsf e + (\tfrac12\alpha_1+\xi \alpha_2)\mathsf Q_1+ (\tfrac12\alpha_2-\xi \alpha_1)\mathsf Q_2\right)\right].
\end{equation}
Now we want to match the right-hand-side of the above equation with the right-hand-side of~\eqref{eq:twist-uni-gen}. We  can do this only if we assume $c\neq 0$, and then
\begin{equation}\label{eq:sol-fact}
c=-\eta \mathcal Q_{\mathsf e},\qquad
\alpha_{\mathsf e}=\eta\mathcal Q_{\mathsf h}/c +i\xi \alpha_1\alpha_2
\end{equation}
and if we also require
\begin{equation}
\tfrac12\alpha_1+\xi \alpha_2=-i\eta\mathcal Q_1/c,\qquad
\tfrac12\alpha_2-\xi \alpha_1=-i\eta\mathcal Q_2/c.
\end{equation}
This system has always a solution except  if $\xi^2=-1/4$, but this possibility is excluded because it must be $\xi\in \mathbb R$ to respect the real form of $\alg{psu}(2,2|4)$. To conclude, we can always solve also for $\alpha_1,\alpha_2,\alpha_e$, and then  the twist can be put in the wanted form
\begin{equation}
W=v^{-1}W'v,\qquad\qquad
W'=\exp(\eta \mathcal Q_{\mathsf e}\mathsf h).
\end{equation}
Let us make a  comment on the assumption $c\neq 0$ made above. Given the solution for $c$ in~\eqref{eq:sol-fact}, there are only two possibilities in which $c$ can vanish. The undeformed limit $\eta\to 0$  is obviously not problematic: when $\eta\to 0$ then $W\to 1$ so the factorisation is trivially true. The only real worry should be the case $\mathcal Q_{\mathsf e}=0$ with $\eta\neq 0$. Let us analyse this situation in the non-unimodular case, when $\mathcal Q_e=-\mathbf Q/\eta$ and $\mathbf Q$ was given in~\eqref{eq:Jordaniancharges}. Then, $\mathcal Q_{\mathsf e}$ can vanish (at finite $\eta$) only if $\mathbf Q=0$. Given~\eqref{eq:jord-twist}, this leads to a trivial twist $W= 1$, unless $\mathbf q$ compensates the zero of $\mathbf Q$ with a divergence. This is possible only if the field configuration is such that 
\begin{equation}
Y_{\mathsf h}(\tau,2\pi)-Y_{\mathsf h}(\tau,0)\neq 0,\qquad
Y_{+}(\tau,2\pi)-Y_{+}(\tau,0)=0.
\end{equation}
In other words, the degrees of freedom in $Y_{\mathsf e}$ are periodic while those in $Y_{\mathsf h}$ are not.
Therefore, in this sector of the theory the non-unimodular twist cannot be put into the form $\exp(c\mathsf h)$, and it is instead equal to $W=\exp(-w\mathsf e)$ with $w=\mathbf Q\mathbf q$  finite and non-zero. This is in fact a non-diagonalisable twist, with all eigenvalues equal to 1. While the leading-order asymptotics of the quasimomenta around $z=1$ (or $x=\infty$) would match those of the untwisted case,  the next-to-leading order will change.
A similar analysis is obviously valid also for the unimodular twist.

We conclude this section by arguing  the spectral equivalence of the unimodular and the non-unimodular models, up to certain caveats that we are about to point out. In the sector where the twists of the unimodular and non-unimodular models are both equivalent to $W'=\exp(\eta \mathcal Q_{\mathsf e}\mathsf h)$,  the fact that the  classical spectra of the two models are the same is rather straightforward. First, in both cases one can implement field redefinitions as explained in~\ref{sec:fact} to obtain boundary conditions controlled just by $W'=\exp(\eta\mathcal Q_{\mathsf e}\mathsf h)$. As remarked above, the explicit expressions for $\mathcal Q_{\mathsf e}$ in the two cases differ, because in the unimodular case there are additional contributions from the fermionic degrees of freedom. However, these vanish when considering classical solutions, and therefore the classical spectral curves of the two models are indistinguishable.
One may worry that the non-diagonalisable sector (where the twist is instead $W=\exp(-w\mathsf e)$) may be problematic from the point of view of the spectral equivalence. But also in this case the twists of the two models would differ only by contributions of the fermionic degrees of freedom, that do not contribute at the classical level.

We can push this argument even to the one-loop level. In fact, one may use the method of~\cite{Gromov:2007aq,Gromov:2008ec} to compute quantum corrections to the classical spectrum, as we have done in the previous section for the non-unimodular case. The advantage of this method is that the only data that is needed to compute the one-loop shift is the classical spectral curve itself. Because the two classical spectral curves agree, the unimodular and non-unimodular models will also have the same 1-loop corrections to the spectrum.
It would be interesting to understand if it is possible to make any statement beyond one-loop, since it is always possible that the spectral equivalence breaks down at higher loops. However, we want to point out that even at  one-loop level, there may be extra subtleties not discussed so far. In fact, the whole argument is done under the assumption that it is possible to reformulate the Yang-Baxter deformed model as the undeformed yet twisted one. While this is certainly true at the classical level because the two $\sigma$-models are equivalent on-shell, it is possible that the equivalence breaks down at the quantum level. An anomaly may occur especially in the non-unimodular case, when the background fields do not satisfy the standard supergravity equations and the $\sigma$-model is not Weyl invariant. In the presence of such an anomaly, even if the (quantum) spectra of the two \emph{twisted} models were the same,  the spectra of the corresponding Yang-Baxter \emph{deformations} (with $\zeta=0$ and with $\zeta=1$) may not be related in an obvious way. 
It would be very interesting to investigate these possible scenarios in more details.

%%%%%%%%%%%%%%%%%%%%%%%%%%%%%%%%%%%%%%%%%%%%%%%%%%%%%%%%%%%%%%%%%%%%%%%%
\section{Conclusions} \label{sec:conclusions}

In this paper, we have considered a Jordanian deformation of the $AdS_5\times S^5$ superstring background, that preserves the (classical) integrability and at least 12 superisometries.\footnote{Interestingly, as already noted in~\cite{vanTongeren:2019dlq} using findings of~\cite{Matsumoto:2014ubv}, this Jordanian deformation can also be obtained  by a sequence of TsT and S-duality transformations. It therefore provides an example of a transformation involving S-duality that preserves integrability. See~\cite{Hoare:2022asa} for more recent examples.} We have identified global coordinates for the deformed background, and in particular, a global time coordinate. This fact is crucial for the correct identification of the energy when considering the spectral problem from first principles, and  we believe that it will be important in order to have insights into an AdS/CFT interpretation of the deformation.

Importantly, we have reformulated the deformed model in terms of an undeformed yet twisted model. In this twisted picture, we were able to obtain a general class of solutions to the $\sigma$-model equations of motion written in terms of Airy functions. Our further study, however, focused on a particular simple solution which we called BMN-like. Given that the difference compared to the standard (periodic) case of $AdS_5\times S^5$ is only in a twist appearing in the boundary conditions, we could borrow several methods of integrability that had already been used in the past. We did this not just at the level of the classical spectral curve but also when including its first quantum corrections since, despite the twist, the methods are essentially unaltered. 

The only place where the construction of the classical spectral curve is changed with respect to the standard case is in the form that the quasimomenta take at large values of the spectral parameter. In particular, we are forced to re-evaluate the ansatz for the asymptotic behaviour of the quantum corrections for some of our quasimomenta. For example, in the standard case, we can extract the contribution to the anomalous correction to the energy from any of the corrections to the quasimomenta associated to the $AdS$ directions. However, in the deformed background we have studied, this contribution does not appear in those quasimomenta which are affected by the twist.

An interesting outcome of our results is that we find no anomalous quantum correction for the charges that control the twisted boundary conditions. The only charge that receives quantum corrections is the energy,  and we interpret this as the spectral problem being well-posed. The situation is reminiscent to the $\beta$-deformation~\cite{Lunin:2005jy}, where the twist appears in the form of phases that enter the Bethe equations~\cite{Beisert:2005if}, for example. Also in that situation, it is  the energy of the string (or the dual anomalous dimension of single-trace operators) that receives quantum corrections, while the charges controlling the twist appear as external data, so that the Bethe equations can be solved consistently. 

The advantage of working in the language of the twisted model resides also in the fact that the conserved charges that label the full spectrum admit \emph{local} expressions in terms of the variables of the twisted model, while they would be non-local in the variables of the deformed one. Moreover, only (the Cartans of) the manifest symmetries of the twisted model appear as labels of the spectrum, and this is crucial because these are only a subset of the isometries of the deformed one.

In the deformed picture, we know that the deformation parameter can be reabsorbed, although this is done in such a way that the undeformed limit $\eta\to 0$ becomes subtle. One can actually think in terms of two possible cases to be considered, namely $\eta=0$ and $\eta\neq 0$. This is related to the fact that Jordanian deformation (at $\eta\neq 0$) are essentially non-abelian T-duality, without any continuous deformation parameter. However, in the twisted picture we do not see evidence for a similar interpretation, because the one-loop spectrum seems to depend continuously on the deformation parameter $\eta$ through the combination $\beta=\frac{\eta}{2 b_z^2}$. It would be interesting to understand this further.

Let us also comment on the fact that we were able to consider both the unimodular Jordanian deformation (giving rise to a type IIB supergravity background) and the non-unimodular one (which does not correspond to a supergravity solution). In this respect, the main message from our results is that the corresponding twisted models share the same spectrum, at least to one-loop. Unless  the argument regarding the equivalence of the deformed model to the twisted one at the quantum level fails, this seems to suggest that the unimodular and non-unimodular deformations share the same spectrum at least to that order. It would be interesting to understand these subtleties further and see whether we can make any statement beyond one-loop. A first non-trivial check was obtained already in appendix \ref{QuadraticFluctuations}, where we found that the bosonic (AdS) frequencies obtained from the analysis of quadratic fluctuations of the \textit{deformed} $\sigma$-model matches with the bosonic (AdS) frequencies obtained from the curve of the twisted $\sigma$-model. In addition, let us mention that the equivalence of the quantum spectrum between a unimodular and a non-unimodular Yang-Baxter deformation has been noted also in~\cite{Seibold:2021rml}, although in that case the \emph{inhomogeneous} Yang-Baxter deformation was considered, for which the reinterpretation as a twist of the boundary conditions in the undeformed model does not hold. 

Our results offer a first step towards the understanding of how to tackle the spectral problem for Jordanian deformations (and more generally diagonalisable Yang-Baxter deformations) of $AdS_5\times S^5$. It would be interesting to start working from the other side of the AdS/CFT duality and construct a corresponding deformation of $\mathcal N=4$ super Yang-Mills. The expectation is that there should be a notion of a deformed/twisted spin-chain in that case, and it is worth exploring the possibility of constructing this spin-chain directly, to obtain insights on the potentially more difficult construction of the deformation of the gauge theory.

From a broader perspective, it would be very interesting to reinterpret the more generic supergravity solution-generating techniques of~\cite{Borsato:2021gma,Borsato:2021vfy} (which include also non-abelian T-duality, and are integrability-preserving) in terms of twisted models within a first-order formulation. 
This may open the possibility of tackling the spectral problem in this more generic class of deformed and dual $\sigma$-models.

\section*{Acknowledgements}
We thank Tristan McLoughlin, Olof Ohlsson Sax,  Roberto Ruiz and Stijn van Tongeren for discussions, and we are grateful to Roberto Ruiz and Stijn van Tongeren for comments on the manuscript.
RB and SD are supported by the fellowship of ``la Caixa Foundation'' (ID 100010434) with code LCF/BQ/PI19/11690019, by AEI-Spain (under project PID2020-114157GB-I00 and Unidad de Excelencia Mar\'\i a de Maetzu MDM-2016-0692),  Xunta de Galicia (Centro singular de investigaci\'on de Galicia accreditation 2019-2022, and project ED431C-2021/14), and by the European Union FEDER. 
LW is funded by a University of Surrey Doctoral College Studentship Award. Additionally, LW is grateful to IGFAE for its hospitality during his research visit under the Turing grant scheme. JMNG is supported by the EPSRC-SFI grant EP/S020888/1 {\it Solving Spins and Strings}. No data beyond those presented and cited in this work are needed to validate this study.

%%%%%%%%%%%%%%%%%%%%%\
\appendix

%%%%%%%%%%%%%%%%%%%%%%%%%%%%%%%%%%%%%%%%%%%%%%%%%%%%%%%%%%%%%%%%%%%%%%%%
\section{Relation to embedding coordinates of $AdS$} \label{a:embedding}
In this appendix, we want to explain how the global coordinates that we use to parameterise the Jordanian-deformed  spacetime are related to the embedding coordinates of $AdS$. This is both to offer additional geometric intuition on the coordinates used and to give an alternative argument to the fact that they are global coordinates, although here this applies only in the undeformed $\eta=0$ limit. 

$AdS_D$ is defined as the hyperboloid  $X_0^2+X_D^2-\sum_{i=1}^{D-1}X_i^2=R^2$ (with $R$ the $AdS$ radius) in $\mathbb R^{2,D-1}$ with metric $ds^2=-dX_0^2-dX_D^2+\sum_{i=1}^{D-1}dX_i^2$. Global coordinates in $AdS$ can be obtained by setting $X_0=R \cosh\rho\cos\tau_G$, $X_D=R\cosh\rho\sin\tau_G$, $X_i=R\sinh\rho\ \Omega_i$ with the constraint $\sum_{i=1}^{D-1}\Omega_i^2=1$. This gives the $AdS$ metric $ds^2=R^2(-\cosh^2\rho\ d\tau_G+d\rho^2+\sinh^2\rho\ d\Omega^2)$, with $d\Omega^2$ the metric of the $(D-2)$-dimensional sphere. For $D>2$, the hyperboloid is covered once if we take $\rho\geq 0, 0\leq\tau_G<2\pi$, and the universal cover is obtained by decompactifying the global time $\tau_G$. The metric diverges at $\rho\to \infty$, but that is not a problematic place: geodesics either reach it in an infinite time, or if they do it in a finite time (as it happens for some null geodesics) then they come back to finite $\rho$ at later time. Therefore,  the spacetime is geodesically complete in these coordinates. 

Alternatively, the identification $X_0=\tfrac12 z(1+z^{-2}(R^2+x^ix^i-t^2))$, $X_D=Rt/z$, $X^i=Rx^i/z (i=1,\dots, D-2)$, $X_{D-1}=\tfrac12 z(1-z^{-2}(R^2-x^ix^i+t^2))$ gives the metric
\begin{equation}
ds^2=R^2(\frac{dz^2+dx^idx^i-dt^2}{z^2}) \ ,
\end{equation}
 in the so-called Poincar\'e coordinates. These coordinates do not cover the whole hyperboloid and in fact, as we will show in the more generic deformed case in appendix \ref{a:poincare}, some geodesics are not complete in this patch (e.g. they reach $z=\infty$ in a finite proper time, and cannot be continued to $z<\infty$). In these coordinates, the boundary of $AdS$ is at $z=0$, where the metric diverges. 

In the coordinates $(T,V,\Theta,P,Z)$ of~\eqref{eq:PtoSchr},  we see that the $AdS$ metric (i.e.~when setting $\eta=0$) diverges at $Z=0$, and because of the relation to the previous coordinates we can identify that as the boundary of $AdS$.
The composition of relations between the different sets of coordinates leads to the relation of the global coordinates $(T,V,\Theta,P,Z)$ to the embedding coordinates.  Setting  $R=1$, we find
\begin{equation} 
\begin{aligned}
X_0&=\frac{\left( 1 + Z^2 + P^2 \right) \cos T - 2V \sin T}{2 Z},\\
   X_1&=\frac{\left(P^2+Z^2-2\right) \sin T+2 V \cos T}{2\sqrt{2} Z},\\
   X_2&=\frac{P \sin \Theta }{Z},\\
   X_3&=\frac{P \cos \Theta}{Z},\\
   X_4&=\frac{\left( -1 + Z^2 + P^2 \right) \cos T - 2V \sin T}{2 Z},\\
   X_5&=\frac{\left(P^2+Z^2+2\right) \sin T+2 V \cos T}{2 \sqrt{2} Z}.
   \end{aligned}
\end{equation}
These expressions  can be inverted as
\begin{equation} \label{eq:SchrToEmb}
\begin{aligned}
P&=
   \frac{\sqrt{2}\sqrt{X_2^2+X_3^2}}{\sqrt{2(X_0-X_4)^2+(X_1-X_5)^2}},\\
\Theta &= \text{arctan}(X_2/X_3),\\
V&=\frac{X_0 (3 X_1+X_5)-X_4 (X_1+3 X_5)}{\sqrt{2}
   \left(2 (X_0-X_4)^2+(X_1-X_5)^2\right)},\\
Z&=
   \frac{\sqrt{2}}{\sqrt{2(X_0-X_4)^2+(X_1-X_5)^2}},\\
T&= \text{arctan}\left(\frac{X_1-X_5}{\sqrt{2}(X_4-X_0)}\right).
      \end{aligned}
\end{equation} 
From these expressions, we see that we can take $P> 0$ and $Z>0$, and moreover\footnote{Given that $\Theta$ and $T$ are obtained from the arctan, normally they would take values in $[-\pi/2,\pi/2]$. In particular, $\arctan(y/x)$ is insensitive to a simultaneous change of sign of $x$ and $y$. To avoid this problem, one can define a function $\arctan(x,y)$ that gives the value of the angle identified by the point in the $(x,y)$ plane taking into account the quadrants in which $x$ and $y$ are placed.} $\Theta\in [0,2\pi[$ and $T\in [0,2\pi[$, which in both cases are periodically identified with periods $2\pi$. For $V$, we have instead $V\in ]-\infty,+\infty[$.
While $\tau_G$ is an angle in the $(X_0,X_5)$ plane, $T$ is instead an angle in the plane $(X_4-X_0,X_1-X_5)$. Also in this case, we decompactify it and take $T\in ]-\infty,+\infty[$.

We will now argue that these coordinates cover the whole hyperboloid.  First, $P$ and $\Theta$ cover all the $(X_2,X_3)$ plane, with a ``warping'' depending on the other 4 coordinates. If we now define  $Y_0^\pm=X_0\pm X_4$ and $Y_1^\pm=(X_1\pm X_5)/\sqrt{2}$, then $Z$ and $T$ take care of the whole $(Y_0^-,Y_1^-)$ plane (notice that $Z$ is the inverse of the radial coordinate in that plane). Finally, we need to understand if we can cover the whole $(Y_0^+,Y_1^+)$ plane. Notice that (unlike other coordinates) $V$ depends also on the combinations $Y_0^+,Y_1^+$, not just $Y_0^-,Y_1^-$. One may worry that we have just one degree of freedom (i.e.~$V$) to cover this $(Y_0^+,Y_1^+)$ plane, but in fact this is not an issue, because we should remember that we are constraining the coordinates to satisfy $1=X_0^2+X_D^2-\sum_{i=1}^{D-1}X_i^2=-X_2^2-X_3^2+Y_0^+Y_0^--2Y_1^+Y_1^-$,  so that if for example $Y_0^+, Y_0^-, Y_1^-, X_2, X_3$ have been fixed (because, for example, one solves for $Y_0^+$ in terms of $V$), then the remaining $Y_1^+$ is uniquely identified. Notice that the hyperboloid constraint is quadratic in $X_M$, but it is linear (for example) in $Y_1^+$, so when solving for $Y_1^+$ the sign is unambiguous. Naively, an issue is present when $Y_1^-=0$, because then $Y_1^+$ is undetermined. But when $Y_1^-=0$ then $V=Y_1^+/Y_0^-$, and  we can just swap $Y_0^+$  and  $Y_1^+$ in the previous reasoning. Then one can solve the hyperboloid constraint for $Y_0^+$ instead. The issue remains when both $Y_0^-=Y_1^-=0$, because then $V,P$ and $Z$ diverge. This is not problematic because $Y_0^-=Y_1^-=0$ is  a codimension-2 subspace, which is not an open set in spacetime.

\section{Geodesic incompleteness of Poincar\'e coordinates} \label{a:poincare}
For completeness, we show in  this appendix that the geodesic incompleteness of the parametrisation by Poincar\'e coordinates  persists also when turning on the deformation $\eta$.
As usual, the strategy is to look for solutions to the geodesic equations that are pathological, i.e.~the boundary values of the coordinates are reached in a finite amount of geodesic time.

To start, let us construct the conserved quantities $Q_{\bar a}=k_a^\mu G_{\mu\nu}\dot X^\nu$ along the geodesics. They are equivalent to the following equations
\begin{equation}\label{eq:cons-kill-poinc}
\begin{aligned}
\dot \theta &= \frac{Q_{\bar 5} z^2}{\rho ^2},\\
\dot x^-&= -\frac{\left(Q_{\bar 2}+Q_{\bar 3}\right) z^2}{\sqrt{2}},\\
\dot x^+&=
   \frac{\eta ^2 \left(Q_{\bar 2}+Q_{\bar 3}\right) \rho ^2-4 \left(Q_{\bar 2}-Q_{\bar 3}\right) z^4+\eta ^2
   \left(Q_{\bar 2}+Q_{\bar 3}\right) z^2}{4 \sqrt{2} z^2},\\
\dot \rho &= \frac{z \left(z \left(\sqrt{2} \left(Q_{\bar 3}-Q_{\bar 2}\right)
   x^-+Q_{\bar 1}\right)-z'\right)}{\rho },\\
\left(Q_{\bar 2}+Q_{\bar 3}\right)
   \left(\rho ^2+z^2\right)&=   Q_{\bar 4}-2 \sqrt{2} Q_{\bar 1} x^-+2 \left(Q_{\bar 2}-Q_{\bar 3}\right) (x^-)^2.
\end{aligned}
\end{equation}
At the same time we also have the conserved quantity 
\begin{equation}\label{eq:geo-l}
\epsilon=\dot X^m \dot X^m G_{mn}
=-\frac{8 z^4 \dot x^- \dot x^++\eta ^2 \left(\dot x^-\right)^2 \left(\rho
   ^2+z^2\right)-4 z^4 \left(\rho ^2 \dot \theta^2+\dot \rho^2+\dot z^2\right)}{4 z^6}.
\end{equation}
Let us consider null geodesics, so that $\epsilon=0$. 
Now the strategy is to look for solutions that at large $z$ go like $z\propto (\tau-\tau_0)^{-A}$ for some $A>0$. These are pathological because $z=\infty$ is reached at finite $\tau$. To simplify the analysis we  fix some of the values of the $Q_{\bar a}$. 
 At the same time, we must be careful and make sure that  the last equation in~\eqref{eq:cons-kill-poinc} does \emph{not} reduce to the form $C_zz^2+C_\rho\rho^2+C_-(x^-)^2=C$ with $C_z,C_\rho,C_-,C>0$: if that happened, the geodesic motion would be  bounded and no pathological behaviour at $z=\infty$ would be possible.  
 To be concrete we take 
\begin{equation}
Q_{\bar 1}=0,\qquad
Q_{\bar 2}\neq 0,\qquad
Q_{\bar 3}=0,\qquad
Q_{\bar 4}= 0,\qquad
Q_{\bar 5}= 0.
\end{equation}
It is of course possible to take more generic situations, but this will already be enough to identify pathological behaviour.
In this case,~\eqref{eq:cons-kill-poinc} reduce to
\begin{equation}\label{eq:eqs-poincare-ansatz}
\begin{aligned}
&\dot \theta=0,\qquad
\dot x^-= -\frac{Q_{\bar 2} z^2}{\sqrt{2}},\qquad
\dot x^+= \frac{Q_{\bar 2} \left(\eta ^2
   \rho ^2-4 z^4+\eta ^2 z^2\right)}{4 \sqrt{2} z^2},\\
&   \dot \rho = \frac{z \left(-\sqrt{2} Q_{\bar 2} z  x^--\dot z\right)}{\rho },\qquad
0=   z^2-2(x^-)^2+\rho^2.
\end{aligned}
\end{equation}
Notice that $\theta$ is constant, and that a solution for $x^+$ can be found once a solution for $z$ is given. Therefore, in the following we focus on the  equations for the remaining coordinates.
At this point we look for geodesic solutions  that at large values of $z$ have
\begin{equation}
x^-\sim\alpha_-z, \qquad \text{for } z\gg 1,
\end{equation}
with $\alpha_-$ a constant. 
Within this ansatz, the algebraic constraint given in the last equation of~\eqref{eq:eqs-poincare-ansatz}  reduces to $(1-2\alpha_-^2)z^2+\rho^2\sim 0$, which is compatible with large $z,\rho$ only if
\begin{equation}
\alpha_-^2>1/2,\qquad\implies\qquad
\rho\sim z\sqrt{2\alpha_-^2-1}.
\end{equation}
Demanding compatibility with the equation for $\dot \rho$ then gives
\begin{equation}
\dot z\sim -\frac{Q_{\bar 2}}{\sqrt{2}\ \alpha_-}z^2.
\end{equation}
This is what we were seeking, since such an equation is solved by\footnote{Importantly, this solution is compatible with~\eqref{eq:geo-l} in the large-$z$ expansion.} $z\propto (\tau-\tau_0)^{-1}$. The pathological behaviour of this solution is enough to conclude that the Poincar\'e coordinates are not global coordinates for the deformed spacetime. 

\section{Identifying the Cartan subalgebra of isometries} \label{a:CSA}

In this appendix, we identify the possible Cartan subalgebras of the algebra of isometries $\mathfrak{k}$ given in \eqref{eq:isoms}.
 Since $M^{23}$ and $p^0+p^1$ are central elements, i.e.~they commute with all generators in $\mathfrak{k}$, we can focus on the remaining generators which span an $\mathfrak{sl}(2,\mathbb{R}) = \mathrm{span}(\mathsf{h},\mathsf{e}_+,\mathsf{e}_-)$ subalgebra with $[\mathsf{h},\mathsf{e}_\pm] = \pm \mathsf{e}_\pm$ and $[\mathsf{e}_+, \mathsf{e}_-] = 2 \mathsf{h}$ as is found by identifying $\mathsf{h}= \tfrac12 (d-M^{01})$, $\mathsf{e}_+= \tfrac12 (p^0-p^1)$, and $\mathsf{e}_-=-\tfrac12 (k^0+k^1)$. Inequivalent choices of the Cartan subalgebra are then obtained by classifying the inequivalent adjoint orbits of $\mathfrak{sl}(2,\mathbb{R}) $ and then requiring that the representative elements have a diagonalisable adjoint action. Although this is a standard exercise, we repeat it explicitly to clarify certain comments.
 
Let us write a generic element of the group $SL(2,\mathbb{R})$ as $g=\exp(c_h \mathsf{h})\exp(c_+ \mathsf{e}_+)\exp(c_- \mathsf{e}_-)$ and of the algebra $\mathfrak{sl}(2,\mathbb{R}) $ as $x=\alpha_h \mathsf{h} +\alpha_+ \mathsf{e}_+ +\alpha_- \mathsf{e}_-$. For each element $x$ of $\mathfrak{sl}(2,\mathbb{R})$, one can then identify its adjoint orbit $\{gxg^{-1}\}$. One finds three inequivalent possibilities. First, in fact, we can transform the generic element $x$ to the element $\mathsf{h}$ whilst preserving reality conditions as
\begin{equation} \label{eq:xtoh}
gxg^{-1}=\sqrt{\gamma}\ \mathsf{h},\qquad 
\text{ with } \qquad 
\gamma=\alpha _h^2+4 \alpha _- \alpha _+ ,
\end{equation}
 by specifying the element $g$ with the coefficients
 \begin{equation}
c_+= \frac{\alpha _+}{\sqrt{\gamma}},\qquad
c_-=
   \frac{\alpha _h \left(\sqrt{\gamma}-\alpha _h\right)-4
   \alpha _- \alpha _+}{2 \alpha _+ \sqrt{\gamma}} , 
\end{equation}
if $\gamma >0$.\footnote{Although the case $\alpha_+=0$ seems singular, it can be analysed on its own (for $\alpha_h\neq 0$) leading to the same conclusion.} The second case is that for $\gamma=0$, which can be obtained when at least $\alpha_+$ or $\alpha_-$ are non-zero (otherwise $x$ would also be zero). For definiteness say that $\alpha_+\neq 0$.
One can then transform $x$ into $\mathsf{e}_+$ as
\begin{equation}
gxg^{-1} = e^{c_h} \alpha _+\ \mathsf{e}_+,\qquad
\text{ with } \qquad
c_-=\frac{\alpha_h}{2\alpha_+}.
\end{equation}
Alternatively, we can take $\alpha_- \neq 0$, which would transform $x$ into $\mathsf{e}_-$. However, $\mathsf{e}_\pm$ are related by an inner automorphism as $\bar g \mathsf{e}_+ \bar g^{-1}=-\mathsf{e}_-$ with $\bar g=\exp(\pi/2(\mathsf{e}_+-\mathsf{e}_-))$  so these two possibilities are in fact equivalent. The third and final case is that for $\gamma<0$, which can be obtained only when both $\alpha_+$ and $\alpha_-$ are non-vanishing. In this case, one can transform $x$ into $\mathsf{e}_+ - \mathsf{e}_-$ as follows: 
\begin{equation}
gxg^{-1} = \frac{1}{2} \alpha _+ \sqrt{-\frac{\gamma}{\alpha _+^2}}\ (\mathsf{e}_+-\mathsf{e}_-),
\end{equation}
by using
\begin{equation}
c_h=\frac{1}{2} \log \left(-\frac{\gamma}{4 \alpha_+^2}\right),\qquad
c_+=0,\qquad
c_-=\frac{\alpha_h}{2\alpha_+}.
\end{equation}
Thus, given a generic $x\in \alg sl(2,\mathbb R)$ which we want to simplify by means of inner automorphisms, we have three inequivalent possibilities\footnote{It is immediate to see that these choices cannot be related by inner automorphisms, given that $\Tr(\mathsf{h}^2)=1, \Tr(\mathsf{e}_+^2)=0,\Tr((\mathsf{e}_+-\mathsf{e}_-)^2)=-2$.}
\begin{equation}
\mathsf{h},\qquad \text{ or }\qquad
\mathsf{e}_+,\qquad\text{ or }\qquad
\mathsf{e}_+-\mathsf{e}_-.
\end{equation}
To identify the possible Cartan subalgebras, we must now verify whether or not the adjoint actions $\Ad_x $  for the above  inequivalent elements are diagonalisable. A simple computation shows that $\Ad_\mathsf{h}$ is diagonalisable with eigenvalues $(0,1,-1)$,  $\Ad_{\mathsf{e}_+}$  is not diagonalisable, and  $\Ad_{\mathsf{e}_+-\mathsf{e}_-}$ is diagonalisable with eigenvalues $(0,2i,-2i)$.\footnote{Notice that $\Ad_{\mathsf{e}_+-\mathsf{e}_-}$ is real, only its diagonalised version has imaginary coefficients.}  To conclude,  up to automorphisms, there are two possible choices of Cartan subalgebras of $\mathfrak{k}$, namely
\begin{equation}
\begin{aligned}
\text{(I)}:&\ \text{span}\{d - M^{01} , \ p^0 +  p^1,\ M^{23}  \},\\
\text{(II)}:&\  \text{span}\{p^0 -  p^1+\alpha (k^0+k^1), \ p^0 +  p^1,\ M^{23} \},
\end{aligned}
\end{equation}
where we leave a possible $\alpha>0$ coefficient for later convenience. The choice $\mathsf{e}_+-\mathsf{e}_-$ would correspond to $\alpha=1$, but in the main text we find it convenient to choose $\alpha=1/2$.

\section{Analytic structure of the perturbations}
\label{StructureFluctuations}

In this appendix, we detail how to compute the correction to the quasimomenta $\delta p_i$ that arise from adding small excitations to our classical solution using only their analytic properties, as described in section~\ref{CorrectionClassical}. Most of our computations mimic those of the $\mathfrak{sl}(2, \mathbb{R})$ circular string in \cite{Gromov:2007aq} due to the presence of a single cut in the $\hat{p}_i$ quasimomenta, but the asymptotic behaviour of our classical quasimomenta and some of their corrections is very different.

\subsection{Excitations associated to the sphere}

Excitations that connect two quasimomenta $\tilde{p}_i$ and $\tilde{p}_j$ are related to considering quadratic fluctuations around a classical solution for the modes associated to the sphere. Of those, the only ones that give rise to physical excitations are those that connect either $\tilde{1}$ or $\tilde{2}$ with either $\tilde{3}$ or $\tilde{4}$. Similarly to the case described in \cite{Gromov:2007aq}, the two remaining possible combinations, $\tilde{1} \tilde{2}$ and $\tilde{3} \tilde{4}$, do not give rise to physical excitations.

For concreteness, here we will focus on the excitation that connects $\tilde{p}_2$ and $\tilde{p}_3$, which we will denote as $\tilde{2} \tilde{3}$, and turn off any other excitation.  The computations for the other three excitations are the same \textit{mutatis mutandis}. Nevertheless, we will make some comments on $\tilde{2} \tilde{4}$, and we will discuss the remaining two in section \ref{a:composition}.

Let us focus first on the correction $\delta \tilde{p}_2$. We know that it has to have poles at $\pm 1$ and at $x^{\tilde{2} \tilde{3}}_n$. Thus, we propose the following ansatz:
\begin{equation}
    \delta \tilde{p}_2=\frac{\delta \alpha_+}{x-1} +\frac{\delta \alpha_-}{x+1}  -  \sum_n \frac{\alpha (x^{\tilde{2} \tilde{3}}_n) N^{\tilde{2} \tilde{3}}_n}{x-x^{\tilde{2} \tilde{3}}_n} \ .
\end{equation}
As there will be no ambiguity, we will drop the superindex of $x_n$ and $N_n$ to alleviate our notation.

We also know that $\delta \tilde{p}_2$ is associated to $\delta \tilde{p}_1$ by inversion symmetry. Thus, we can use the information about their behaviour at large values of $x$ to fix the residues of the poles. On the one hand, from equation~(\ref{asymptotics}), we have
\begin{align}
    \delta \tilde{p}_1 &\approx \mathcal{O} (x^{-2}) \ , &     \delta \tilde{p}_2 &\approx -\frac{4\pi}{\sqrt{\lambda}} \sum_n \frac{N_n}{x} + \mathcal{O} (x^{-2}) \ .
\end{align}
On the other hand, from our ansatz and inversion symmetry~(\ref{inversioncorrections}), we have
\begin{align}
    \delta \tilde{p}_1 &\approx \left( \delta \alpha_+ - \delta \alpha_- - \sum_n \frac{\alpha (x_n) N_n}{x_n} \right) + \frac{\delta \alpha_+ + \delta \alpha_- - \sum_n \frac{\alpha (x_n) N_n}{x_n^2}}{x}+ \mathcal{O} (x^{-2}) \ , \\
    \delta \tilde{p}_2 &\approx \frac{\delta \alpha_+ + \delta \alpha_- - \sum_n \alpha (x_n) N_n}{x} + \mathcal{O} (x^{-2}) \ .
\end{align}
Consistency between the two expansions gives us enough information to fix the three residues
\begin{align}
    {\alpha} (x_n) &= \frac{4\pi}{\sqrt{\lambda}} \frac{x_n^2}{x_n^2-1} \ , \label{eq:residue} \\
    2\delta \alpha_\pm &= \sum_n {\alpha} (x_n) N_n \left( \frac{1}{x_n^2} \pm \frac{1}{x_n} \right) \ . \label{eq:dapm-23-sphere}
\end{align}
As we anticipated in section~\ref{CorrectionClassical}, the function $\alpha(x)$ does not need to be fixed beforehand. The consistency between the positions and signs of the poles with the asymptotic behaviour of the corrections is enough to fix its form as a function of the position of the pole. As expected, the function $\alpha (x)$ is unchanged from the undeformed case \cite{Gromov:2007aq}. This is consistent with the fact that it is related to the weight appearing in the definition of the filling fractions \eqref{ffAdS}. In particular, we require that
\begin{equation}
    -N_n=-\frac{\sqrt{\lambda}}{8\pi^2i}\oint_C dx\left(1-\frac{1}{x^2}\right) \delta \tilde{p}_2 =-\frac{\sqrt{\lambda}}{4\pi} N_n \alpha (x_n) \left( 1- \frac{1}{x_n^2} \right) \ ,
\end{equation}
where $C$ is the integration contour that encircles the small cut we are putting in. As the expressions of the filling fractions are not affected by the deformation, neither are the residues $\alpha (x)$.

If we now use the condition \eqref{PolePosition} to find the position of the cut, we get
\begin{equation}
    \tilde{p}_2 (x_n) - \tilde{p}_3 (x_n) =2\pi n  \quad
     \Longrightarrow \quad x_n= \frac{\omega \pm \sqrt{\omega^2 +n^2}}{n} \ .
\end{equation}
Here, we are only interested in the solution with the plus sign, as it is the one in the physical region. Substituting the explicit expression for $x_n$ in ${\alpha}(x_n)$,   we find that \eqref{eq:dapm-23-sphere} gives
\begin{equation}\label{eq:dapm-23-sphere-new}
2\delta \alpha_\pm = \frac{4\pi}{\sqrt{\lambda}} \sum_n N_n \left(\frac{1\pm x_n}{x_n^2-1} \right) =  \frac{4\pi}{\sqrt{\lambda}} \sum_n \frac{n N_n }{2 \omega} \left(\frac{n}{\omega - \sqrt{\omega^2 + n^2}}  \pm 1 \right) \ .
\end{equation}
The second term in this expression actually vanishes when we impose the level-matching condition $\sum_n n N_n =0$. This means that $\delta \alpha_+ = \delta \alpha_-$, although we will not make use of this relation in the following equations.

Now that we have computed $\delta \alpha_\pm$, we can consider the effect that the excitation has on the quasimomenta associated to the (deformed) $AdS$ part of our space and compute $\delta \Delta$. The ansätze for these corrections are a bit more involved, as we have to consider the possibility of a shift of the branch points arising from a backreaction of the excitation. Thus, we will assume the following ansätze:
\begin{equation}
	\delta \hat{p}_1 (x) = f(x) +\frac{g(x)}{K(1/x)}  \ , \qquad \delta \hat{p}_4 (x) = f(x) -\frac{g(x)}{K(1/x)} \ ,
\end{equation}
where $K(x)=\sqrt{ 1 -  x^2\beta^2}$ and the functions $f(x)$ and $g(x)$ are functions to be determined. The second term of our ansatz is inspired by the fact that $\partial_\beta K(x) \propto \frac{1}{K(x)}$ and  by  eq.~\eqref{eq:res-p2}, as the quasimomentum $\hat{p}_4$ still has to be the analytic continuation of $\hat{p}_1$ through a square-root cut.

Let us start by constraining the function $f(x)=\frac{\delta \hat{p}_1 (x) +\delta \hat{p}_4 (x)}{2}$. As we are considering a bosonic excitation, the synchronisation of the poles at $\pm 1$ forces $\delta \hat{p}_1 (x) $ and $\delta \hat{p}_4 (x) $ to have residues with opposite signs, meaning that $f(x)$ has no poles at $\pm 1$.  In addition,  $\delta \hat{p}_1 \approx -\delta \hat{p}_4 \approx \frac{4\pi}{x \sqrt{\lambda}} \frac{\delta \Delta}{2} + \mathcal{O} (x^{-2})$ for large values of $x$. As we have not added any new poles to the $AdS$ directions, this means that $f(x)$ is a holomorphic function that approaches zero at infinity. According to Liouville's theorem, the only function that fulfils these requirements is $f(x)=0$.

Now, we can fix the function $g(x)$ by demanding that $\delta \hat{p}_1$ has the correct properties. As $\delta \hat{p}_1 (x)$ only has poles at $\pm 1$, we propose the following ansatz for $g(x)$
\begin{equation}
    g(x)= \frac{K(1) \delta \alpha_+}{x-1} + \frac{K(1) \delta \alpha_-}{x+1} \ ,
\end{equation}
where we have used the synchronisation of the poles at $\pm 1$ as in eq.~\eqref{eq:poles-synch} to fix the residues. Matching the asymptotic behaviour of $\delta \hat{p}_1 (x)$ with its expected behaviour gives us
\begin{equation}
   \frac{\delta \Delta}{2}= \frac{\sqrt{\lambda}}{4\pi} \frac{K(1)}{K(0)} (\delta \alpha_+ + \delta \alpha_-) = \sqrt{1-\beta^2} \sum_n \frac{N_n}{x_n^2-1} \ .
\end{equation}
Notice that $\delta \Delta$ can only be real if $|\beta|\leq 1$, which is exactly what we found from reality conditions implied by the Virasoro constraint.

We should remark that this construction is blind to having the second pole in the sheet $\tilde{3}$ or the sheet $\tilde{4}$, as we did not use that information at any point of the reconstruction. This implies that the contributions to $\delta \Delta$ of $\tilde{2} \tilde{3}$ and $\tilde{2} \tilde{4}$ have the same form as a function of $x_n$. We will later use the inversion symmetry to show that the other two give us the same result.

\subsection{Excitations associated to the deformed $AdS$ space}

Excitations that connect two quasimomenta $\hat{p}_i$ and $\hat{p}_j$ are related to modes associated to the deformed $AdS$ space. Similarly to the case of the sphere, the only ones that give rise to physical excitations are those that connect either $\hat{1}$ or $\hat{2}$ with either $\hat{3}$ or $\hat{4}$. Due to the presence of the twist, the excitations associated to $\hat{1} \hat{4}$ and $\hat{2} \hat{3}$ behave differently to those associated to $\hat{1} \hat{3}$ and $\hat{2} \hat{4}$, and thus we will treat them separately. We will consider only one case for each respective set because the other one can be obtained through inversion symmetry, as we will discuss in \ref{a:composition}.

Let us  start with the case associated to $\hat{1} \hat{4}$ and turn off any other excitation. First, note that we are not modifying the quasimomenta associated to the sphere. This means that all $\delta \tilde{p}_i$ decay at infinity as ${\cal O}(x^{-2})$. The only possible configuration with this property is the one with $\delta \alpha_\pm =\delta \beta_\pm =0$, which allows us to set all the sphere corrections to zero,  $\delta \tilde{p}_i=0$. Consequently, none of the corrections $\delta \hat{p}_i$ have poles at $\pm 1$ due to their synchronisation.

Similarly to what happened for the excitation $\tilde{2} \tilde{3}$, the residues at $x_n$ of $\delta \hat{p}_1$ and $\delta \hat{p}_4$ have the opposite value. Furthermore, for large values of $x$, we have that $\delta \hat{p}_1 \approx -\delta \hat{p}_4 \approx \frac{4\pi}{x \sqrt{\lambda}} \frac{\delta \Delta + 2 \sum_n N_n}{2}  + \mathcal{O} (x^{-2})$. If we consider the ansatz
\begin{equation}
	\delta \hat{p}_1 (x) = f(x) +\frac{g(x)}{K(1/x)}  \ , \qquad \delta \hat{p}_4 (x) = f(x) -\frac{g(x)}{K(1/x)} \ ,
\end{equation}
these considerations imply that $f(x)=0$ and fix $g(x)$ to
\begin{equation}
    g(x)=\sum_n \frac{K(1/x_n) \alpha (x_n) N_n}{x-x_n} \ .
\end{equation}
Here we have used the insight from our sphere computation to argue that the residue at $x_n$ has to be the same as the undeformed one, so we can use the function $\alpha (x)$ we computed above  in \eqref{eq:residue}. Matching the asymptotic behaviour of our ansatz with its expected behaviour gives us
\begin{equation}
    \delta \Delta=  \sum_n \left(2\frac{x_n^2 K(1/x_n) }{x_n^2 -1 } -2 \right) N_n  \ .
\end{equation}

Let us consider now the case associated to $\hat{1} \hat{3}$ and turn off any other excitation. Again, we are not modifying  the quasimomenta associated to the sphere, implying that none of the corrections $\delta \hat{p}_i$ have poles at $\pm 1$. 

Similarly to our previous cases, we start by considering the ansatz
\begin{equation}
	\delta \hat{p}_1 (x) = f(x) +\frac{g(x)}{K(1/x)}  \ , \qquad \delta \hat{p}_4 (x) = f(x) -\frac{g(x)}{K(1/x)} \ .
\end{equation}
However, in contrast to previously, $\delta \hat{p}_1 (x)$ has a pole at $x_n$ but $\delta \hat{p}_4 (x)$ does not, meaning that the function $f(x)$ is not holomorphic and thus does not vanish in this case. Notice that, instead, $\delta \hat{p}_4 (x)$ has a pole at $1/x_n$ inherited from $\delta \hat{p}_3 (x)$ due to the inversion symmetry. In fact, assuming that $\delta \hat{p}_3 (x)$ has a pole at $x_n$ with residue $-\alpha (x_n)$,  inversion symmetry forces $\delta \hat{p}_4 (x)$ to have a pole at $1/x_n$ with residue $-\frac{\alpha (x_n)}{x_n^2}=\alpha \left( \frac{1}{x_n} \right)$.\footnote{Notice that, although $\delta \hat{p}_2$ and $\delta \hat{p}_4$ have a pole, they are not in the physical region and they do not contribute to $N_n^{\hat{2} \hat{4}}$.} Substituting this information into an ansatz of the form $f(x) = a + \frac{b}{x-x_n} + \frac{c}{x-1/x_n}$, we arrive to
\begin{equation}
    f(x)=\sum_n \frac{4\pi}{\sqrt{\lambda}} \frac{x N_n}{2 (x-x_n) (x-1/x_n)} \ ,
\end{equation}
which fulfils all our requirements, including consistency with the asymptotic behaviour of $\delta \hat{p}_1 (x)$ and $\delta \hat{p}_4 (x)$, which reads $f(x) \approx \sum_n  \frac{4\pi}{x \sqrt{\lambda}} \frac{N_n}{2} +\mathcal{O} (x^{-2})$.
Similarly to the function $f(x)$, we can assume that $g(x)$ can be written as $g(x) = a + \frac{b}{x-x_n} + \frac{c}{x-1/x_n}$. These constants can be fixed by demanding that $\delta \hat{p}_1$ has the properties we want, i.e.~it has a pole with residue $\alpha (x_n)$ at $x_n$ and no pole at $1/x_n$. After some algebra, we get that
\begin{equation}
    g(x) = \sum_n \frac{4\pi}{\sqrt{\lambda}} \frac{K(x_n) (x-x_n) +x_n^2 K(1/x_n) (x-1/x_n)}{2(x_n^2 -1) (x-x_n) (x-1/x_n)} N_n \ .
\end{equation}
Matching the asymptotic of our ansatz with the expected asymptotic behaviour of $\delta \hat{p}_1$ gives us
\begin{equation}
    \delta \Delta= \sum_n \left( \frac{x_n^2 K(1/x_n) + K(x_n)}{x_n^2-1}-1 \right) N_n \ .
\end{equation}

\subsection{Fermionic excitations}

Let us move finally to fermionic excitations. Excitations that connect a quasimomentum $\hat{p}_i$ and a quasimomentum $\tilde{p}_j$ or vice-versa are related to fermionic modes. Here, we will focus only on the excitation associated to $\hat{1} \tilde{3}$ and turn off all other excitations. We will make some comments on the remaining excitations at the end of this section.

We should first focus on the corrections to the quasimomenta associated to the sphere. As we are adding an excitation to the sheet $\tilde{3}$, we will consider the following ansatz:
\begin{align}
	\delta \tilde{p}_3 (x) &=\frac{\delta \beta_+}{x-1} + \frac{\delta \beta_-}{x+1} + \sum_n \frac{\alpha (x_n) N_n}{x- x_n}=-\delta \tilde{p}_4 (1/x) \ , \\
	\delta \tilde{p}_1 (x) &= \delta  \tilde{p}_2 (x) =0 \ .
\end{align}
Notice that, as we are considering a fermionic excitation, these corrections fulfil the relaxed synchronisation condition \eqref{eq:poles-synch}. If we impose the  correct value of the residues and the correct asymptotic behaviour, we get exactly the same equations as the ones for the $\tilde{2}\tilde{3}$ excitations (up to replacing $\delta\alpha_\pm $ with $\delta\beta_\pm$ and $N_n$ with $-N_n$). Thus, we can borrow the results we found for $\delta \beta_\pm$ and the residue $\alpha (x_n)$ given in \eqref{eq:residue} and \eqref{eq:dapm-23-sphere-new}.

For the correction to the $AdS$ quasimomenta, we will assume the same ansatz as in the previous cases. Similarly to the case $\hat{1} \hat{3}$, we cannot set the function $f(x)$  to zero because $\delta \hat{p}_1$ has a pole at $x_n$ but $\delta \hat{p}_4$ does not. In addition, $\delta \hat{p}_4$ has poles at $\pm 1$ from the synchronisation condition, while $\delta \hat{p}_1$ does not. After imposing these restrictions, as well as the required asymptotic behaviour, we get
\begin{equation}
	f(x)=\sum_n \frac{4\pi}{\sqrt{\lambda}} \frac{x^2 N_n}{2(x-x_n) (x^2-1)} \ .
\end{equation}
With this information, we can now compute the function $g(x)$ by demanding that $\delta \hat{p}_1 (x)$ has the correct residues at $\pm 1$ and $x_n$, which gives us
\begin{equation}
    g(x)=\sum_n \frac{4\pi}{\sqrt{\lambda}} \frac{x_n^2 (x^2-1) K(1/x_n) +(x^2 - x_n^2) K(1)}{2(x-x_n) (x^2-1) (x_n^2 -1)} N_n \ .
\end{equation}
Finally, matching $\delta \hat{p}_1 (x)$ with its required asymptotic behaviour, we obtain that
\begin{equation}
	\delta \Delta = \sum_n \left( \frac{x_n^2 K(1/x_n) +K(1)}{x_n^2-1} -1 \right) N_n \ .
\end{equation}

In the next section we will discuss how to obtain the contribution of $\hat{1} \tilde{4}$ from this one, but it is easy to see that the computation has to be exactly the same as the one presented above, giving us the same $\delta \Delta$. The same happens for the excitations $\tilde{1} \hat{4}$ and $\tilde{2} \hat{4}$.

The steps to compute the contribution of $\hat{2} \tilde{3}$ are relatively similar. In fact, by analysing the pole structure, it is easy to reach the conclusion that the expression of $\delta \hat{p}_2$ is exactly the same one as the expression of $\delta \hat{p}_1$ for the $\hat{1} \tilde{3}$ excitation after substituting $K(1/x)$ by $K(x)$ and adding a constant contribution to $g(x)$. Applying the inversion symmetry to get $\delta \hat{p}_1$ and matching it with its required asymptotic behaviour, we obtain that
\begin{equation}
	\delta \Delta = \sum_n \left( \frac{K(x_n) +K(1)}{x_n^2-1}  \right) N_n \ .
\end{equation}

Although at this point we would have to study also the fermionic excitations that involve either $\hat{3}$ or $\hat{4}$, we can argue that this is not necessary. This happens because our classical solution has pairwise symmetric quasimomenta
\begin{align}
    \hat{p}_1 &= -\hat{p}_4 \ , & \hat{p}_2 &= -\hat{p}_3 \ , & \tilde{p}_1 &= -\tilde{p}_4 \ , & \tilde{p}_2 &= -\tilde{p}_3 \ . \label{symmetricquasimomenta}
\end{align}
This implies that, after all the appropriate computations, we will find that
\begin{align}
    \Omega^{\hat{1} \tilde{4}} & =\Omega^{\tilde{1} \hat{4}} \ , & \Omega^{\hat{2} \tilde{4}} & =\Omega^{\tilde{1} \hat{3}} \ , & \Omega^{\hat{1} \tilde{3}} & =\Omega^{\tilde{2} \hat{4}} \ , & \Omega^{\hat{2} \tilde{3}} & =\Omega^{\tilde{2} \hat{3}} \ ,
\end{align}
both for $\Omega$ understood as a function of the position of the poles and as a function of the mode number $n$.

\subsection{Composition and inversion}\label{a:composition}

Up to this point, we have computed $\delta \Delta$ for some selected excitations. This is enough for us, as the contribution of the other excitations can be computed using properties such as composition of poles and the inversion property.

\begin{figure}
\centering
\includegraphics[width=0.5\textwidth]{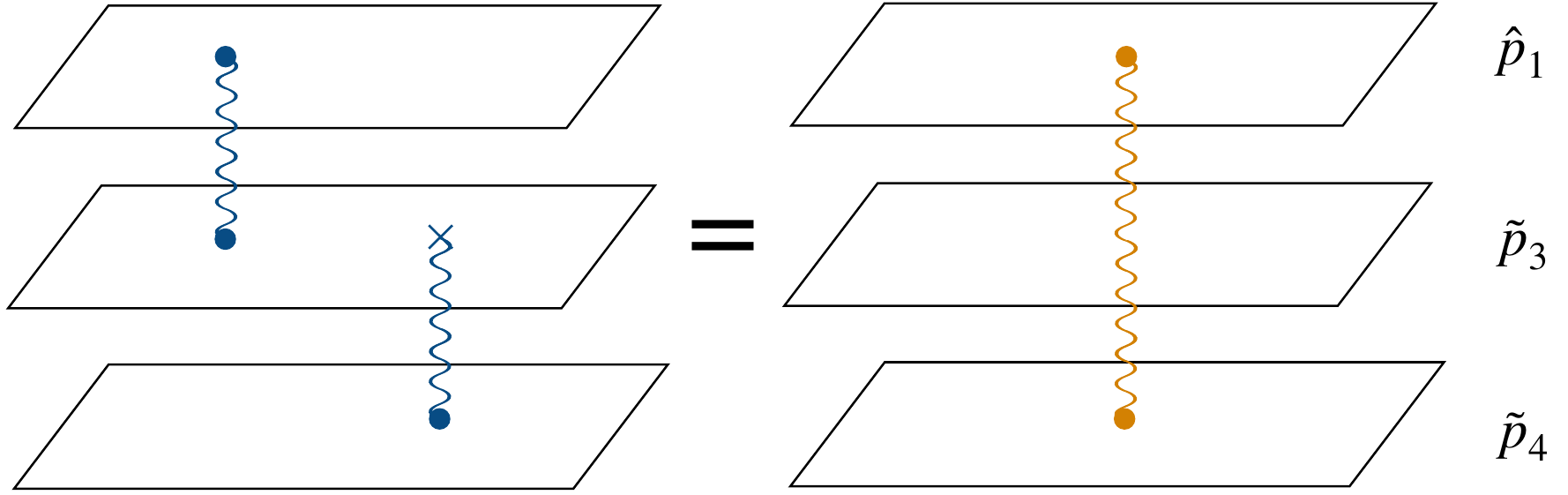}
\caption{\label{fig:composition}\footnotesize 
Illustration of the composition rule \eqref{eq:composition}. The fermionic frequency $\Omega^{\hat{1} \tilde{3}} (x) $ and the bosonic frequency $\Omega^{\tilde{3} \tilde{4}} (y)$ can be composed, because the two excitations have a pole with opposite residue on sheet $\tilde{p}_3$. In particular, the residue on sheet $\tilde{p}_3$ vanishes in the limit $y\rightarrow x$, giving us the pole structure (and asymptotic behaviour) associated with the fermionic excitation $\hat{1} \tilde{4}$. Thus, $\Omega^{\hat{1} \tilde{3}} (x) +\Omega^{\tilde{3} \tilde{4}} (x)=\Omega^{\hat{1} \tilde{4}} (x)$  off-shell.}
\end{figure}

The idea is that the procedure we used to compute $\delta \Delta$ does not care about the exact position $x_n$ of the poles/microscopic cuts we are adding. Before substituting the expressions of $x_n$ obtained from \eqref{PolePosition} in the frequency of the corresponding excitation, we can thus think of $\Omega(x)$ as formal, off-shell, functions of $x$. Once we substitute $x=x_n$ the frequency becomes on-shell. We can use the off-shell expressions of $\Omega(x)$ in our advantage in two different ways. The first one is the inversion symmetry: if we consider an excitation connecting a given pair of sheets, we can move the pole associated to  it to the interior of the unit circle, making it non-physical, and a new physical pole will emerge in the sheets connected to the original ones by inversion symmetry. The second one is composing two excitations that share a pole with opposite residue on the same sheet, say $j$, but connect to different sheets, say $i$ and $k$ respectively. The pole on  sheet $j$ will cancel, making the composition of the two excitations $ij$ and $jk$ indistinguishable from having only one excitation $ik$. We illustrate the latter in figure~\ref{fig:composition}. In mathematical terms, these two ideas imply
\begin{align}
	\Omega^{\hat{1} \hat{4}} (x) &= -\Omega^{\hat{2} \hat{3}} \left( \frac{1}{x} \right) -2 \ , \\
	\Omega^{\tilde{1} \tilde{4}} (x) &= -\Omega^{\tilde{2} \tilde{3}} \left( \frac{1}{x} \right)  +\Omega^{\tilde{2} \tilde{3}} (0) \ , 
\end{align}
and 
\begin{equation} \label{eq:composition}
	\Omega^{ij} (x) \pm \Omega^{jk} (x) = \Omega^{ik} (x) \ ,
\end{equation}
where the $\pm$ is chosen appropriately to cancel the residue in sheet $j$. We will not reproduce here the derivation of these expressions, which can be found in \cite{Gromov:2008ec}.
Let us stress again that these relations between the different contributions only hold when they are understood as functions of $x$, not as functions of $n$.

Let us now compute some of the remaining frequencies using the ones we obtained above. The process is relatively similar for most of them, so we will consider only some specific cases: the excitations $\hat{2} \hat{3}$, $\tilde{1} \tilde{4}$ and $\hat{1} \tilde{4}$. 
Both $\hat{2} \hat{3}$ and $\tilde{1} \tilde{4}$  can be obtained by immediate application of the above formulas
\begin{align}
    &\Omega^{\hat{2} \hat{3}} (x_n) = -\Omega^{\hat{1} \hat{4}} \left( \frac{1}{x_n} \right) -2= \left( -\frac{2 K(x_n)}{1-x_n^2} +2 \right) -2 = \frac{2 K(x_n)}{x_n^2 - 1}   \ , \\
    &\Omega^{\tilde{1} \tilde{4}} (x_n) = -\Omega^{\tilde{2} \tilde{3}} \left( \frac{1}{x_n} \right)  +\Omega^{\tilde{2} \tilde{3}} (0)= -2 \frac{\sqrt{1-\beta^2}}{x_n^{-2}-1} -2 \sqrt{1-\beta^2}= 2 \frac{\sqrt{1-\beta^2}}{x_n^2-1} \ . 
\end{align}
The last one is obtained by considering first the composition of the excitations $\tilde{2} \tilde{3}$ and $\tilde{2} \tilde{4}$. This gives us that the excitation $\tilde{3} \tilde{4}$ has $\Omega^{\tilde{3} \tilde{4}}=0$, which is consistent with the fact that it is an unphysical excitation. This result can then be used to show, e.g., that $\Omega^{\hat{1} \tilde{4}}=\Omega^{\hat{1} \tilde{3}}+\Omega^{\tilde{3} \tilde{4}}=\Omega^{\hat{1} \tilde{3}}$.

\section{Bosonic sector of the quadratic fluctuations} \label{QuadraticFluctuations}

In order to cross-check the results we obtained using the classical spectral curve method, we can compute the contribution of the bosonic excitations to the one-loop correction to the energy by considering the effective Lagrangian of small fluctuations around the classical solution  we are interested in. We will perform this computation in the deformed periodic picture, rather than the undeformed twisted one. The expansion of the deformed $AdS_5$ sector of the Lagrangian is given by\footnote{The term $\epsilon(\text{EOM})$ in the expansion vanishes upon the equations of motion (EOM).}
\begin{equation}
\begin{aligned}
	&\mathcal{L} (x_{\text{clas.}} + \epsilon x) \approx \mathcal{L} (x_{\text{clas.}}) + \epsilon (\text{EOM}) + \frac{\epsilon^2}{2 b_Z^6} \left[ 4 b_Z^4 (\dot{P}^2 -P^{\prime 2} + \dot{Z}^4 - Z^{\prime 2}) \right. \\ & -(4b_Z^6 + \eta^2 b_Z^2) (\dot{T}^2 - T^{\prime 2}) -8 b_Z^4 (\dot{T} \dot{V} - T' V')+4 a_T b_Z Z (\eta^2 \dot{T} +4 b_Z^2 \dot{V} + 3 \eta Z') \\ &\left. -a_T^2 (4b_Z^4 + \eta^2) P^2 -4 a_T^2 \eta^2 Z^2   - 4 \eta a_T b_Z^2 P P' + 4 b_Z \eta ( T' \dot{Z} - \dot{T} Z' ) \right]+ \mathcal{O} (\epsilon^3)  \ ,
\end{aligned}
\end{equation}
where, by a slight abuse of notation, we have denoted the fluctuation around the coordinates with the same symbols as the coordinates. Here, the dot and prime represent derivatives with respect to $\tau$ and $\sigma$, respectively. Notice that  neither the coordinate $\Theta$ nor the Kalb-Ramond field terms contribute at quadratic order. The fact that $\Theta$ does not appear is simply a consequence of the non-canonical kinetic term that it has, which would appear at higher orders in the field expansion.\footnote{If we want to do the counting of degrees of freedom correctly, we should change from $P$ and $\Theta$ to Cartesian coordinates $X^2,X^3$. However, as we are only interested in independently checking our results in appendix \ref{StructureFluctuations}, this computation is sufficient.}

The next step is to compute the equations of motion of the fluctuations. We will use the fact that we do not want to spoil the periodicity condition of our classical solution by considering the following ansätze for each of the coordinates:
\begin{equation}
	X^m = \sum_n A_m \cos (\Omega_n \tau + n \sigma) + B_m \sin (\Omega_n \tau + n \sigma) \ .
\end{equation}
After some algebra, the equations of motion become the  system of linear equations $Mv=0$ with $v$ the vector given by $v=(A_Z \
A_T \
A_V \
A_P \
B_Z \
B_T\
B_V \
B_P)$ and $M$ is the matrix
\begin{equation} \scalebox{0.5}{\ensuremath{
\begin{pmatrix}
	\sqrt{\frac{\eta^3}{2 \beta^3}} (n^2 + 4 a_T^2 \beta^2 -\Omega_n^2) & 0 & 0 & 0 & 0 & -a_T \eta^2 \Omega_n & -\frac{2 a_T \eta \Omega_n}{\beta} & 0 \\
	0 & -\frac{1+\beta^2}{\beta^2} \eta^2 (n^2-\Omega_n^2) & -\frac{2\eta}{\beta} (n^2 - \Omega_n^2) & 0 & \sqrt{2^3 \eta^3 \beta} a_T \Omega_n & 0 & 0 & 0 \\
	0 & -(n^2-\Omega_n^2) & 0 & 0 & \sqrt{\frac{2^3 \beta}{\eta}} a_T \Omega_n & 0 & 0 & 0 \\
	0 & 0 & 0 & \sqrt{\frac{2 \eta^3}{\beta^3}} (a_T^2 (1+ \beta^2) +n^2 - \Omega_n^2) & 0 & 0 & 0 & 0 \\
	0 & a_t \eta^2 \Omega_n & \frac{2 a_T \eta \Omega_n}{\beta} & 0 & \sqrt{\frac{\eta^3}{2 \beta^3}} (n^2 + 4 a_T^2 \beta^2 -\Omega_n^2) & 0 & 0 & 0 \\
	-\sqrt{2^3 \eta^3 \beta} a_T \Omega_n & 0 & 0 & 0 & 0 & -\frac{1+\beta^2}{\beta^2} \eta^2 (n^2-\Omega_n^2) & -\frac{2\eta}{\beta} (n^2 - \Omega_n^2) & 0 \\
	-\sqrt{\frac{2^3 \beta}{\eta}} a_T \Omega_n & 0 & 0 & 0 & 0 & -(n^2-\Omega_n^2) & 0 & 0 \\
	0 & 0 & 0& 0 & 0 & 0 & 0 & \sqrt{\frac{2 \eta^3}{\beta^3}} (a_T^2 (1+ \beta^2) +n^2 - \Omega_n^2)
\end{pmatrix} }} \ .
\end{equation}
If we want this system of equations to have a non-trivial solution, we need the matrix of coefficients to have vanishing determinant. This gives us the condition
\begin{equation}
	(n^2-\Omega_n^2)^2 (n^2 +a_T^2 (1 + \beta^2 )-\Omega_n^2 )^2 (n^4 -4 a_T^2 \beta^2 n^2 -2(2a_T^2 +n^2) \Omega_n^2 +\Omega_n^4 )^2=0 \ .
\end{equation}
We can check that the solutions for $\Omega_n$ to this equation, divided by $Q_T=-a_T$, perfectly match (\ref{AdSFrec1}), (\ref{AdSFrec2}) and (\ref{AdSFrec3}) up to a constant term. Additionally, we also get solutions of the form $\Omega_n=\pm n$. The modes associated to these are not physical and are cancelled by conformal ghosts \cite{Park:2005ji}.

Instead of repeating the same computation for the $S^5$ sector of the Lagrangian, we can argue that this part of the space is blind to the deformation and hence the final result has to be the same as the one for the undeformed background. In fact, we can check that
\begin{equation}
    \Omega = \frac{ \omega \pm \sqrt{\omega^2+n^2}}{Q_T} = -\sqrt{1 - \beta^2} \mp \sqrt{1 - \beta^2 +\frac{n^2}{a_T^2}} \ ,
\end{equation}
which perfectly matches (\ref{SFrec}).

The mismatch by a constant between the contributions computed using the quadratic fluctuations and the ones computed using the classical spectral curve may be uncomfortable. However, they are also present for undeformed $AdS_5 \times S^5$ and arise due to each method describing the perturbations around the classical solution using a different frame of reference \cite{Gromov:2007aq}.

Even though these shifts have a non-physical origin, this does not mean at all that they are harmless, as they can give rise to ambiguities. To discuss those ambiguities, first we have to distinguish between two kinds of shifts that may appear: a constant shift of $\Omega$ (usually proportional to the energy or an angular momentum), and a shift of the mode number $n$ (usually proportional to a winding number). The ones of the first kind are completely harmless, but the ones of the second kind give rise to an ambiguity. To see that, we can compare what would be the contribution to the one-loop energy we obtain with and without performing a shift of the mode number. Although the sum $\sum_{n=-\infty}^\infty \left[ 2 \Omega (x_n) - \Omega (x_{n-m}) - \Omega (x_{n+m}) \right]$ might initially seem to vanish because the terms cancel each other after a relabelling, this is not entirely correct if we consider a partial sum up to a large enough value, $\Lambda$. Using that $\Omega (x_n) \approx n$ for large $n$ we see that, up to subleading orders in $\Lambda$,
\begin{equation}
    \sum_{n=-\Lambda}^\Lambda \left[ 2 \Omega (x_n) - \Omega (x_{n-m}) - \Omega (x_{n+m}) \right] = 2\sum_{n=\Lambda - m +1}^\Lambda \left[ \Omega (x_n) - \Omega (x_{n-m}) \right] \approx 2m \ .
\end{equation}
Luckily, we do not have shifts of this second kind in our problem. In fact,  they are usually associated to winding in the classical solution, which ours does not possess. We therefore conclude that these shifts do not create an ambiguity in our final result for $E_{1-\text{loop}}$.

%%%%%%%%%%%%%%%%%%%%%%%%%%%%%

\bibliographystyle{nb}
\bibliography{biblio}{}

\end{document}